\documentclass[aps,superscriptaddress,amsmath,amssymb,eqsecnum,nofootinbib,11pt]{revtex4-1}
\allowdisplaybreaks
\usepackage{graphicx}
\usepackage{epstopdf}
\usepackage{epsfig}
\usepackage{amssymb}
\usepackage{mathrsfs}
\usepackage{epstopdf}
\usepackage{subfigure}
\usepackage{bm}
\usepackage{float}
\usepackage{ulem}

\begin{document}

\title{{\bf Lecture notes on the Skyrme model}}

\author{Yong-Liang Ma}
\email{ yongliangma@jlu.edu.cn}
\affiliation{Center for Theoretical Physics and College of Physics, Jilin University, Changchun, 130012, China}

\author{Masayasu Harada}
\email{ harada@hken.phys.nagoya-u.ac.jp}
\affiliation{Department of Physics, Nagoya University, Nagoya, 464-8602, Japan.}

\date{\today}
\begin{abstract}
This lecture provides a pedagogical instruction to the basic concepts of the Skyrme model and its some applications. As the preliminary for understanding the Skyrme model, we first briefly explain the large $N_c$ expansion, chiral symmetry and its breaking. Next we give a brief review of nonlinear sigma model including the power counting scheme of the chiral perturbation theory, starting from the linear sigma model. We then give an exhaust explanation of the Skyrme model and its applications. After the presentation of the Skyrme model for baryons in free space, we introduce how to study the baryonic matter and medium modified hadron properties by using the Skyrme model. Finally we discuss a way to incorporate the lowest-lying vector mesons into the Skyrme model based on the hidden local symmetry. Some possible further developments are also covered.
\end{abstract}

\maketitle

\newpage

\tableofcontents

\newpage

\section{Motivation}

\label{sec:mot}

Now, it is believed that quantum chromodynamics (QCD) is a fundamental theory of the strong interaction. In QCD, the strong interaction is described by an ${\rm SU}(3)_c$ gauge theory of quarks and gluons with the Lagrangian
\begin{eqnarray}
{\cal L}_{\rm QCD} & = & \sum_{l}\left\{\bar{q}_{l,\alpha}
i \left(\partial_\mu \gamma^\mu -
ig_sA_\mu^aT^a\gamma^\mu\right)_{\alpha\beta}q_{l,\beta}-m_l\bar{q}_lq_l
\right\} - {1\over4}G^a_{\mu\nu}G^{a\mu\nu}, \label{eq:lagrQCD}
\end{eqnarray}
where $T^a$ is the generator of ${\rm SU}(3)_c$ group satisfying ${\rm Tr}(T^a T^b)=(1/2)\delta^{ab}$, the subscripts $\alpha, \beta$ stand for the color indices, the summation is over the flavor index $l$, $G_{\mu\nu}^a=\partial_\mu A_\nu^a-\partial_\nu A^a_\nu-ig_sf_{abc}A_\mu^bA_\nu^c$ is the field strength tensor of gluon fields, $g_s$ is the strong coupling constant and $f_{abc}$ is the structure constant of ${\rm SU}(3)_c$ group. In QCD, the fundamental parameters are the coupling constant $g_s$ (or $\alpha_s = g^2_s/4\pi$ ) and the current quark masses $m_l$.

Theoretically, it is proved that the coupling constant $\alpha_s$ satisfies the following renormalization group equation at one-loop order~\cite{Politzer:1973fx,Gross:1973id,Gross:1973ju}
\begin{eqnarray}
\mu_R^2 \frac{d \alpha_s}{d\mu_R^2} & = & \beta(\alpha_s) ={} -
\frac{\left(33 - 2n_f\right)}{12\pi}\alpha_s^2, \label{eq:qcdrge}
\end{eqnarray}
where $n_f$ is the number of flavors. This implies that in case of $33 - 2n_f > 0$, i.e., $n_f < 33/2$, which is satisfied by the present observation $n_f = 6$, $\alpha_s$ decreases with the increase of energy transfer, that is, QCD is an asymptotically free theory.

Because of the asymptotic freedom, the coupling constant $g_s$ is large at low energy scale so that one cannot analytically calculate the low energy strong processes using the standard perturbative theory of QCD where the coupling constant $g_s$ is regarded as an expansion parameter. In addition, although the numerical simulation of hadron properties based on lattice technique got great progresses recently, it suffers from the notorious sign problem when one attempts to apply this technique to the chiral symmetry restoration region in the dense system. So that, to study the low energy processes of strong interaction, some effective theories or models are necessary. This Lecture is devoted to introduce one of the effective models for baryons, the Skyrme model~\cite{Skyrme:1961vq}. The advantageous of Skyrme model which we will learn in the lecture is that, using this model, we can describe the meson and baryon dynamics in free space, the baryonic matter and also medium modified hadron properties~\cite{Lee:2003aq} in a unified manner.

The Skyrme model is constructed based on a nonlinear mesonic theory possessing a non-trivial topological field configuration (soliton) which can be identified with baryon. This model was proposed more than 50 years ago by T.H.R. Skyrme~\cite{Skyrme:1961vq}. However it was not taken seriously until the compelling arguments proposed by E. Witten~\cite{Witten:1979kh} which combined the 't Hooft large $N_c$ expansion~\cite{'tHooft:1973jz} with current algebra, and from then on the Skyrme's theory was widely applied in baryon and baryonic matter physics (for reviews, see~\cite{Zahed:1986qz,Holzwarth:1985rb,Balachandran:1985pa,BReditor}).

This lecture note is organized as follows:

Considering the importance of the large $N_c$ expansion in understanding the intrinsic charactors of Skyrme model and chiral symmetry of QCD in the study of low energy hadron physics, we will briefly discuss them in Chapter~\ref{sec:prelim}.

In Chapter~\ref{sec:nonlinearsigma}, starting from the linear sigma model we discuss the nonlinear realization of the chiral symmetry. The power counting mechanism of the chiral perturbation theory (ChPT) is briefly introduced. Following this, we discuss the topology of the nonlinear sigma model and show that a non-trivial topological configuration of the nonlinear sigma model has the following properties shared by baryons in the large $N_c$ limit:
\begin{enumerate}
  \item It carries a conserved topological charge which can be identified with the conserved baryon number in QCD.

  \item It is a heavy object and interacts strongly with another configuration. These properties are consistent with the qualitative argument of baryon properties based on the Large $N_c$ expansion of QCD.

  \item It yields a rich quantum sector by collectively quantizing the static soliton so one can identify the quantized soliton with baryon.

\end{enumerate}

In chapter \ref{sec:skyrmodel} we discuss the Skyrme model and its applications. The basics of the Skyrme model, such as its static solution and quantization, is involved. After this, we discuss the applications of the Skyrme model to the baryon properties. The main points in the calculations of the axial coupling, charge radii and magnetic moment of baryons are briefly explained.

We discuss in chapter \ref{sec:matter} the applications of Skyrme model to nuclear physics starting from the exploration of the two-body nuclear force by using the product ansatz of skyrmions which is proper when the two interacting skyrmions are far away from each other. This exploration tells us how to arrange the nearest skyrmions to get the strongest attractive interaction. By putting the skyrmion onto the crystal lattice we investigate the nuclear matter properties by regarding the skymion matter as nuclear matter~\cite{Klebanov:1985qi}. We discuss three kinds of crystal structures used so far in the literature in the study of nuclear matter, i.e., cubic crystal, body-centered cubic crystal and face-centered cubic crystal. As a typical example, we explicitly show how the nuclear matter properties and the medium modified hadron properties could be explored based on the face-centered cubic crystal.

Nuclear physics tells us that the vector mesons are crucial for understanding nuclear force, we discuss this aspect in Chaper~\ref{sec:skyrhls}. In this chapter, we first introduce the basic of a chiral effective model of vector mesons, the hidden local symmetry (HLS) approach~\cite{Bando:1984ej,Bando:1987br,Harada:2003jx}. Then we study the skyrmion properties, such as soliton mass and moment of inertia, including the vector meson effect from the leading order HLS Lagrangian.

The last chapter is for a discussion of the recent developments and remarks of some possible further applications of the Skyrme model.

\newpage

\section{Preliminary}

\label{sec:prelim}

In this chapter, we give the preliminary of our lecture: (1) Large $N_c$ expansion of QCD and (2) the chiral symmetry and chiral symmetry breaking in QCD.

\subsection{Large $N_c$ expansion}

\label{subsec:largeNc}

As pointed in the last chapter, skyrmion and baryon share several properties in the sense of large $N_c$ limit. This motivates us to start this lecture from a discussion of the main idea of large $N_c$ expansion~\cite{'tHooft:1973jz}. Following Ref.~\cite{Witten:1979kh}, after an explanation of the basic idea of the large $N_c$ expansion we will discuss baryon properties from the large $N_c$ expansion which is
essential for understanding the baryon dynamics by using the Skyrme model.

\subsubsection{The idea of large $N_c$ expansion}

The essential point in the large $N_c$ expansion is to generalize the number of colors from $3$ to $N_c$ and regard the latter as a parameter in the gauge theory~\cite{'tHooft:1973jz,Witten:1979kh}. Then, in the case of very large $N_c$, in a Feynman diagram, the number of possible intermediate states carrying different colors may be so large that the summing over the possible intermediate states gives rise to a large combinational factor. This combinational factor is responsible for the nature of the large $N_c$ expansion and the expansion is a power series in $1/N_c$.

\begin{figure*}[htbp]
\includegraphics[scale=0.5]{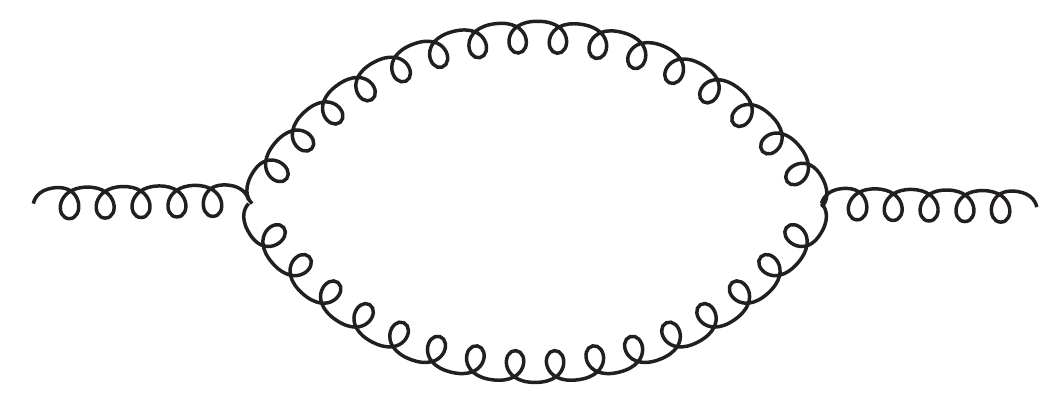}
\caption[]{Diagram contributing to the gluon vacuum polarization.} \label{fig:gluonvp}
\end{figure*}

To illustrate how $N_c$ enters a Feynman diagram, we consider a diagram contributing to the vacuum polarization of gluon depicted in Fig.~\ref{fig:gluonvp}.

From QCD Lagrangian \eqref{eq:lagrQCD} one concludes that this diagram is of the quadratic order of the QCD coupling constant $g_s$, i.e., $O(g_s^2)$. In addition, with respect to the structure constant of the non-Abelian gauge group SU($N_c$), this diagram is easily deduced to be $\propto N_c$. Therefore, we finally conclude that Fig.~\ref{fig:gluonvp} scales as $g_s^2 N_c$. This tells us that we should consider the large $N_c$ limit with $g_s^2N_c$ fixed, otherwise, the corrections to the gluon propagator would diverge for a large number of colors and this divergence hinders us to construct a self-consistent QCD theory for a large number of colors. With respect to the above argument, one may define an effective coupling constant $g_{\rm eff}$ which is a smooth function of $N_c$, i.e., $O(N_c^0)$, and relates to QCD coupling constant through
\begin{eqnarray}
g_s & = & \frac{g_{\rm eff}}{\sqrt{N_c}}.
\end{eqnarray}
Since $g_{\rm eff} \sim O(N_c^0)$, when the number of colors $N_c$ is large, QCD becomes a weakly coupled gauge theory. In terms of $g_{\rm eff}$, each QCD vertex receives a combinational factor $1/\sqrt{N_c}$. Therefore the Feynman diagram which survives under $N_c \to \infty$ limit must have a large combinational factor to compensate the factor $1/\sqrt{N_c}$ coming from QCD vertex.

\begin{figure*}[htbp]
\includegraphics[scale=0.7]{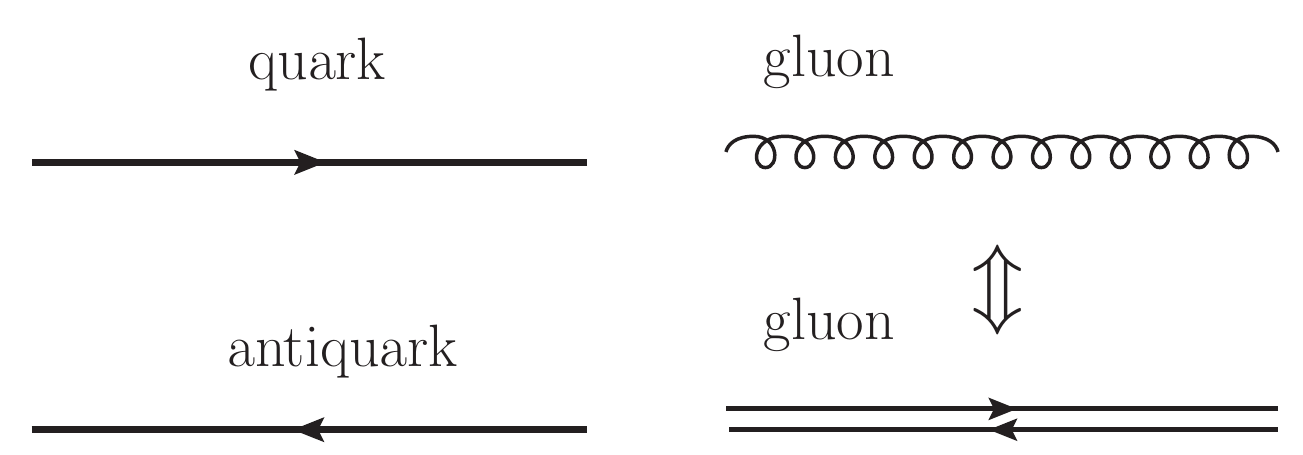}
\caption[Double-line notations for quark and gluon fields.]{%
The double-line notations for quark and gluon fields.} \label{fig:doublelgluon}
\end{figure*}

The combinational factor can be easily calculated by using the double-line notations for the quark and gluon fields~\cite{'tHooft:1973jz}. In quantum field theory, we represent a quark field with color index $i$, $q^i$, by an arrowed line and an anti-quark field with color index $i$, $\bar{q}_i$, by an arrowed line with arrow direction opposite to that of the quark field. Now, we explicitly write down the color indices of the gluon field $(A_\mu)^i_j = (A_\mu ^a T^a)^i_j$. Then we can think the gluon field as a quark-antiquark field $q^i\bar{q}_j$ which suggests that, similarly as the representing of quark propagator with an arrowed line, we could represent the gluon propagator as a doubly arrowed line with one carrying color index and the other carrying anticolor index. These line expressions can be illustrated by Fig.~\ref{fig:doublelgluon}.

\begin{figure*}[htbp]
\includegraphics[scale=0.5]{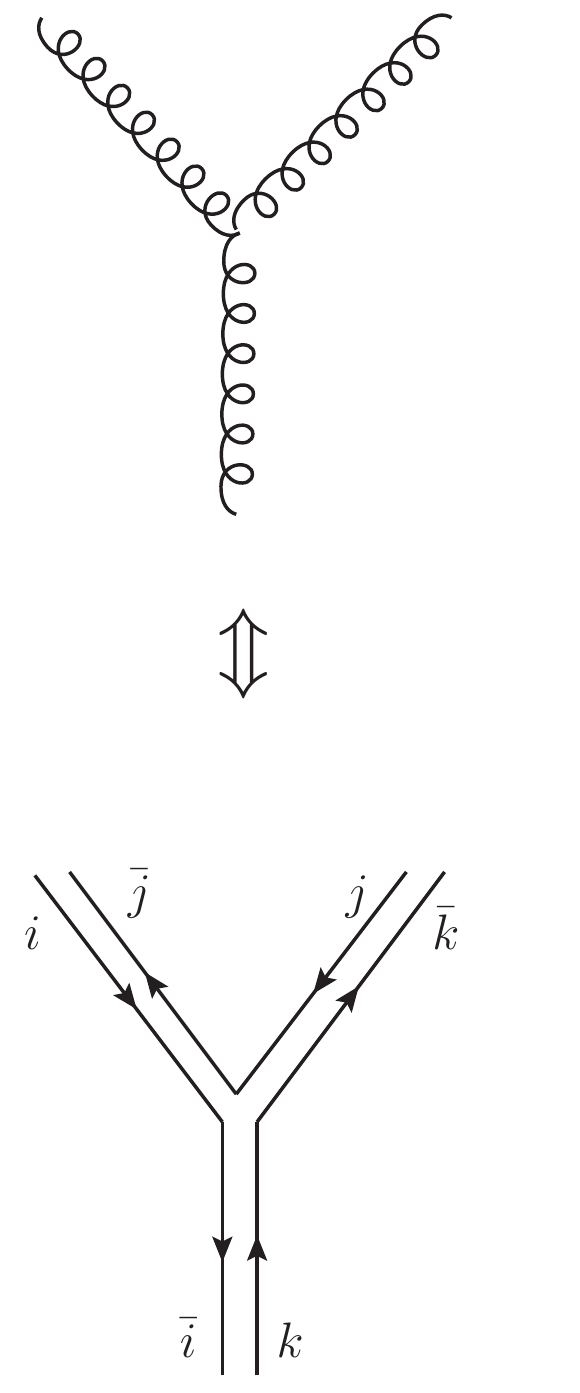}\includegraphics[scale=0.5]{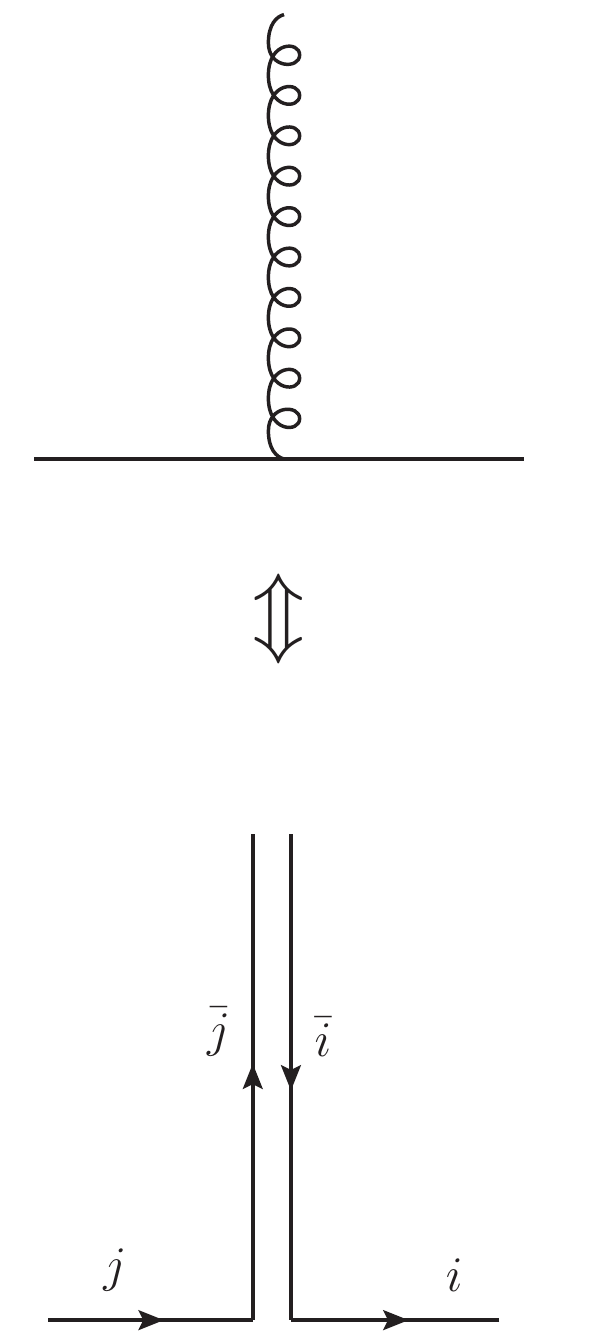}\includegraphics[scale=0.5]{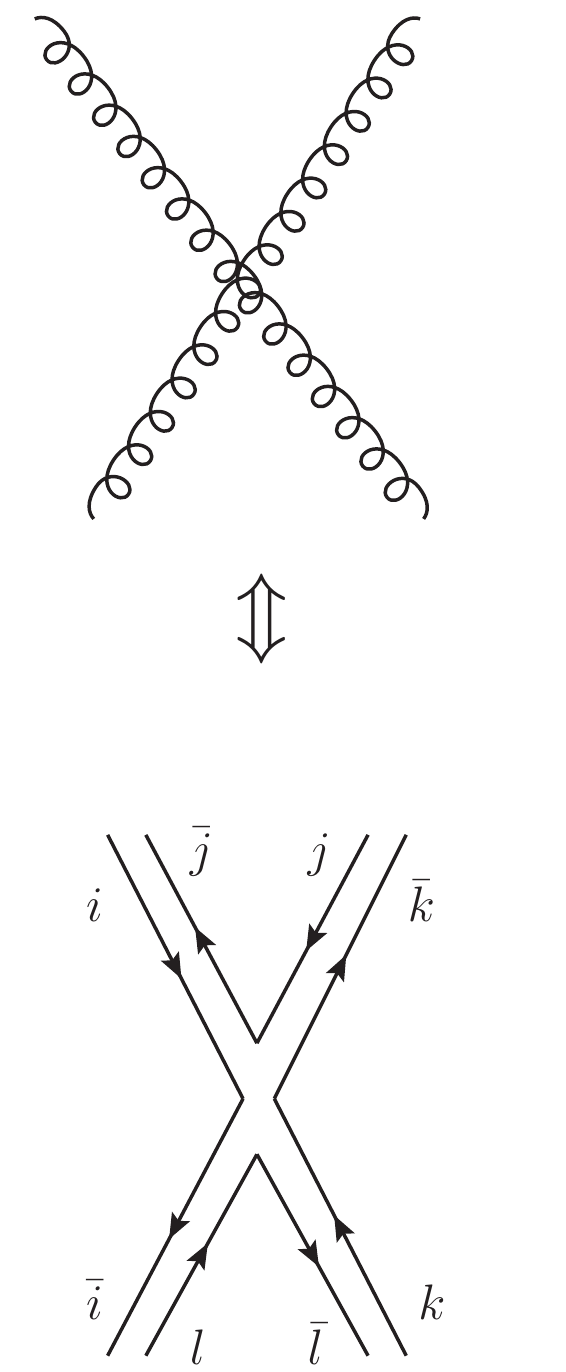}
\caption[]{%
Double-line notation of the QCD interaction.}
\label{fig:doublelvertex}
\end{figure*}

With the double-line notation of gluon fields, one can express the QCD interaction vertex
\begin{eqnarray}
{\rm Tr}A_\mu A_\nu \partial_\mu A_\nu & = & A_{\mu;j}^i A_{\nu;k}^j
\partial_\mu A_{\nu;i}^k \nonumber \\
\bar{q}\gamma_\mu A_\mu q & = & \bar{q}_i\gamma_\mu q^j A_{\mu;j}^i \nonumber \\
{\rm Tr}A_\mu A_\nu A_\mu A_\nu & = & A_{\mu;j}^i A_{\nu;k}^j
A_{\mu;l}^k A_{\nu;i}^l
\end{eqnarray}
as in terms of Fig.~\ref{fig:doublelvertex}. And in this notation, the color conservation is simply expressed by the fact that each color line that enters the diagram also leaves it.

\begin{figure}[htbp]
\includegraphics[scale=0.6]{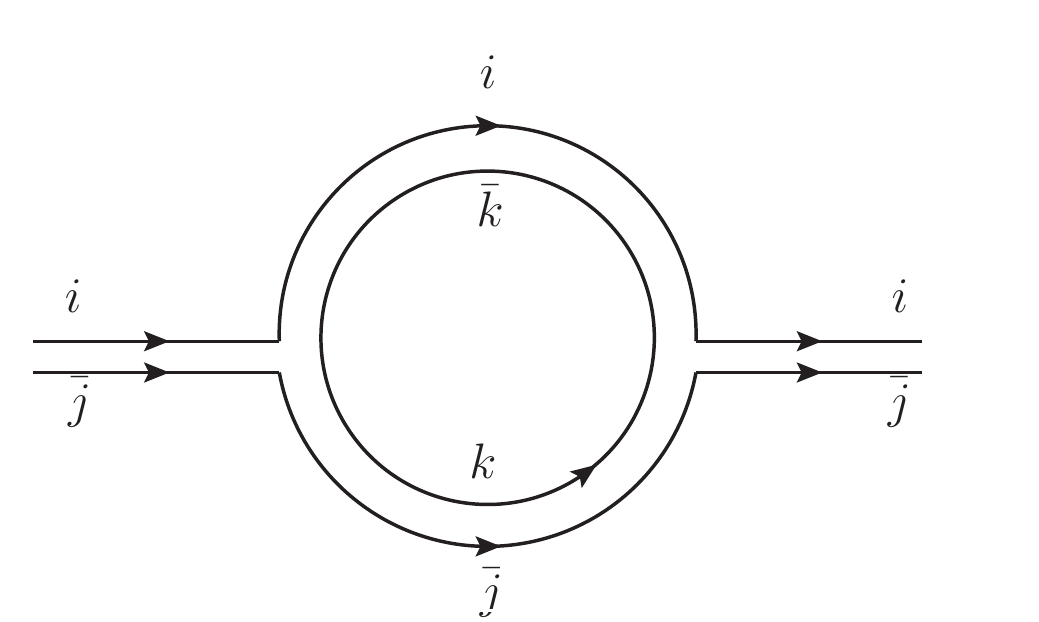}
\caption[The lowest-order gluon vacuum polarization in the double-line notation.]{%
The lowest-order gluon vacuum polarization in the double-line notation.} \label{fig:doublelgvp}
\end{figure}

By using the double-line notation, the $N_c$ counting of a Feynman diagram can be easily determined. For example, the gluon vacuum polarization can be illustrated by the double-line notation as in Fig.~\ref{fig:doublelgvp}. This shows that, in the center, there is a closed circle which has a color index $k$ so that the sum over $k$ gives a factor $N_c$. Consequently, Fig.~\ref{fig:doublelgvp} is of order $1$.
One can easily arrive at the following conclusion of a Feynman diagram from the double-line notation: {\it The $N_c$ counting of a Feynman diagram with $n$ $g_s$ and $m$ closed loops is $N_c^{m-n/2}$. As a result, in the limit $N_c \to \infty$, the diagram is divergent for $m > n/2$ while for $m = n/2$, it's $N_c^0$. Diagrams in these two cases survive in the $N_c \to \infty$ limit. However, a diagram with $m < n/2$ is suppressed by positive power of $1/N_c$ and therefore vanishes in the limit $N_c \to \infty$.}

\subsubsection{Meson properties from the large $N_c$ expansion}

The discussion of meson properties in the large $N_c$ expansion could be made by introducing the gauge invariant quark bi-linear operators with the consistent quantum numbers for generating the interested mesons from vacuum. Since mesons are color neutral, the interpolated current should be color singlet. In the following, the relevant currents are denoted by $J = \bar{q}\Gamma q ~(\Gamma = 1, \gamma_5, \gamma^\mu, (\partial_\mu - i g_s A_\mu), \cdots,) $ with $A_\mu$ being the gluon field.

\begin{figure}[h]
\includegraphics[scale=0.6]{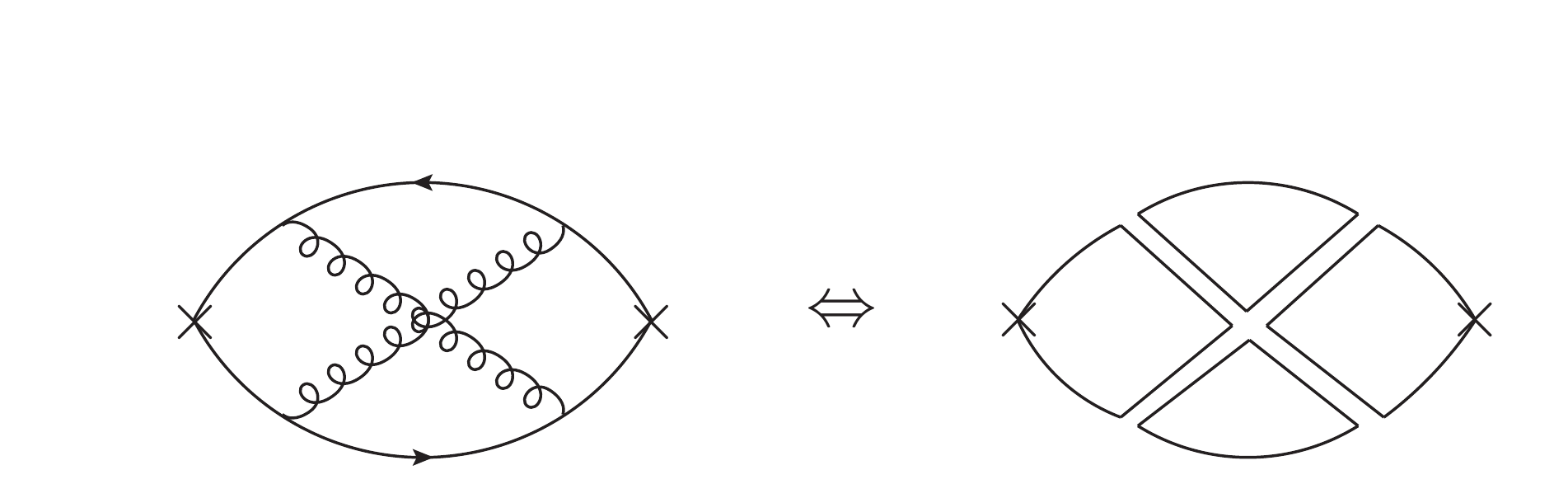}
\caption[]{%
Typical QCD correction to quark bilinear operatror (indicatd by ``$\times$'') and its double-line notation.} \label{fig:quarkbilinear}
\end{figure}

We first explore the meson mass and its decay constant by considering the current correlator $\langle J(k)J(-k)\rangle$. Since the current $J$ is a color singlet, in terms of the double-line notation, one can easily check that there should have at least one quark loop at the edges thus the diagram is of $O(N_c)$ at leading order. A typical diagram was shown in Fig.~\ref{fig:quarkbilinear}. By inserting a complete meson intermediate state $\sum_n |M_n\rangle \langle M_n| = 1$ into the correlator\footnote{Note that we only consider one-meson intermediate state here. The multi-meson intermediate state is non-leading $N_c$ contribution}, one has
\begin{eqnarray}
\langle J(k)J(-k)\rangle & =
& \sum_n\frac{\langle J(k)|M_n\rangle \langle M_n|J(-k)\rangle}{k^2 - m_n^2} = \sum_n\frac{f_{M_n}^2}{k^2-m_n^2} \, ,\label{eq:2pointmeson}
\end{eqnarray}
with the sum running over all meson states. Here $m_n$ is the mass of the $n$th meson, and $f_{M_n}=\langle0|J|n\rangle$ is the $n$th meson decay constant which denotes the amplitude for creating meson $M_n$ from the vacuum by the current $J$. Since the two-point function \eqref{eq:2pointmeson} is of order $N_c$,
\begin{eqnarray}
f_{M_n} & = & \langle 0|J|n\rangle \quad \mbox{is of order
$\sqrt{N_c}$}. \label{eq:ncordermesoncreation}
\end{eqnarray}
The same as the left-hand side, the right-hand side of (\ref{eq:2pointmeson}) should have a smooth limit for large $N_c$, consequently the meson masses have smooth limits. To guarantee that the left hand side of Eq.~\eqref{eq:2pointmeson} has the same large momentum scaling behaviour as the right hand side which are calculated as $\propto \ln k^2$ at large momentum $k^2$, the number of meson states should be infinite.

One can easily discuss the leading $N_c$ order of an $n$-meson vertex by using the $N_c$ scaling of the meson decay constant. In the spectral decomposition the $n$-point quark bilinear correlation function has contributions of the form
\begin{eqnarray}
\langle J_1 J_2 \cdots J_n\rangle & \sim & \sum_{i_1}\frac{\langle
0|J_1|i_1\rangle}{k_{i_1}-m_{i_1}^2} \sum_{i_2}\frac{\langle
0|J_2|i_2\rangle}{k_{i_2}-m_{i_2}^2} \cdots  \sum_{i_n}\frac{\langle
0|J_n|i_n\rangle}{k_{i_n}-m_{i_n}^2} \Gamma_{i_1, \cdots, i_n}^{(n)} \nonumber\\
& = & \mathcal{O}\left( N_c^{n/2}\right)\times \Gamma_{i_1, \cdots,
i_n}^{(n)} = \mathcal{O}(N_c),
\end{eqnarray}
where $\Gamma_{i_1, \cdots, i_n}^{(n)}$ is the $n$-meson vertex function. Since the correlation function scales like $N_c$, the $N_c$ scaling of $\Gamma_{i_1, \cdots, i_n}^{(n)}$ is
\begin{eqnarray}
\Gamma_{i_1, \cdots, i_n}^{(n)} & = &
\mathcal{O}\left(N_c^{1-n/2}\right) .
\end{eqnarray}
In the case of $n=3$ one concludes that $\Gamma^{(3)} \sim 1/\sqrt{N_c}$. This implies that {\it in the framework of large $N_c$ expansion, an effective meson model becomes a weakly coupled model and the meson decay is forbidden in the large $N_c$ limit}.

The above discussions on mesons made of quark-antiquark can be extended to glueball states. By considering the relevant correlation functions of current $J_G ~(J_G = {\rm Tr}G_{\mu\nu}G^{\mu\nu}, {\rm Tr}G_{\mu\nu}\tilde{G}^{\mu\nu})$ which creates a glueball field one can deduce that the glueball decay constant $f_{G_n} \equiv \langle 0|J_G|{G_n}\rangle$ is of $O(N_c)$ and glueball mass $m_{G_n}$ is $O(N_c^0)$. {\it In the $N_c \to \infty$ limit, gluball states are free, stable, non-interacting, and infinite in number.}

\begin{figure}[htbp]
\includegraphics[scale=0.6]{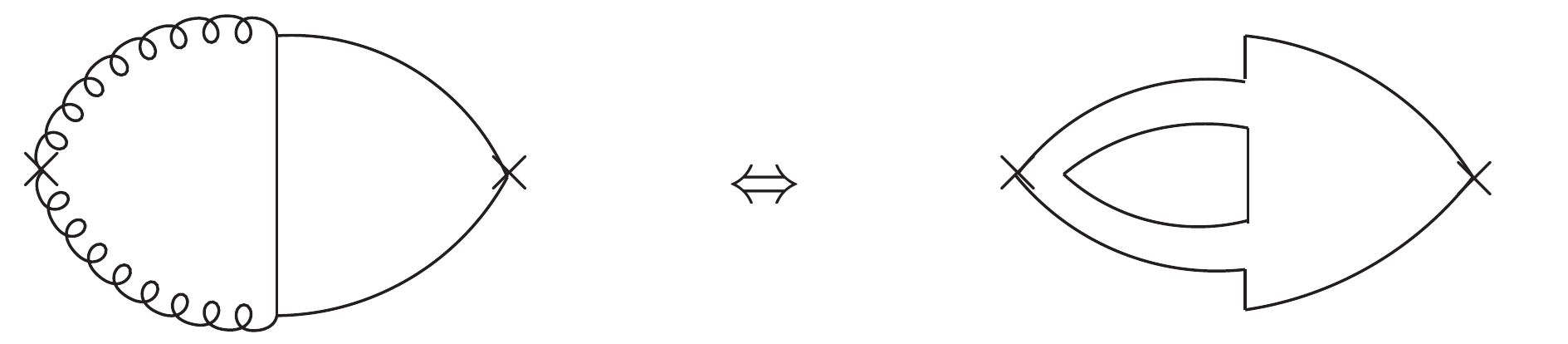}
\caption[Leading order diagram for the glueball-meson mixing and its double-line notation.]{%
The leading order diagram for the glueball-meson mixing and its double-line notation.}
\label{fig:glueballmesonmix}
\end{figure}

Next, we consider the $N_c$ order of the glueball and meson mixing. This could be achieved by considering the following spectral representation of the corresponding correlator
\begin{eqnarray}
\langle J_G(k)J(-k)\rangle & \sim & \sum_{i}\frac{\langle
0|J_{G_i}|G_i\rangle}{k_{G_i}-m_{G_i}^2} \sum_{j}\frac{\langle
0|J|j\rangle}{k_{j}-m_{j}^2}\Gamma_{\rm mix} = \mathcal{O}\left(
N_c^{3/2}\right)\times \Gamma_{\rm
mix}.\label{eq:glueballmesonmixing}
\end{eqnarray}
From Fig.~\ref{fig:glueballmesonmix} one sees that, because there are two gluon-quark vertices, the $N_c$ power of the left hand side of Eq.~(\ref{eq:glueballmesonmixing}) is
\begin{eqnarray}
N_c^2 \times \left(\frac{1}{\sqrt{N_c}}\right)^2 & = & N_c.
\end{eqnarray}
Consequently we have
\begin{eqnarray}
\Gamma_{\rm mix} & \sim &  N_c \times N_c^{-3/2} =
\frac{1}{\sqrt{N_c}},
\end{eqnarray}
which means that the amplitude for mixing a glueball state to a diquark meson is of order $1/\sqrt{N_c}$. {\it Therefore, one concludes that, in the large $N_c$ limit, the glueball states are decoupled from mesons}.

By using the same procedure, one can arrive at the following conclusion: {\it The amplitude
for a glueball state decaying to two glueball states or to two
mesons is of order $1/N_c$. The amplitudes for glueball-glueball and
glueball-meson elastic scattering are of order $1/N_c^2$.}

{\it In summary, to the leading order in $1/N_c$, amplitudes of diagrams of interactions with arbitrary numbers of meson and glueball
states can be obtained by summing over the tree
diagrams and, in these diagrams the general local vertex with $k$
mesons and $l$ glueball states is of order $N_c^{-l- k/2 + 1}$ (except $k = 0$ in which case it is of order $N_c^{-l+2}$).} For example, the diagram
with one glueball and one meson interaction is of order
$N_c^{-1-1/2+1} = 1/ \sqrt{N_c}$ which agrees with the above
discussion of the glueball-meson mixing.

\subsubsection{Baryon properties from the large $N_c$ expansion}

In the case that there are $N_c$ colors, the lowest lying baryons must be the composite states of $N_c$ quarks and must have wave functions which are totally antisymmetric in color indices and therefore symmetric in all other indices. It is because of this structure the large $N_c$ behavior of baryon is more subtle than that of meson since, in the Feynman diagrams for baryons, both the combinational factors and the shape of the diagrams depend on $N_c$.

If one naively considers the corrections to the baryon propagator from $m$-gluon exchanging among the $N_c$ constituent quarks, the $N_c$ counting of the corrections is
\begin{eqnarray}
\left(\frac{1}{\sqrt{N_c}}\right)^{2m}
\left[\frac{1}{2}N_c(N_c-1)\right]^{m} & \sim & \mathcal{O}(N_c^m).
\end{eqnarray}
This equation tells us that, as $N_c$ tends to infinity, the perturbative expansion of $g_s$ (here, number of gluon exchanging) is divergent which contradicts to the baryon properties which have a smooth limit in $N_c$ . This contradiction indicates that the $N_c$ counting from the diagram representation is not convenient to explore the $N_c$ behaviours of baryon properties.

E.\,Witten invented a proper way to sum all contributions based on the many-body techniques~\cite{Witten:1979kh}. He considered the case that the quarks in a baryon are so heavy that can be treated as non-relativistic objects. In such a case, the many-body problem can be reduced to a two-body problem in the Hartree approach in which a single quark is sitting in an average potential generated by the remaining $N_c -1$  quarks. Then the Hamilton operator reads\footnote{Since here we are considering the non-relativistic limit, spin dependent forces are not necessary to be included.}
\begin{eqnarray}
H & = & N_c M + \sum_{i=1}^{N_c}\frac{-\partial^2}{2M} -
\frac{g_{\rm eff}^2}{N_c}\sum_{i<j}^{N_c}\frac{1}{|\mathbf{r}_i -
\mathbf{r}_j|},
\end{eqnarray}
with $M$ being the mass of a single quark in the baryon. The wave function of the ground state baryon can be construct from the quark constituents which are arranged in the S-wave as
\begin{eqnarray}
\Psi(\mathbf{r}_1,\cdots,\mathbf{r}_{N_c}) & = &
\prod_{i=1}^{N_c}\phi(\mathbf{r}_i).
\end{eqnarray}
Thus we have the following eigenvalue equation
\begin{eqnarray}
\langle \Psi| H-E|\Psi\rangle & = &{} - N_c\epsilon + N_c M +
\frac{N_c}{2M}\int d^3r
\partial\phi^{\ast}(\mathbf{r})\cdot\partial\phi(\mathbf{r}) \nonumber\\
& &{} - \frac{N_c(N_c-1)}{2}\frac{g_{\rm eff}^2}{N_c}\int d^3r_1 \int d^3r_2
\frac{|\phi(\mathbf{r}_1)|^2|\phi(\mathbf{r}_2)|^2}{|\mathbf{r}_1 -
\mathbf{r}_2|}\, , \label{eq:variatbaryon}
\end{eqnarray}
where $E = N_c \epsilon $ with $\epsilon$ being the energy of a baryon carried by each quark. Eq.~\eqref{eq:variatbaryon} tells us that baryon masses are $\mathcal{O}(N_c)$. In addition, one can obtain the charge radius of a nucleon as
\begin{eqnarray}
\langle r^2\rangle \equiv \frac{1}{N_c} \langle
\Psi|\sum_{i}\mathbf{r}_i^2|\Psi\rangle = \frac{1}{N_c} N_c \int d^3r_1 r_1^2 |\phi(\mathbf{r}_1)|^2 \sim \mathcal{O}(N_c^0),
\end{eqnarray}
because $\phi$ has a smooth large $N_c$ limit.

From \eqref{eq:variatbaryon} one sees that the average potential carried by one quark in a baryon is $\mathcal{O}(N_c^0)$. This conclusion is still intact even when the three- and four-body forces arising from the self-interaction of gluons are included because the increasing of $N_c$ power from the combination of quarks is compensated by the increasing of the $g_s = g_{\rm eff}/\sqrt{N_c}$ power. Thus no matter how complicated the relativistic Hartree problem is, baryon masses are $\mathcal{O}(N_c)$ and the radii of baryons are $\mathcal{O}(N_c^0 )$.

For the baryon-baryon scattering there are $N_c^2$ possibilities to exchange one gluon between two quarks in the $2N_c$ constituents of the two baryons. Since there is a coupling constant $g_s$ at each end of the exchanged gluon, the contribution from these diagrams to the energy of the two-baryon system is $\mathcal{O}(N_c)$. However, for the meson-baryon scattering, the situation is different. Since we can only pick a single quark from the meson, the one gluon exchange contribution to the system energy is $\mathcal{O}(N_c^0)$.  This means that in the meson-baryon scattering process, in the large $N_c$ limit, the baryon stays as a static source and only the meson reacts. In summary, we have the following conclusions for baryon behaviors~\cite{Witten:1979kh}:
\begin{enumerate}
  \item Baryon masses are proportional to $N_c$.
  \item Baryon radii are $\mathcal{O}(N_c^0)$.
  \item Baryon-baryon scattering amplitudes are
  $\mathcal{O}(N_c)$.
  \item Meson-baryon scattering amplitudes are
  $\mathcal{O}(N_c^0)$.
\end{enumerate}
In the following, one will see that these $N_c$ behaviors of baryon properties are shared by the soliton configurations. {\it Therefore, in the sense of large $N_c$ limit, baryons could be regarded as solitons in a bosonic (here meson) field theory.}

\subsection{Chiral symmetry and chiral symmetry breaking  of QCD}

Chiral symmetry and chiral symmetry breaking have played important roles in the low energy dynamics of QCD. In the light quark sector, there exists the approximate chiral symmetry at the level of the Lagrangian, which is spontaneously broken by the strong interaction of QCD. Accordingly, the pion is regarded as the approximate Nambu-Goldstone boson associated with the spontaneous symmetry breaking which obeys the low energy theorems derived from the symmetry properties.
These aspects will be discussed in this part. For comprehensive reviews, see, e.g., Refs.~\cite{Scherer:2002tk,Ecker:1994gg}.

\subsubsection{The chiral symmetry of QCD}

Let us start our discussion of chiral symmetry from the following solutions of the Dirac
equation of a massless fermion
\begin{eqnarray}
u^\pm(p) & = & \sqrt{E}\left(%
\begin{array}{c}
  \xi^\pm \\
  \pm\xi^\pm \\
\end{array}%
\right)\equiv\hat{u}^\pm(p);\;\;\;\; v^\pm(p) = \sqrt{E}\left(%
\begin{array}{c}
  \pm \xi^\pm \\
  \xi^\pm \\
\end{array}%
\right)\equiv\hat{v}^\pm(p). \label{eq:4spinm0}
\end{eqnarray}
where we have used
\begin{eqnarray}
\vec{\sigma}\cdot\hat{p}~\xi^\pm=\pm\xi^\pm
\, .
\end{eqnarray}
When $\hat{p} = ( 0 , 0,1)$ these $\xi^\pm$ become the eigenstate of $\sigma^3$ as $(\xi^+)^T=(1,0)$ and $(\xi^-)^T=(0,1)$ corresponding to the
spin of the fermion. From Eq.~(\ref{eq:4spinm0}) one obtains
\begin{eqnarray}
\vec{\sigma}\cdot\hat{p}~\hat{u}^\pm(p) = \pm \hat{u}^\pm(p);
\;\;\;\;\;\; \vec{\sigma}\cdot\hat{p}~\hat{v}^\pm(p)=\pm
\hat{v}^\pm(p),
\end{eqnarray}
which means that $\hat{u}^\pm(p)$ are the eigenstates of the
helicity operator $\vec{\sigma}\cdot\hat{p}$.

Using the Dirac matrix $\gamma_5$, one can define the projection
operators
\begin{eqnarray}
P_R & \equiv & {1\over2}(1+\gamma_5); \;\;\;\;\; P_L \equiv
{1\over2}(1-\gamma_5) , \label{eq:definechiralproj}
\end{eqnarray}
which explicitly have the properties
\begin{eqnarray}
P_R^2 & = & P_R,~~~~P_L^2 = P_L,~~~~P_RP_L = P_LP_R = 0,~~~~P_R +
P_L =
1 ,\nonumber\\
P_R^\dag & = & P_R,~~~~P_L^\dag = P_L. \label{complete-P}
\end{eqnarray}
By using the explicit expression of $\gamma_5$ in the Dirac
representation
\begin{eqnarray}
\gamma_5 = \gamma^5 = \left(%
\begin{array}{cc}
  0 & I_{2\times2} \\
  I_{2\times2} & 0 \\
\end{array}%
\right),
\end{eqnarray}
one can easily check the following identities:
\begin{eqnarray}
P_R \hat{u}^+(p) & = & {1\over2}\left(%
\begin{array}{cc}
  I_{2\times2} & I_{2\times2} \\
  I_{2\times2} & I_{2\times2} \\
\end{array}%
\right)\sqrt{E}\left(%
\begin{array}{c}
  \xi^+ \\
  \xi^+ \\
\end{array}%
\right)=\sqrt{E}\left(%
\begin{array}{c}
  \xi^+ \\
  \xi^+ \\
\end{array}%
\right)=\hat{u}^+(p),\nonumber\\
P_L \hat{u}^-(p) & = & \hat{u}^-(p) ,\nonumber\\
P_R \hat{u}^-(p) & = & 0 ;\;\;\;\;\; P_L \hat{u}^+(p) = 0.
\end{eqnarray}
Similar relations hold for the spinor $\hat{v}^{\pm}$. These
relations indicate that, for a massless fermion, $P_R$ and $P_L$
project out the positive and negative helicity states, respectively. Corresponding to the
eigenvalues of the helicity operators, we name $P_R$ and $P_L$ as the
right- and left-handed projection operators, respectively and the
massless limit as the chiral limit.

Using the projection operators $P_L$ and $P_R$, one can decompose a
fermion field as~\footnote{This decomposition is general and has nothing to do with the chiral limit.}
\begin{eqnarray}
\psi & = & \psi_R+\psi_L,\;\;\;\; \bar{\psi} =
\bar{\psi}_R+\bar{\psi}_L,
\end{eqnarray}
with
\begin{eqnarray}
\psi_R(x) = P_R\psi(x),~~~~\psi_L(x) = P_L\psi(x), \nonumber\\
\bar{\psi}_R(x) = \bar{\psi}(x)P_L,~~~~\bar{\psi}_L(x) =
\bar{\psi}(x)P_R,\nonumber
\end{eqnarray}
where $\psi_{R}$ and $\psi_L$ are called right- and left-handed fermion fields, respectively.
In terms of the right- and left- handed quark fields, the fermion part of the QCD Lagrangian \eqref{eq:lagrQCD} can be
rewritten as
\begin{eqnarray}
{\cal L}_{\rm QCD}^{\rm fermion} & = & \sum_l \left[\bar{q}_{R,l,\alpha}
iD\hspace{-0.25cm}\slash_{\alpha\beta}q_{R,l,\beta}+\bar{q}_{L,l,\alpha}
iD\hspace{-0.25cm}\slash_{\alpha\beta}q_{L,l,\beta} + m_l \bar{q}_{L,l,\alpha}
q_{R,l,\alpha} + m_l \bar{q}_{R,l,\alpha}
q_{L,l,\alpha} \right], \label{eq:QCDfermionchiral}
\end{eqnarray}
where $D_\mu=\partial_\mu-ig_sG_\mu^aT^a$ is the covariant derivative
in the color space, $l$ is the flavor index. In this literature, we will focus on the two flavor case, i.e., $l = u,d$~\footnote{The extension to three-flavor case, i.e., $l = u, d, s$, is straightforward.}.  Lagrangian \eqref{eq:QCDfermionchiral} shows that the current quark mass breaks chiral symmtry explicitly and in the chiral limit the left- and right- hand components of quark fields decouple from each other.

For simplicity, omitting the flavor and color indices, QCD Lagrangian in the chiral limit is reexpressed as
\begin{eqnarray}
{\cal L}_{\rm chiral} & = & \bar{q}_R iD\hspace{-0.25cm}\slash
q_R+\bar{q}_L iD\hspace{-0.25cm}\slash q_L ,
\label{chiral-QCD}
\end{eqnarray}
where $q^T=(u,d)$ and the pure gluon part has been omitted. Since the covariant derivative $D_\mu$ is defined in the color space, Lagrangian (\ref{chiral-QCD}) is invariant under the following ${\rm U}(2)_L \times {\rm U}(2)_R$ flavor transformation:
\begin{eqnarray}
\left(%
\begin{array}{c}
  u_R \\
  d_R \\
\end{array}%
\right)& \mapsto & U_R\left(%
\begin{array}{c}
  u_R \\
  d_R \\
\end{array}%
\right)=\exp\left(-i\sum_{a=1}^3\Theta_a^RT^a\right)e^{-i\Theta^R}\left(%
\begin{array}{c}
  u_R \\
  d_R \\
\end{array}%
\right)\label{Trans-R},\\
\left(%
\begin{array}{c}
  u_L \\
  d_L \\
\end{array}%
\right)& \mapsto & U_L\left(%
\begin{array}{c}
  u_L \\
  d_L \\
\end{array}%
\right)=\exp\left(-i\sum_{a=1}^3\Theta_a^LT^a\right)e^{-i\Theta^L}\left(%
\begin{array}{c}
  u_L \\
  d_L \\
\end{array}%
\right),\label{Trans-L}
\end{eqnarray}
where $T^a=\tau^a/2$ are the generators of $\mbox{SU}(2)$ group with $\tau^a$ being the Pauli matrix, $\Theta^L_a$ and $\Theta^R_a$ are the transformation parameters corresponding to the generator $T^a$ and $\Theta^L$ and $\Theta^R$ are the transformation parameters of the $\mbox{U}(1)$ group.

Corresponding to the transformations (\ref{Trans-R},\ref{Trans-L}) one has the Noether currents
\begin{eqnarray}
J_L^{\mu,a} & = & \bar{q}_L\gamma^\mu T^aq_L,\;\;\;\; J_R^{\mu,a} =
\bar{q}_R\gamma^\mu T^aq_R, \;\;\;\;\; a = 1,2,3,\nonumber\\
J_L^{\mu} & = & \bar{q}_L\gamma^\mu
q_L,\;\;\;\;\;\;\;\;\; J_R^{\mu} = \bar{q}_R\gamma^\mu q_R,
\label{def-current-R}
\end{eqnarray}
and their combinations
\begin{eqnarray}
J^{\mu,a} & = & J_R^{\mu,a}+J_L^{\mu,a} = \bar{q}\gamma^\mu T^aq,
\;\;\;\;\;  J_5^{\mu,a} = J_R^{\mu,a}-J_L^{\mu,a} =
\bar{q}\gamma^\mu\gamma_5
T^aq\ ,
\\
J^{\mu} & = & J_R^{\mu}+J_L^{\mu}=\bar{q}\gamma^\mu
q,\;\;\;\;\;\;\;\;\;\;\;\;\;\;\; J_5^{\mu} =
J_R^{\mu}-J_L^{\mu}=\bar{q}\gamma^\mu\gamma_5q.
\end{eqnarray}
All these currents are conserved at the classical level but the
conservation of the axial-vector $J_{\mu5}$ is explicitly broken by the anomaly
due to the quantum corrections~\cite{Adler:1969gk,Bell:1969ts} which in the chiral limit is expressed as:
\begin{eqnarray}
\partial_\mu J_5^{\mu} & = & \frac{e^2}{16\pi^2}\epsilon^{\mu\nu\alpha \beta}F_{\mu\nu}F_{\alpha\beta}.
\end{eqnarray}
So that, in the chiral limit ($m_q = 0$), QCD Lagrangian
(\ref{chiral-QCD}) has a global
$\mbox{SU(2)}_L\times\mbox{SU(2)}_R\times\mbox{U(1)}_V$ symmetry,
the same as QCD Hamiltonian, $H_{\rm QCD}^0$.

Note that the $\mbox{SU}(2)_L\times \mbox{SU}(2)_R$ chiral symmetry is a global symmetry, so that at QCD level it does not correspond to to any gauge boson. However, in chiral effective theories and models, once one wants to study the electroweak interaction of hadrons, the chiral symmetry could be gauged and, after appropriate combinations, the gauge bosons can be related to the electroweak bosons $W^\pm, Z^0$ and photon $A_\mu$.

After the space coordinate integral, one obtains the charges of the left- and right-handed transformations as~\footnote{Strictly speaking, the charges $Q_L$ and $Q_R$ are not well-defined due to the divergence caused by the existence of the Nambu-Goldstone boson when the chiral symmetry is spontaneously broken. }
\begin{eqnarray}
Q^a_L(t) & = & \int d^3x\, q^\dagger_L(\vec{x},t)T^a
q_L(\vec{x},t),\quad a=1,2,3,\label{charge-La}\\
Q^a_R(t) & = & \int d^3x\, q^\dagger_R(\vec{x},t)T^a
q_R(\vec{x},t),\quad a=1,2,3,\label{charge-Ra}\\
Q_V(t) & = & \int d^3x\, \left[q^\dagger_L(\vec{x},t)q_L(\vec{x},t)+
q^\dagger_R(\vec{x},t)q_R(\vec{x},t)\right],\label{charge-V}
\end{eqnarray}
and all of them are conserved quantities, i.e.,
\begin{equation}
[Q_L^a,H^0_{\rm QCD}] = [Q_R^a,H^0_{\rm QCD}] = [Q_V,H^0_{\rm QCD}]
= 0.
\end{equation}
Using the commutation relations of the pauli matrices and field operators, one can prove that the charges of the chiral currents satisfy the following $\mbox{SU(2)}_L\times\mbox{SU(2)}_R\times\mbox{U(1)}_V$ Lie algebra:
\begin{eqnarray}
[Q_L^a,Q_L^b] & = & if_{abc}Q_L^c,\label{com-LL}\\
{[Q_R^a,Q_R^b]} & = & if_{abc}Q_R^c,\label{com-RR}\\
{[Q_L^a,Q_R^b]} & = & 0,\label{com-LR}\\
{[Q_L^a,Q_V]} & = & [Q_R^a,Q_V]=0,\label{com-LV}
\end{eqnarray}
which means that these charges can be regarded as the generators of
the transformation
$\mbox{SU(2)}_L\times\mbox{SU(2)}_R\times\mbox{U(1)}_V$. Similarly to the current, one can combine the charges corresponding to the left-
and right-handed transformations to get the following charges
\begin{eqnarray}
Q_V^a = Q_R^a+Q_L^a,~~~~~~Q_A^a=Q_R^a-Q_L^a.
\end{eqnarray}
From the commutation relations (\ref{com-RR},\ref{com-LR}) one
obtains
\begin{eqnarray}
[Q_V^a,Q_V^b] = i\epsilon_{abc}Q_V^c ,~~~~~~[Q_A^a,Q_A^b] = i\epsilon_{abc}Q_V^c
,~~~~~~[Q_V^a,Q_A^b] = i\epsilon_{abc}Q_A^c , \label{com-QA-QV}
\end{eqnarray}
which show that $Q_V^a$ forms a complete $\mbox{SU}(2)$ algebra but this is not the case for $Q_A^a$.

Under parity transformation, one can prove
\begin{eqnarray}
Q_R^a & \rightarrow & Q_L^a ; \;\;\;\;\;\; Q_L^a \rightarrow Q_R^a ,\\
Q_V^a & \rightarrow & Q_V^a ; \;\;\;\;\;\; Q_A^a \rightarrow - Q_A^a
,
\end{eqnarray}
which indicate that the left-handed and right-handed charges are
exchanged, the vector charge is invariant but the axial-vector
charge changes its sign.

\subsubsection{Chiral symmetry breaking}

Next, let us study what will happen if the chiral symmetry is not broken. In the case of exact chiral symmetry, because the ground state of QCD is invariant under chiral transformation one must have
\begin{eqnarray}
Q_V^a|0\rangle & = & 0 , \;\;\;\;\;\; Q_A^a|0\rangle = 0.
\end{eqnarray}
And, since the vector charge $Q_V^a$ and the axial-vector charge
$Q_A^a$ are conserved under the chiral transformation, one has
\begin{eqnarray}
[Q_V^a, H_{\rm QCD}^{0}] = [Q_A^a, H_{\rm QCD}^{0}] =  0.
\label{com-H-Q}
\end{eqnarray}
We introduce a hadron state $|i, +\rangle$ satisfying
\begin{eqnarray}
H_{\rm QCD}^{0} |i,+\rangle & = & E_i|i,+\rangle , \;\;\;\;\;\;\;  P
|i,+\rangle = |i,+\rangle,
\end{eqnarray}
where $E_i$ is the energy eigenvalue and $P$ is the parity operator. $i$ is an index corresoponding to the representation under the symmetry group. For a hadron state could be generated by rotation $|\phi^+\rangle = Q_V^a|i, +\rangle$,
~\footnote{ Here, we symbolically write $|\phi^+\rangle = Q_V^a|i, +\rangle$ although the rotation $Q_V^a|i, +\rangle$ could generate multi-hadron states.
}
one can show
\begin{eqnarray}
H_{\rm QCD}^{0} |\phi^+\rangle & = & H_{\rm QCD}^{0}
Q_V^a|i,+\rangle = E_i |\phi^+\rangle , \nonumber\\
P |\phi^+\rangle & = & P Q_V^a|i,+\rangle = Q_V^a|i,+\rangle =
|\phi^+\rangle ,\label{eq:chiralpositiveparity}
\end{eqnarray}
where in the first equation the commutation relation (\ref{com-H-Q}) has been applied. This relation indicates that the hadron state $|\phi^+\rangle$ is also an eigenstate of $H_{\rm QCD}^{0}$ and has positive parity. On the other hand, if we define another hadron state by rotation $|\phi^-\rangle = Q_A^a|i,+\rangle$, we can obtain
\begin{eqnarray}
H_{\rm QCD}^{0} |\phi^-\rangle & = & H_{\rm QCD}^{0}
Q_A^a|i,+\rangle = E_i |\phi^-\rangle, \nonumber\\
P |\phi^-\rangle & = & P
Q_A^a P^{-1} P|i,+\rangle = - Q_A^a|i,+\rangle = - |\phi^-\rangle, \label{eq:chiralnigativeparity}
\end{eqnarray}
which means that the state $|\phi^-\rangle$ is also an eigenstate of $H_{\rm QCD}^{0}$ with energy $E_i$ but negative parity. Then we draw the conclusion that {\it if chiral symmetry is an exact symmetry of QCD, there must be degenerate states in the hadron spectrum carrying opposite parity. This strongly indicates that the chiral symmetry must be broken dynamically since we do not have such a phenomena in Nature.}

Hadron spectrum tells us that chiral symmetry must be broken and QCD vacuum should preserve the vector part of the chiral symmetry. The spontaneous breakdown of the axial charge is defined as
\begin{eqnarray}
{}^\exists \Phi(y) \ \mbox{s.t.} \
\int d^3x \langle 0 \vert [ J^{\mu=0}_{a}(x) , \Phi(y) ] \vert 0 \rangle = \langle 0 \vert \delta \Phi(y) \vert 0 \rangle \neq 0 \ , \label{eq:breakQA}
\end{eqnarray}
where $\Phi(y)$ is an operator which might be a composite operator.  $\langle 0 \vert \delta \Phi(y) \vert 0 \rangle $ is called the order parameter. Now, a natural question is what is the order parameter of chiral symmetry breaking in terms of the intrinsic QCD quantity in the chiral limit. To answer this, we consider the following scalar and pseudoscalar quark-antiquark densities
\begin{eqnarray}
S_a(y)&=&\bar{q}(y)\tau_a q(y),\quad a=0,1,2,3,\\
P_a(y)&=&i\bar{q}(y)\gamma_5\tau_a  q(y), \quad a=0,1,2,3,
\end{eqnarray}
where $\tau_0 = I_{2\times 2}$ and $\tau_a (a = 1,2,3)$ is the Pauli matrix.

Under the $\mbox{SU}(2)_V$ transformation and using the expression of the vector charge \eqref{charge-V}, these scalar densities
transform as
\begin{eqnarray}
[Q^a_V(t),S_0(y)] & = & 0,\quad a=1,2,3,\label{comu-QV-S0}\\
{[Q^a_V(t),S_b(y)]}&=&i\sum_{c=1}^3\epsilon_{abc}S_c(y), \quad
a,b=1,2,3. \label{comu-QV-Sa}
\end{eqnarray}
When the $\mbox{SU}(2)_V$ symmetry is not broken, the QCD ground state $|0\rangle$ has an $\mbox{SU}(2)_V$ symmetry, that is
$Q_V^a|0\rangle=0$, so that
\begin{equation}
\langle 0|S_a(y)|0\rangle =\langle 0|S_a(0)|0\rangle \equiv\langle
S_a\rangle =0,\quad a=1,2,3,
\end{equation}
which means that the triplet component of the scalar density
vanishes. In the case $ a = 3 $ we have
\begin{equation}
\langle\bar{u}u\rangle-\langle\bar{d}d\rangle = 0. \label{eq:quarkvevt3}
\end{equation}
On the other hand, if the iso-singlet current $S_0$ is not zero, then
\begin{equation}
0\neq \langle \bar{q}q\rangle = \langle\bar{u}u+\bar{d}d\rangle = 2\langle\bar{u}u\rangle =
2\langle\bar{d}d\rangle
\end{equation}
in combination with \eqref{eq:quarkvevt3}.

For the pseudoscalar density one can obtain
\begin{equation}
i[Q_a^A, P_b(y)\delta_{ab}] = \bar{u}u+\bar{d}d, \quad a =1,2,3, \label{eq:comuQaPa}
\end{equation}
which leads to~\footnote{Strictly speaking the charge operator $Q_A^a$ is not well defined.  The above argument is a schematic, and correctly it is defined as
$\int d^3 x \langle0| [ j_0^a(x) , P^b(y) ] |0\rangle = \delta _{ab} \langle 0| \bar{q} q |0\rangle$.}
\begin{eqnarray}
\langle 0|i[Q_a^A,P_a(y)]|0\rangle = \langle\bar{q}q\rangle,\quad a=1,2,3 .
\label{eq:comuQaPavev}
\end{eqnarray}
To explore the implication of $\langle\bar{q}q\rangle$ on the chiral symmetry, we consider the following completeness relation made of states $\phi^a_\lambda$~\footnote{Here, we assume that the one-particle states form a complete set in the sense of large $N_c$ limit.
}
\begin{eqnarray}
\mathbf{1} & = & \sum_{a,\lambda}\int \frac{d^3p}{(2\pi)^3}\frac{1}{2E_{\bf{p}}}|\phi_\lambda^a\rangle\langle \phi^a_\lambda|
\, ,\label{eq:complete}
\end{eqnarray}
where $a$ is the isospin index and $\lambda$ is the index of the QCD mass eigenstate. Inserting the completeness relation \eqref{eq:complete} one has
\begin{eqnarray}
\langle\bar{q}q\rangle & = &i \int d^3 x \left[ \langle 0 | J_0^a(x) P_b(y) | 0 \rangle  - \langle 0 | P_b(y) J_0^a(x) | 0 \rangle \right] \nonumber\\
& = & i\sum_{b,\lambda}\int d^3 x \int\frac{d^3p}{2E_{\bf{p}}}\left\{ \langle
0|J_0^a(x)|\phi_\lambda^b(\vec{p})\rangle\langle
\phi_\lambda^b(\vec{p})|P_a(y)|0\rangle-\langle
0|P_a(y)|\phi_\lambda^b(\vec{p})\rangle\langle
\phi_\lambda^b(\vec{p})|J_0^a(x)|0\rangle \right\}. \nonumber\\
\end{eqnarray}
In the case of $\langle\bar{q}q\rangle \neq 0$ there must be a state $|\phi_\lambda^b(\vec{p})\rangle$ satisfying $\langle \phi_\lambda^b (\vec{p}) \vert A_0^a(x) \vert 0 \rangle \neq 0$  and $\langle 0|P_a(y)|\phi_\lambda^b(\vec{p})\rangle \neq 0$. {\it This means that the existence of the nonvanishing quark condensation is the sufficient but not necessary condition for the
chiral symmetry breaking}. This is because, if $\langle\bar{q}q\rangle = 0$ one can not conclude $Q_a^A \vert 0 \rangle
= 0$ since this can be realized by $\langle 0|P_a(y)|\phi_\lambda^b(\vec{p})\rangle = 0$.~\footnote{Atually, it was discussed in the literature (see,.e.g., Ref.~\cite{Kogan:1998zc})that, even the quark-antiquark condensate vanishes, chiral symmetry can still be broken by multiquark condensate such as tetraquark condensate.}

When the chiral symmetry is spontaneously broken we obtain (we ommit the mass index $\lambda$ in the following)
\begin{eqnarray}
\langle 0|A_0^a(x)|\phi^b(\vec{p})\rangle \equiv p_0f_\pi\delta^{ab} e^{ip\cdot x} \neq
0, \label{eq:pidecayconstant}
\end{eqnarray}
where $f_\pi$ is the decay constant of the Nambu-Goldstone boson. Because of the Lorentz invariance we can express
(\ref{eq:pidecayconstant}) in a covariant form as
\begin{equation}
\langle 0|A^a_\mu(0)|\phi^b(p)\rangle=ip_\mu f_\pi
\delta^{ab}.\label{eq:PCAC}
\end{equation}

\subsubsection{Pions as Nambu-Goldstone bosons}

In the above we learned that the reality tells us that chiral symmetry should be broken down to the flavor symmetry. Next, we discuss the chiral symmetry breaking from the quantum field theory point of view.

\begin{figure}[htbp] \centering
\subfigure[ Wigner-Weyl phase.]
{ \label{subfig:PlotWigner2d}
\includegraphics[width=0.35\columnwidth]{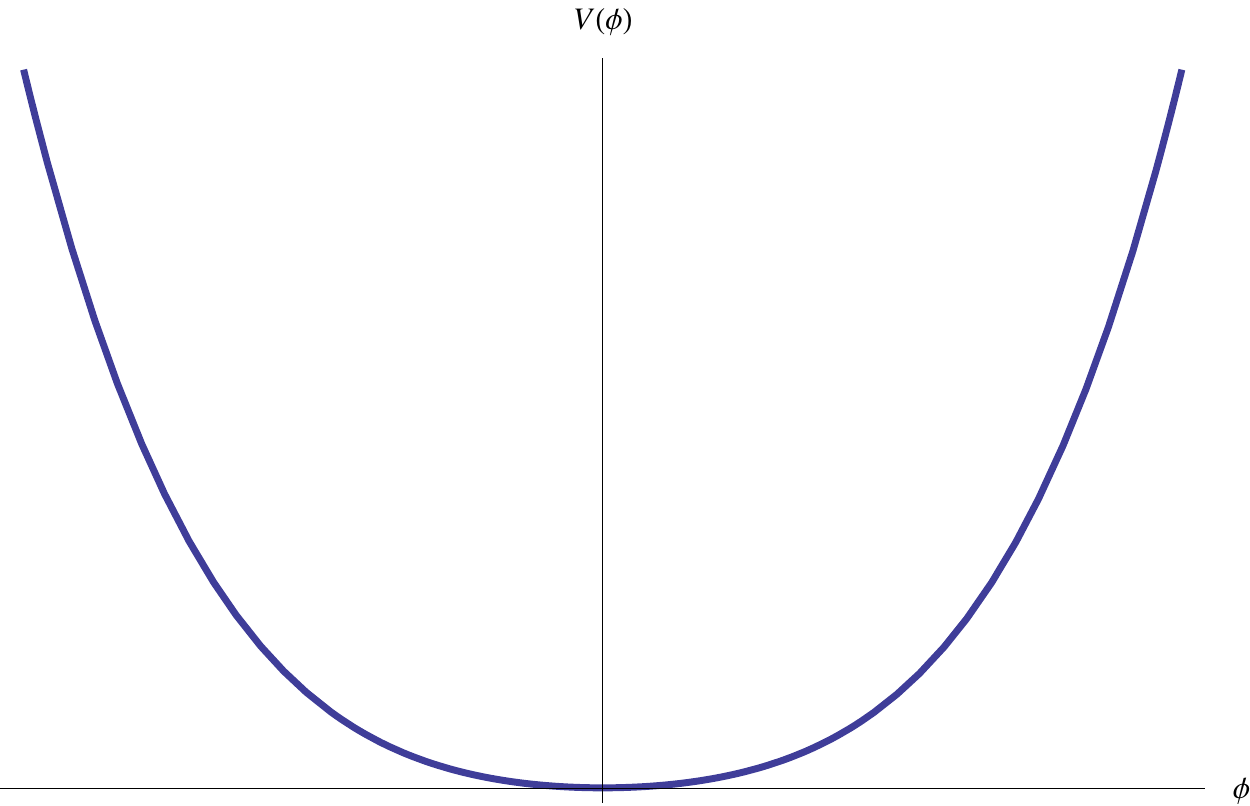}
}
\qquad\qquad
\subfigure[Nambu-Goldstone phase]
{ \label{subfig:PlotNG2d}
\includegraphics[width=0.35\columnwidth]{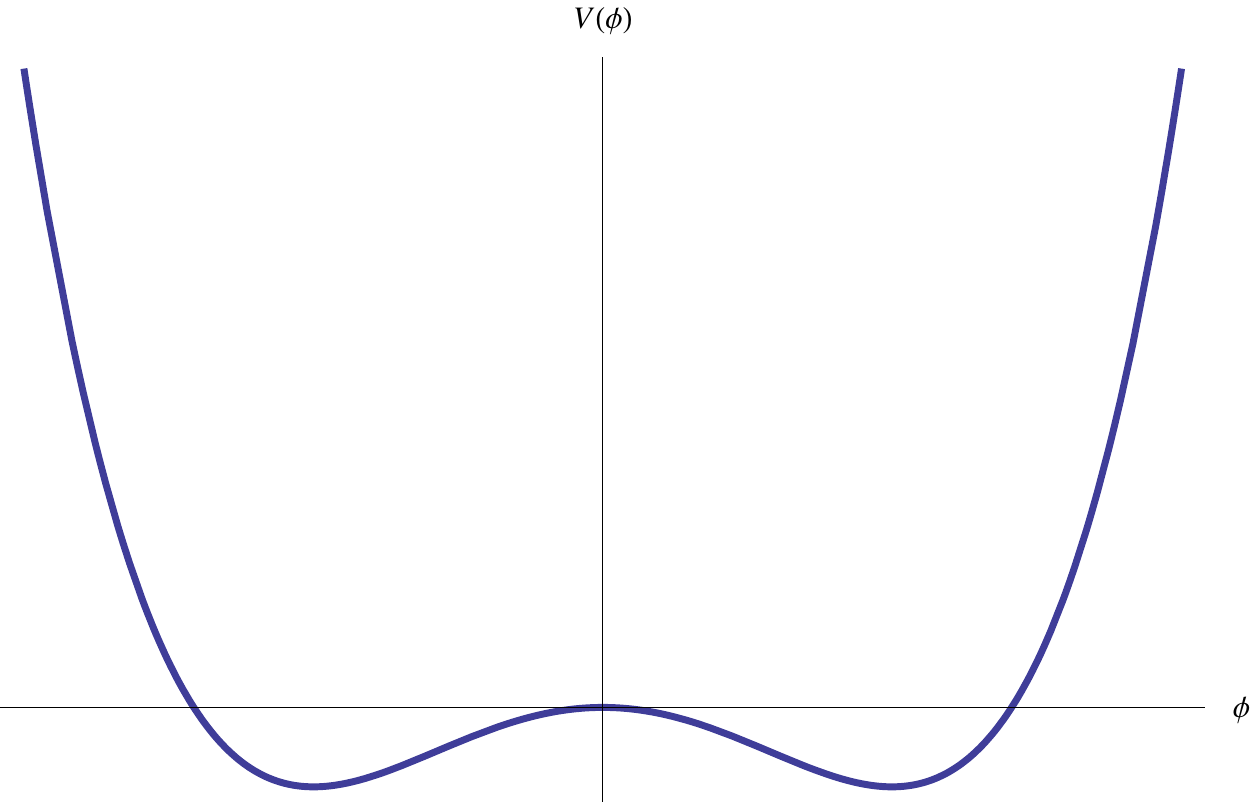}
}
\caption[]{Realization of discrete symmetry.}
\label{fig:potentialplot2d}
\end{figure}

For our purpose, we first consider a simple model including only a real scalar field $\phi(x)$, the $\lambda \phi^4$ theory,
\begin{eqnarray}
{\cal L}(\phi,\partial_\mu\phi) & = & \frac{1}{2}\partial_\mu \phi
\partial^\mu \phi
-\frac{m^2}{2}\phi^2-\frac{\lambda}{4!}\phi^4,\label{eq:sbreal}
\end{eqnarray}
where we choose $\lambda > 0$. One can easily see that the Lagrangian \eqref{eq:sbreal} has a discrete symmetry $R: \phi \to{} -\phi$. From the Lagrangian
(\ref{eq:sbreal}), the potential of the system is obtained as~
\footnote{For considering the minimum of the energy, it is sufficient to explore the minimum of the potential since the kinetic energy vanishes for the constant field which gives the minimum of the kinetic energy.
}
\begin{eqnarray}
{\cal V}(\phi) & = & \frac{m^2}{2}\phi^2+\frac{\lambda}{4}\phi^4.
\end{eqnarray}
We now consider two cases:
\begin{itemize}
\item $m^2 > 0$ (see Fig.~\ref{subfig:PlotWigner2d}): The potential
$\cal V$ has its minimum at $\phi = \phi_0 = 0$. In the quantized theory this minimum associates a unique ground state $|0\rangle$.
In literature, this symmetry realization is referred to as the Wigner-Weyl mode.

\item $m^2<0$ (see Fig.~\ref{subfig:PlotNG2d}): In this case the potential exhibits
two distinct minima. In literature, this mode is referred to as the
Nambu-Goldstone realization of the symmetry.
\end{itemize}

In the Nambu-Goldstone mode, at the minima, the VEV of $\phi$ field becomes
\begin{eqnarray}
\phi_0^\pm & = & \pm \sqrt{\frac{-m^2}{\lambda}}\equiv \pm \, v.
\end{eqnarray}
By expanding the field $\phi$ with respect to its value at the minima, $\phi = \phi_0^\pm + \tilde{\phi}$, the Lagrangian \eqref{eq:sbreal} becomes
\begin{eqnarray}
{\cal L}(\tilde{\phi},\partial_\mu \tilde{\phi})
=\frac{1}{2}\partial_\mu\tilde{\phi}\partial^\mu \tilde{\phi} - \frac{1}{2}(2\lambda v)\tilde{\phi}^2 \mp \lambda
v\tilde{\phi}^3 - \frac{\lambda}{4}\tilde{\phi}^4 +\frac{\lambda}{4}v^4.
\end{eqnarray}
One can easily see that in terms of the variable $\tilde{\phi}$, because of the third term, the discrete symmetry $R$ is no longer manifest. This simple example shows that {\it selecting one of the ground states has led to a
spontaneous breaking of the discrete $R$ symmetry}.

\begin{figure*}[htbp]
\includegraphics[scale=0.5]{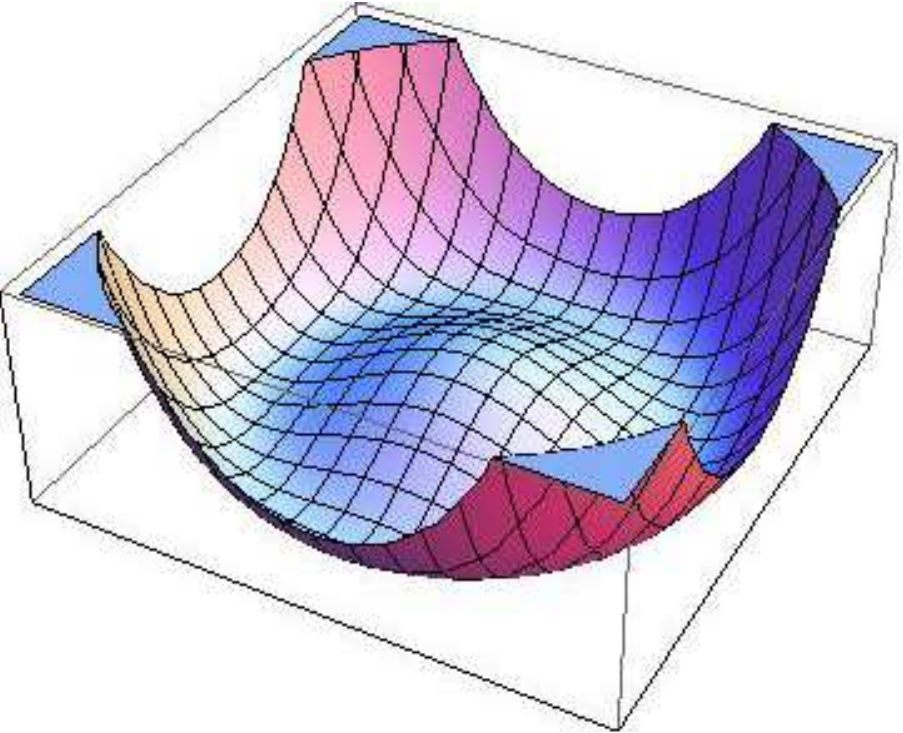}
\caption{\label{fig:hat}
The potential of the model \eqref{eq:SSBNAbe} in the Nambu-Goldstone phase.}
\end{figure*}

We next generalize the above discussion to a system with a continuous symmetry.
For this purpose, we consider the following Lagrangian with $O(3)$ symmetry:
\begin{eqnarray}
{\cal L}(\vec{\Phi},\partial_\mu\vec{\Phi}) & = &
\frac{1}{2}\partial_\mu \Phi_i\partial^\mu
\Phi_i-\frac{m^2}{2}\Phi_i\Phi_i -\frac{\lambda}{4}(\Phi_i\Phi_i)^2,
\label{eq:SSBNAbe}
\end{eqnarray}
where $m^2<0$, $\lambda>0$, with real fields $\Phi_i (i =1, 2, 3)$. Since $m^2<0$, the symmetry is realized in the Nambu-Goldstone mode. The potential of the system \eqref{eq:SSBNAbe} is illustrated by Fig.~\ref{fig:hat}. The Lagrangian (\ref{eq:SSBNAbe}) is invariant under a global rotation,
\begin{eqnarray}
g\in \mbox{O(3)}:\,\,\Phi_i\to\Phi_i'=D_{ij}(g)\Phi_j=
(e^{-i\alpha_k T_k})_{ij}\Phi_j.
\end{eqnarray}
The matrices $T_k$ are the generators of the so(3) Lie algebra and satisfy
the commutation relations $[T_i,T_j]=i\epsilon_{ijk} T_k$.

In the Nambu-Goldstone phase, the potential of the system has its minimum at
\begin{eqnarray}
|\vec{\Phi}_{\rm min}| & = & \sqrt{\frac{-m^2}{\lambda}}\equiv v,
\quad |\vec{\Phi}|=\sqrt{\Phi_1^2+\Phi_2^2+\Phi_3^2}.
\end{eqnarray}
Since $\vec{\Phi}_{\rm min}$ can point in any direction in the $O(3)$
space, we have an uncountable infinite number of degenerate
vacua. Without loss of generality, we can select a particular direction of $\Phi_{\rm min}$ as
\begin{eqnarray}
\vec{\Phi}_{\rm min} = v \hat{e}_3, \label{eq:goldstonevevnonab}
\end{eqnarray}
which is not invariant under the full group $G=\mbox{O(3)}$
because rotations around the $1$ and $2$ axes change $\vec{\Phi}_{\rm
min}$ although it is invariant under the rotation around $3$ axis. Specifically, if
\begin{eqnarray}
\vec{\Phi}_{\rm min} =
v\left(\begin{array}{r}0\\0\\1\end{array}\right),
\end{eqnarray}
we obtain
\begin{eqnarray}
T_1 \vec{\Phi}_{\rm min}=
v\left(\begin{array}{r}0\\-i\\0\end{array}\right),\quad T_2
\vec{\Phi}_{\rm min}=
v\left(\begin{array}{r}i\\0\\0\end{array}\right), \quad T_3
\vec{\Phi}_{\rm min}=0.
\end{eqnarray}
Note that, because the set of transformations which do not leave
$\vec{\Phi}_{\rm min}$ invariant does not contain the identity, it does not form a group.
However, there is subgroup $H$ of $G$ which leaves $\vec{\Phi}_{\rm min}$ invariant, namely, the rotations about the $3$ axis:
\begin{eqnarray}
h\in H:\quad \vec{\Phi}'=D(h)\vec{\Phi} = e^{-i\alpha_3
T_3}\vec{\Phi}, \quad D(h)\vec{\Phi}_{\rm min} = \vec{\Phi}_{\rm
min},
\end{eqnarray}
which is the $O(2)$ symmetry.
As before, we expand $\Phi_3$ with respect to $v$,
\begin{eqnarray}
\Phi_3 & = & v+\tilde{\Phi}_3,
\end{eqnarray}
and express the potential as
\begin{eqnarray}
\tilde{\cal V} & = & \frac{1}{2}(2\lambda v)\tilde{\Phi}_3^2 +\lambda v\tilde{\Phi}_3
(\Phi_1^2+\Phi_2^2+\tilde{\Phi}_3^2)
+\frac{\lambda}{4}(\Phi_1^2+\Phi_2^2+\tilde{\Phi}_3^2)^2-\frac{\lambda}{4}v^4.
\end{eqnarray}
From this potential one finds that, after spontaneous symmetry breaking, two bosons $\Phi_1$ and $\Phi_2$ become massless while one boson, $\tilde{\Phi}_3$, is massive with mass square $m_{\tilde{\Phi}_3}^2 = 2\lambda v$.

The above analysis shows that for each of the two generators $T_1$ and $T_2$ which does not annihilate the ground state one obtains a massless Nambu-Goldstone boson $\Phi_1$ and $\Phi_2$ but for the generator $T_3$ one obtains a massive field $\tilde{\Phi}_3$. From Fig.~\ref{fig:hat} one can understand the present situation as follows: When one makes an infinitesimal variation orthogonal to the circle of the vacuum, one suffers a restoring forces linear in the variation but a variation tangent to the circle of the vacuum suffers restoring forces of higher orders.

The above discussion can be straightforwardly generalized to a model with an arbitrary compact
Lie group $G$. One finally arrives at the following Nambu-Goldstone theorem:
\begin{enumerate}
\item A continuous global symmetry breaking will generate massless bosons, Nambu-Goldstone bosons (NGBs).

\item The number of NGBs is determined by the pattern of the symmetry breaking.
Let $G$ denotes the symmetry group of the
Lagrangian, with $n_G$ generators and $H$ the subgroup with $n_H$
generators which leaves the ground state after spontaneous symmetry
breaking invariant. The total number of NGBs equals $n_G-n_H$.

\item The NGBs generated by the spontaneous symmetry breaking have the same quantum numbers as that of the generators of the symmetry which is broken since these NGBs can be generated by $Q^a |0 \rangle, Q^a \in G/H$.
\end{enumerate}

Since Nature tells us that chiral symmetry must be broken dynamically, the
Nambu-Goldstone theorem implies that there must exist massless NGBs which have the same quantum numbers as that of the broken current. From the hadron spectrum one can see that
the lowest-lying pseudoscalar mesons are much lighter than other
hadrons, so that it is reasonable to regard them as the
Nambu-Goldstone bosons and the small masses of the pseudoscalar
mesons arise from the explicit chiral symmetry breaking due to the
small light quark masses. In other words, the physical spectrum
demands that the chiral symmetry must be broken and the breaking pattern
should be $\mbox{SU}(2)_L\times \mbox{SU}(2)_R \rightarrow
\mbox{SU}(2)_V$ which indicates
\begin{eqnarray}
Q_A^a | 0 \rangle & \neq & 0,
\end{eqnarray}
with $| 0 \rangle$ being the QCD vacuum. Then, by using Eq.~\eqref{eq:PCAC} and the conservation of the axial-vector current in the chiral limit one can prove that the state by rotation $Q_A^a | 0 \rangle$ is massless. With respect to the parity transformation property, one can easily conclude that the state generated by rotation $Q_A^a | 0 \rangle$ is odd. In addition, by using the transformation of the axial-vector charge given by Eq.~\eqref{com-QA-QV} one arrives at the conclusion that under the $\mbox{SU}(2)_V$
transformation, the pseudoscalar triplet transforms as the adjoint
representation of $\mbox{SU}(2)_V$.

{\it In a word, the state created by
rotation $Q_A^a | 0 \rangle$ is a massless, pseudoscalar particle with negative parity which transforms as the adjoint
representation of $\mbox{SU}(2)_V$}.

Since the pseudoscalar mesons fill in the
$\mbox{SU}(2)_V$ adjoint representation matrix, they can be classified by using the third component of isospin $T_3$. So that we have particle identification in Table.~\ref{table:idenpi} (for simplicity, we express the state generated by rotation $Q_A^a | 0 \rangle$ by
$\pi^a$).
\begin{table}
\begin{center}
\begin{tabular}{c|c|c|c}
  \hline
Combined state  & $|\pi_1 \rangle + i|\pi_2 \rangle$ & $|\pi_1 \rangle - i|\pi_2 \rangle$ & $|\pi_3 \rangle $ \\
  \hline
$T_3$ & $ +1 $ & $ -1 $ & $ 0 $ \\
  \hline
Meson state & $\pi^+$ & $\pi^-$ & $\pi^0$ \\
  \hline
\end{tabular}
\caption{\label{table:idenpi} Identification of the lowest-lying pseudoscalar mesons.}
\end{center}
\end{table}
Then we finally write the pseudoscalar meson matrix as
\begin{eqnarray}
\pi=\sum_{a=1}^3T_a\pi_a & \equiv & \frac{1}{2}\left(%
\begin{array}{ccc}
  \pi^0 & \sqrt{2}\pi^-  \\
  \sqrt{2}\pi^+ &  -\pi^0  \\
\end{array}%
\right)\, , \label{matrix-pseudoscalar}
\end{eqnarray}
where the coefficients are from the normalization. In addition,
considering the charge conjugation and parity transformation
properties of pseudoscalar mesons, we impose the following
transformation properties of $\pi^a$ fields
\begin{eqnarray}
&&P:\pi(\vec{x},t)\rightarrow -\pi(-\vec{x},t) ,\\
&&C:\pi(\vec{x},t)\rightarrow\pi^{T}(\vec{x},t).
\end{eqnarray}

\newpage

\section{The nonlinear sigma model of pseudoscalar mesons}

\label{sec:nonlinearsigma}

In this section we introduce the nonlinear realization of the chiral symmetry and the basic idea of  the chiral perturbation theory, especially the power counting mechanism. We also discuss the topology of the nonlinear sigma model which is essential for understanding the baryon dynamics using a mesonic theory.

\subsection{From the linear sigma model to nonlinear sigma model}

We introduce the matrix $M$ describing the mesons as quark-antiquark
bound states with the schematic structure~\footnote{Therefore the mesons in the present model are two quark states. For a discussion of the linear sigma model including tetraquark mesons, see, e.g., Refs.~\cite{Fariborz:2005gm,Black:2000qq,Harada:2012km}.}
\begin{eqnarray}
M_{ba} & = & \left(q_{b\alpha}\right)^\dag \gamma_0
\frac{1-\gamma_5}{2}q_{a\alpha} = \bar{q}_{R,b\alpha} q_{L,a\alpha},
\end{eqnarray}
where $a$ and $\alpha$ are, respectively, the flavor and color indices.
Under the chiral transformation the matrix $M$ transforms as
\begin{eqnarray}
M & \rightarrow & g_L M g_R^\dag,\label{eq:chirallight}
\end{eqnarray}
where $g_{L,R} \in {\rm SU}(2)_{L,R}$. We can decompose the matrix
$M$ in terms of the isosinglet field $\sigma$ and the isotriplet
pseudoscalar meson $\bm{\pi}$ as
\begin{eqnarray}
M & = & \sigma \mathbb{I} + i {\bm\tau} \cdot {\bm\pi},
\label{eq:decompM}
\end{eqnarray}
with $\tau_i$ as the Pauli matrix. Using the meson matrix $M$ one can
write down a linear sigma model with Lagrangian
\begin{eqnarray}
{\cal L}_{\rm L\sigma M} & = & \frac{1}{4}{\rm Tr}(\partial _\mu M
\partial^\mu
M^{\dag})- V_0(M, M^\dag) - V_{SB} \nonumber\\
& = & \frac{1}{2}\left(\partial_\mu \sigma\partial_\mu \sigma +
\partial_\mu \bm{\pi} \cdot \partial_\mu \bm{\pi} \right)- V_0(\sigma^2 + \bm{\pi}^2) -
V_{SB},
\label{eq:lagrlight}
\end{eqnarray}
where $V_0(M)$ is the potential term which is invariant under ${\rm SU}(2)_L
\times {\rm SU}(2)_R \times {\rm U}(1)_V$ transformation, and
$V_{SB}$ stands for the explicit chiral symmetry breaking term due
to the current quark mass. Since ${\rm SU}(2)_L\times {\rm SU}(2)_R \sim
{\rm SO}(4)$, the Lagrangian (\ref{eq:lagrlight}), except the
$V_{SB}$ term, has an ${\rm SO}(4)$ symmetry with $(\sigma,
\bm{\pi})$ as its four-vector~\footnote{ Actually, there exists full $O(4)$ symmetry.}.

Now, let us introduce two parameters $\alpha_L$ and $\alpha_R$ for parameterizing the chiral transformation matrices
as
\begin{eqnarray}
g_L = e^{ \bm{\alpha}_L\cdot\bm{\tau}/2 } \ , \;\;\;\;\;
g_R = e^{  \bm{\alpha}_R\cdot\bm{\tau}/2 } \ . \label{eq:chiraltrans}
\end{eqnarray}
Under the infinitesimal vector transformation, $\alpha_L = \alpha_R = \alpha$, the matrix $M$ transforms as
\begin{eqnarray}
M & \to & M^\prime = g_L M g_R^\dag\Big|_{\alpha_L = \alpha_R =
\alpha} = \left[ 1 - i \frac{1}{2}\bm{\alpha}\cdot\bm{\tau}\right] M
\left[1 +
i\frac{1}{2}\bm{\alpha}\cdot\bm{\tau}\right] + \cdots \nonumber\\
& = & M - \left[ i \frac{1}{2}\bm{\alpha}\cdot\bm{\tau},  M \right]
+ \cdots = M - \left[ i \frac{1}{2}\bm{\alpha}\cdot\bm{\tau},
i\bm{\tau}\cdot \bm{\pi} \right] + \cdots,\label{eq:transvectorM}
\end{eqnarray}
which, upto $\mathcal{O}(\alpha)$, leads to
\begin{eqnarray}
\delta_V\sigma & = &0, \;\;\;\;\;\;\;\;\;\;\;\;\;\;\;\;\;\;\;\;
\delta_V\bm{\pi} = (\bm{\alpha} \times
\bm{\pi})\,.
\label{eq:transvectorsp}
\end{eqnarray}
This equation means that $\sigma$ is a scalar under the vector transformation but $\pi^a$ is a vector. Similarly, under the axial-vector transformation, $\alpha_R = -\alpha_L =\beta$,
the meson matrix
transforms as
\begin{eqnarray}
M & \to & M^\prime = g_L M g_R^\dag\Big|_{\alpha_R = -\alpha_L =
\beta} = \left[ 1 + i \frac{1}{2}\bm{\beta}\cdot\bm{\tau}\right] M
\left[1 +
i\frac{1}{2}\bm{\beta}\cdot\bm{\tau}\right] + \cdots \nonumber\\
& = & M + \left\{ i \frac{1}{2}\bm{\beta}\cdot\bm{\tau},  M \right\}
+ \cdots = M + i \bm{\beta}\cdot\bm{\tau}\sigma + \left\{ i
\frac{1}{2}\bm{\beta}\cdot\bm{\tau}, i\bm{\tau}\cdot \bm{\pi}
\right\} + \cdots,\label{eq:transaxialM}
\end{eqnarray}
which, upto $\mathcal{O}(\beta)$,  yields
\begin{eqnarray}
\delta_A \sigma & = & - \bm{\beta}\cdot\bm{\pi}, \;\;\;\;\;\;
\delta_A \bm{\pi} = \bm{\beta}\sigma,\label{eq:transaxialsp}
\end{eqnarray}
which shows that, in contrast to the vector transformation
(\ref{eq:transvectorM}), the axial transformation does not form any
group (not a symmetry).

Using the transformation property in Eq.~(\ref{eq:chirallight}), one can derive the classically
conserved Noether's currents associated with the left- and right-handed
transformations from Lagrangian (\ref{eq:lagrlight}) as
\begin{eqnarray}
J_{L,\mu}^i & = & \frac{\partial {\cal L_{\rm L\sigma M}}}{\partial
\partial_\mu\alpha_{L,i}} = - \frac{i}{8}{\rm Tr}\left[ \tau_i M\partial_\mu M^\dag - \tau_i \partial_\mu MM^\dag \right], \nonumber\\
J_{R,\mu}^i & = & \frac{\partial {\cal L_{\rm L\sigma M}}}{\partial
\partial_\mu \alpha_{R,i}} =
- \frac{i}{8}{\rm Tr}\left[ \tau_i M^\dag\partial_\mu M - \tau_i
\partial_\mu M^\dag M \right]. \label{eq:currentlr}
\end{eqnarray}
And, by using these equations, the currents
associated with the vector and axial-vector transformations can be derived as
\begin{eqnarray}
J_\mu^i & = & J_{R,\mu}^i + J_{L,\mu}^i = \epsilon^{ijk}\pi^j\partial_\mu \pi^k, \nonumber\\
J_{\mu5}^i & = & J_{R,\mu}^i - J_{L,\mu}^i = \sigma\partial_\mu
\pi^i - \pi^i\partial_\mu \sigma\, . \label{eq:currentva}
\end{eqnarray}

For a special choice of the potential $V_0$ in Eq.~\eqref{eq:lagrlight}, for example $V_0 ={} - \mu^2 (\sigma^2 + \pi^2) + (\sigma^2 + \pi^2)^2$ with $\mu^2$ being a positive parameter with the dimension of mass square, in which the potential of the system is in the Nambu-Goldstone mode, the vacuum expectation value of the sigma field will be non-zero. In such a case, one can deduce
\begin{eqnarray}
\langle 0|J_{\mu5}^i(x)|\pi^k(p)\rangle & = & \langle
0|\sigma\partial_\mu \pi^i - \pi^i\partial_\mu
\sigma|\pi^k(p)\rangle = \langle 0|\sigma\partial_\mu \pi^i
|\pi^k(p)\rangle = \langle 0|\sigma |0\rangle ip_\mu e^{ip\cdot
x}\delta^{ik}. \label{eq:pcaccurrent}
\end{eqnarray}
In this derivation, we have considered that in the vacuum, $\partial_\mu \sigma = 0$ and used the normalization of the pion field
\begin{eqnarray}
\langle 0| \pi^i(x) |\pi^k(p)\rangle & = &  e^{ip\cdot
x}\delta^{ik}.
\end{eqnarray}
Combing Eq.~(\ref{eq:pcaccurrent}) with Eq.~(\ref{eq:PCAC}) one has
\begin{eqnarray}
f_\pi & = & \langle 0|\sigma |0\rangle. \label{eq:fpivevsigma}
\end{eqnarray}
This shows that in the present model, $f_\pi$ is proportional to the two-quark condensate and thus the two-quark condensate could be regarded as the order parameter of chiral symmetry breaking~\footnote{Note that in the case of multi-quark state, such as four-quark state, is included, not only the two-quark condensate but also multi-quark condensate is the order parameter of chiral symmetry breaking.}.

The lowest energy of the model~\eqref{eq:lagrlight} can be obtained by requiring that the fields
$\sigma$ and $\bm{\pi}$ are constants in space-time with values minimizing the potential $V_0(\sigma^2 + \bm{\pi}^2)$. By a suitable choice of the potential $V_0(\sigma^2+\pi^2)$, chiral symmetry can be realized as the Nambu-Goldstone mode and, in such a mode, the minimum energy could be obtained at some finite values of $c^2= \sigma^2 + \bm{\pi}^2$. Since the potential $V_0$ is chiral invariant, there are infinitely many degenerate states in the ground state. These
infinitely many degenerate states in the ground state are related with each other by chiral rotations in the $(\sigma,\bm{\pi})$ space with keeping $\sigma^2 + \bm{\pi}^2 = c^2$.
For determining the vacuum, we select one state from infinitely degenerated states with $\sigma^2 + \pi^2 = c^2$ , e.g., $\sigma = c$. Because the vacuum with $\bm{\pi} = 0$ is not invariant with respect to the chiral transformations, the chiral symmetry is spontaneously broken.

When the chiral symmetry is realized as Nambu-Goldstone mode, three Nambu-Goldstone bosons appear, which can be described by the pion fields. The sigma field provides a massive field which can be integrated out in the low-energy region.  At the tree level of the sigma, this integrating out is easily done by using the following constraint:
\begin{eqnarray}
\bm{\pi}^2(\mathbf{x},t) + \sigma^2(\mathbf{x},t) & = & f_\pi^2
\label{eq:constraintlsm}
\end{eqnarray}
for all $\mathbf{x}$ and $t$. In Eq.~(\ref{eq:constraintlsm}) we
have replaced the constant $c$ with $f_\pi$ concerning $c = \langle 0| \sigma
| 0\rangle$ and Eq.~(\ref{eq:fpivevsigma}). Equation~(\ref{eq:constraintlsm}) implies that, at the lowest energy of the system, $\sigma$ is a simple but
nonlinear function of $\bm{\pi}$ and therefore it is enough to include one of them as a dynamical field. In such a case, the Lagrangian (\ref{eq:lagrlight}) will be simplified considerably
because the combination $\sigma^2 + \bm{\pi}^2$ becomes a constant so can be omitted.

From Eq.~(\ref{eq:constraintlsm}) one can get the following
equation of motion of $\sigma$ field
\begin{eqnarray}
\sigma & = & \sqrt{f_\pi^2 - \bm{\pi}^2} ,
\label{eq:eomsigma}
\end{eqnarray}
which yields
\begin{eqnarray}
\partial_\mu \sigma = -\frac{\bm{\pi}\cdot \partial_\mu \bm{\pi}}{\sqrt{f_\pi^2 -
\bm{\pi}^2}} \, .
\end{eqnarray}
Substituting this relation to the linear sigma
model Lagrangian (\ref{eq:lagrlight}) and neglecting the constant
contribution from the potential term $V_0$ we have
\begin{eqnarray}
{\cal L}_{\rm L\sigma M} & = & \frac{1}{2}\partial_\mu
\sigma\partial_\mu \sigma + \frac{1}{2}\partial_\mu \bm{\pi} \cdot
\partial_\mu \bm{\pi} = \frac{1}{2}\frac{\left(\bm{\pi}\cdot \partial_\mu
\bm{\pi}\right)\left(\bm{\pi}\cdot \partial_\mu
\bm{\pi}\right)}{f_\pi^2 - \bm{\pi}^2} + \frac{1}{2}\partial_\mu
\bm{\pi} \cdot
\partial_\mu \bm{\pi} \nonumber\\
& = & \frac{1}{2}\partial_\mu \bm{\pi} \cdot
\partial_\mu \bm{\pi} + \frac{1}{2f_\pi^2}\left(\bm{\pi}\cdot \partial_\mu
\bm{\pi}\right)\left(\bm{\pi}\cdot \partial_\mu \bm{\pi}\right) +
\cdots . \label{eq:nonlinearsigmapi}
\end{eqnarray}
This Lagrangian tells us that, after integrate out the scalar meson field, the linear sigma model becomes nonrenormalizable in four dimensional space-time.

By using Eq.~\eqref{eq:eomsigma}, the $M$ field in the linear sigma model~\eqref{eq:lagrlight} with decomposition~\eqref{eq:decompM} can be rewritten as
\begin{eqnarray}
M & = & \sigma \mathbb{I} + i {\bm\tau} \cdot {\bm\pi} = \mathbb{I}\sqrt{f_\pi^2 - \bm{\pi}^2} + i {\bm\tau} \cdot {\bm\pi} = f_\pi \left[\mathbb{I}\sqrt{1 - \frac{\bm{\pi}^2}{f_\pi^2}} + i \frac{{\bm\tau} \cdot {\bm\pi}}{f_\pi}\right].
\end{eqnarray}
By defining new field variables $\phi_i$ relating to
the $\pi_i$ fields through
\begin{eqnarray}
\frac{\tau_i \pi_i}{f_\pi} & = & \sin\left(\frac{\tau_i
\phi_i}{f_\pi}\right),\label{eq:piphi}
\end{eqnarray}
one can express the meson field $M$ in terms of $\phi_i$ as
\begin{eqnarray}
M & = & \sigma + i\tau_i \pi_i = f_\pi \left(\cos\left(\frac{\tau_i
\phi_i}{f_\pi}\right) + i \sin\left(\frac{\tau_i
\phi_i}{f_\pi}\right)\right) = f_\pi \exp\left(i\frac{\tau_i
\phi_i}{f_\pi}\right).
\label{eq:linearMnlU}
\end{eqnarray}
We introduce a new variable $U(x)$ through definition
\begin{eqnarray}
U(x) & \equiv & 
\exp\left(i\frac{\tau_i
\phi_i}{f_\pi}\right),\label{eq:defineUofM}
\end{eqnarray}
which, under chiral transformation, transforms as $U(x) \to g_L
U(x)g_R^\dag$ and is unitary $U^\dag U = UU^\dag = 1$. And, due to the intrinsic odd parity of the pion, we have the parity transformation
\begin{eqnarray}
U(\mathbf{x},t) & \rightarrow & \exp\left({} - i\frac{\tau_i
\phi_i(- \mathbf{x},t)}{f_\pi}\right) = U^\dagger(-\mathbf{x},t).
\label{eq:defineUP}
\end{eqnarray}
Using the field $U(x)$, we finally rewrite the kinetic term of the linear sigma
model as~\footnote{In the present approach, there is no $O(p^4)$ term. The potential term just provides a constant.}
\begin{eqnarray}
{\cal L}_{\rm L\sigma M} & = & {\rm Tr}(\partial _\mu M
\partial^\mu M^{\dag}) =
\frac{f_\pi^2}{4}{\rm Tr}\left(\partial _\mu U(x)
\partial^\mu U^{\dag}(x) \right). \label{eq:lagrnls}
\end{eqnarray}
Note that the unitary matrix $U(x)$ does not define a vector space of chiral group because the sum of two $SU(2)$ matrices is not an $SU(2)$ matrix. The realization of chiral symmetry through $U(x)$ is called a nonlinear realization and after substituting \eqref{eq:linearMnlU} into the linear sigma model the model is called nonlinear sigma model.

One possible choice of the explicit chiral symmetry breaking term $V_{SB}$ in Eq.~\eqref{eq:lagrlight}
is $V_{SB} = -c\sigma$ with $c$ as a parameter. By using Eqs.~\eqref{eq:transvectorsp} and \eqref{eq:transaxialsp} one can easily check that such a choice indeed breaks the chiral symmetry explicitly. By using Eq.~\eqref{eq:eomsigma} and expanding the pion fluctuations with respect to the QCD vacuum, one has
\begin{eqnarray}
V_{SB} & = & -c\sigma = -cf_\pi \cos\left(\frac{\tau_i
\phi_i}{f_\pi}\right) = -cf_\pi +
\frac{c}{2}\frac{\bm{\phi}^2}{f_\pi} +
\mathcal{O}\left(\frac{1}{f_\pi^3}\right),
\end{eqnarray}
which yields $c = m_\pi^2 f_\pi$. Including this $V_{SB}$ term in
the Lagrangian (\ref{eq:lagrnls}) one gets the equation of motion of
pion as
\begin{eqnarray}
\partial_\mu \partial^\mu \phi^i = m_\pi^2 \phi^i,
\end{eqnarray}
therefore, by using (\ref{eq:currentva}), we have
\begin{eqnarray}
\partial_\mu J_{5}^{\mu \, i} & = & \sigma \partial_\mu \partial^\mu \pi^i
= f_\pi \cos\left(\frac{\tau_i
\phi_i}{f_\pi}\right) \partial_\mu \partial^\mu \pi^i \simeq
f_\pi \partial_\mu \partial^\mu \pi^i =  m_\pi^2 f_\pi \pi^i,
\end{eqnarray}
which is the standard partially conserved axial-vector current (PCAC) relation.

So far, the predictions from the nonlinear sigma model agree with the low energy requirement of the meson dynamics based on the chiral symmetry. The nonlinear sigma model can be extended to include fermions, such as baryons with preserving the chiral invariance, but we will not discuss this aspect here. Below, we will focus on the phenomena noted by Skyrme long time ago: The nonlinear sigma model consists an intrinsic topological structure which yields non-perturbative field configurations that can be regarded classical baryons.

In order to study the electroweak processes of pseudoscalar mesons, one should include the electroweak gauge bosons in the nonlinear sigma model. In the nonlinear sigma model, the source for the electroweak gauge bosons is the chiral symmetry. The interaction between the electroweak gauge bosons and pseudoscalar mesons can be obtained by gauging the chiral symmetry of the nonlinear sigma model and relating the flavor symmetry of the nonlinear sigma model to the flavor symmetry of QCD. From the Lagrangian (\ref{eq:lagrnls}) one has
\begin{eqnarray}
{\cal L}_{\rm NL\sigma}^{\rm gauged} & = & \frac{F_\pi^2}{4} {\rm
Tr}\left[D_\mu U D_\mu U^\dag \right],\label{gaugenonlinearsigma}
\end{eqnarray}
where the covariant derivative is defined as
\begin{eqnarray}
D_\mu U & = & \partial _ \mu U - i \mathscr{L}_{\mu} U + i U
\mathscr{R}_{\mu}, \label{eq:exsourceu}
\end{eqnarray}
with $\mathscr{L}_{\mu}$ and $\mathscr{R}_{\mu}$ being the gauge fields
corresponding to the gauged left- and right-handed chiral symmetries,
respectively.

By matching the chiral symmetry of QCD to the transformation of the
field $U(x)$ one can express these gauge fields in terms of the
electroweak gauge bosons as
\begin{eqnarray}
\mathscr{R}_{\mu} & = & -e Q A_\mu - g \frac{\sin^2\theta_W}{\cos\theta_W}Q Z_\mu , \nonumber\\
\mathscr{L}_{\mu} & = & -e Q A_\mu + g Q_Z Z_\mu +
\frac{g}{\sqrt{2}}(W_\mu^+ Q_W + W_\mu^- Q_W^\dag),
\label{def:gauge}
\end{eqnarray}
where $Q$ is the charge matrix of quarks which in the two flavor
case $Q = {\rm diag}(2/3,-1/3)$ and $e = g \sin\theta_W$. The
matrices $Q_W$ and $Q_Z$ are defined as
\begin{eqnarray}
Q_W & = & \left(
            \begin{array}{cc}
              0 & V_{ud} \\
              0 & 0 \\
            \end{array}
          \right); \;\;\;\;\;\;\; Q_Z = \frac{1}{\cos\theta}\left(
            \begin{array}{cc}
              1/2 & 0 \\
              0 & -1/2 \\
            \end{array}
          \right) - \frac{\sin^2\theta_W}{\cos\theta_W}Q,
\end{eqnarray}
where $V_{ud}$ is the appropriate Cabibbo-Kobayashi-Maskawa matrix
elements. In \eqref{eq:exsourceu} $g$ is the coupling constant of the ${\rm SU}(2)_L$
weak gauge group in the standard model and at the lowest order
perturbation theory it is determined by the Fermi constant and the W
boson mass via the relation
\begin{eqnarray}
G_F & = & \sqrt{2}\frac{g^2}{8m_W^2} = 1.16637(1) \times 10^{-5}
{\rm GeV}^{-2}.
\end{eqnarray}
As an example, from (\ref{gaugenonlinearsigma}) one can get the
following $W^+\pi^-$ interaction Lagrangian
\begin{eqnarray}
{\cal L}_{W^+\pi^-} & = & - \frac{gF_\pi
}{2}V_{ud}W_\mu^+\partial_\mu \pi^-.
\end{eqnarray}

From the above discussions one concludes that the nonlinear sigma
model possesses the following properties of low energy QCD dynamics:
\begin{enumerate}
  \item It covers the chiral symmetry and the spontaneous chiral symmetry breaking
  of low energy QCD.

  \item It shows the origin of pseudoscalar meson mass by including
  the explicit chiral symmetry term.

  \item The chiral symmetry of QCD can be regarded as a source of the electroweak gauge boson, i.e., the electroweak gauge boson can be included in the nonlinear sigma model by gauging the chiral symmetry.
\end{enumerate}

\subsection{Power counting mechanism, loop correction and higher order terms of the chiral perturbation theory.}

\label{ssec:PowerCounting}

The nonlinear sigma model discussed above is the leading order term of the chiral perturbation theory (ChPT) which is a powerful effective field theory for the processes of pions in the low-energy QCD. Here, we briefly discuss the power counting mechanism of the ChPT. An effective field theory in particle physics should have two properties: The scale below which the theory is applicable and the consistent power counting mechanism which can be used to order various terms. In the ChPT, the scale of the theory can be estimated through some physical processes, such as $\pi$-$\pi$ scattering to one-loop. In such a way, the scale of the ChPT is found to be arround $\Lambda_\chi \simeq 4\pi F_\pi \simeq 1.1~$GeV~\cite{Manohar:1983md}. Next, we consider the second property, the power counting mechanism.

The Lagrangian of the ChPT, as an effective theory of the strong processes including only pseudoscalar mesons, due to the Lorentz invariance, has the general form
\begin{eqnarray}
{\cal L}_{\mbox{ChPT}}=\sum_{n=1}^\infty {\cal
L}^{2n}_{eff}.\label{eq:genChPT}
\end{eqnarray}
Since in the practical calculation, it is impossible to exhaust all
the terms in the effective Lagrangian, we should find a criteria to
estimate the weight of the contributions from different terms in
(\ref{eq:genChPT}).

Compared to the chiral symmetry breaking scale $\Lambda_\chi \simeq 1.1$~GeV, the
pseudoscalar meson mass $m_\pi \simeq 140$~MeV is a small quantity. Therefore it is
reasonable to regard $m_\pi/\Lambda_\chi$ as an expansion parameter
in the ChPT. So that, in the ChPT we take the derivative on the
pseudoscalar field as $\mathcal{O}(p)$. Since, when we consider the
explicit chiral symmetry breaking induced by the light current quark mass which is proportional to the pseudoscalar meson mass
square, it is counted as $\mathcal{O}(p^2)$. Once the external sources $\mathscr{L}_\mu$ and $\mathscr{R}_\mu$ are
introduced in the way of Eq.~(\ref{eq:exsourceu}), because they
always appear in company with the derivative, they can be taken as
$\mathcal{O}(p)$. Sometimes, scalar $\mathscr{S}$ and pseudoscalar
$\mathscr{P}$ sources are introduced in company with the light quark
mass, therefore they can be regarded as $\mathcal{O}(p^2)$. With
this criteria, all the terms in the chiral perturbation can be
arranged. We summarized the counting rules of the operators in the ChPT in Table.~\ref{table:chiralorderopera}.
\begin{table}%
\caption{Counting rules of the operators and fields in ChPT.
\label{table:chiralorderopera}}
\begin{tabular}{llllllll}
\hline \hline  Operator/field \, & \, $U$ \, & \, $\partial_\mu$ \, & \, $\mathscr{L}_\mu$  \, & \, $\mathscr{R}_\mu$ \, & \, $\mathscr{S}$  \, & \, $\mathscr{P}$  \, & \, $m_q$  \\
\hline
 Counting rule & \, $\mathcal{O}(1)$ \, & \, $\mathcal{O}(p)$ \, & \, $\mathcal{O}(p)$ \, & \, $\mathcal{O}(p)$  \, & \, $\mathcal{O}(p^2)$  \, & \, $\mathcal{O}(p^2)$  \, & \, $\mathcal{O}(p^2)$ \\
\hline \hline
\end{tabular} 
\end{table}

Using the power counting mechanism discussed above, the ChPT can be
constructed to any order. We will not discuss the details of the
construction (see, e.g., Ref.~\cite{Wei:79,Gasser:1983yg,Gas:85a}) but only list the next to leading order $\mathcal{O}(p^4)$
terms related to the $4$-point vertices of Nambu-Goldstone bosons
here
\begin{eqnarray}
{\cal L}_4 & = & L_1 \mbox{Tr}\left[ D_{\mu}U D^{\mu}U^{\dag}
D_{\nu}U D^{\nu}U^{\dag} \right ] + L_2 \mbox{Tr}\left[D_{\mu}U
D_{\nu}U^\dag D^{\mu}U D^{\nu}U^{\dag} \right] ,\label{eq:LP4}
\end{eqnarray}
where the covariant derivative $D_\mu$ is define by
Eq.~(\ref{eq:exsourceu}). In this Lagrangian, the coefficients $L_i$ include the information
of the fundamental QCD.

So far, the coefficients of ChPT are mainly fixed from model calculation or phenomena. For example, $L_1$ and $L_2$ are found to be $L_1 = ( -1.7 \pm 3.8)\times\times 10^{-3} $ and $L_2 = (1.3 \pm 0.7)\times 10^{-3}$ at $m_\rho$ scale~\cite{Harada:2003jx} and the leading order anomalous part of ChPT fixed from topological consideration~\cite{WZ,Veneziano:1977zk,Rosenzweig:1979ay,Witten:1983tw}. However, since ChPT is a low energy effective theory of QCD, its low energy constants should, in principle, be determined from fundamental QCD. Such explorations are perfomed in, e.g., Refs.~\cite{Wang:1999cp,Yang:2002hea,Yang:2002re,Wang:2002rb,Jiang:2009uf,Ma:2003uv,Jiang:2010wa}.

After the establishment of the counting rules of the operators and fields appearing in
the ChPT, following Refs.~\cite{Wei:79,Harada:2003jx}, we next discuss the chiral order of a matrix element $M$ with $N_e$ external $\pi$ lines. The dimension of this matrix element is given by
\begin{equation}
D_1 \equiv \mbox{dim} ( M ) = 4 - N_e \ .
\end{equation}

In ChPT, due to the relation between quark mass matrix and the Lorentz invariance, the interaction Lagrangian with $d$ derivatives, $k$ pion fields and $j$ quark mass matrices is symbolically expressed as
\begin{equation}
g_{d,j,k} (m_\pi^2)^j (\partial)^{2d} (\pi)^k  \ ,
\label{eq:Lchptcount}
\end{equation}
where the dimension of the coupling constant $g_{d,j,k}$ is
\begin{equation}
\mbox{dim} ( g_{d,j,k} ) = 4 - 2d - 2 j - k \ .
\end{equation}
Let $\bar{N}_{d,j,k}$ denote the number of the above interaction included in a diagram for the matrix element $M$. Then the total dimension carried by all the coupling constants in the matrix element is given by
\begin{equation}
D_2 = \sum_{d,j,k}\bar{N}_{d,j,k} ( 4 - 2d - 2 j - k) \ .
\end{equation}
By simply counting the number of pion fields, one can easily show
\begin{equation}
\sum_k \bar{N}_{d,j,k} k  = 2 N_i  + N_e \ ,
\end{equation}
with $N_i$ being the total number of internal $\pi$ lines. So that we can obtain
\begin{equation}
D_2 = \sum_{d,j} N_{d,j} ( 4 - 2d - 2 j ) - 2 N_i - N_e \ ,
\end{equation}
where $ N_{d,j} \equiv \sum_k \bar{N}_{d,j,k}$. Since each loop in a diagram corresponds to an independent momentum, the vertex number, internal line number and loop number $N_L$ has the relation
\begin{equation}
N_L = N_i - \sum_{d,j} N_{d,j} + 1 \ ,
\end{equation}
then $D_2$ becomes
\begin{equation}
D_2 = 2 - 2 N_L + N_e + \sum_{d,j} N_{d,j} ( 2 - 2d - 2 j ) \ .
\end{equation}

Generally, the matrix element $M$ can be expressed as
\begin{equation}
M = E^D m_\pi^{D_3} f\left( E/\mu ,\, M_\pi/\mu \,\right) \ ,
\label{M D D3}
\end{equation}
where $\mu$ is a common renormalization scale and $E$ is a common
energy scale. From Lagrangian \eqref{eq:Lchptcount}, the value of $D_3$ is determined by counting the number of vertices
with $m_\pi$ as
\begin{equation}
D_3 = \sum_{d,j} N_{d,j} (2 j) \ .
\end{equation}
$D$ is given by subtracting the dimensions carried by the coupling
constants and $m_\pi$ from the total dimension of the
matrix element $M$, i.e.,
\begin{equation}
D = D_1 - D_2 - D_3 =
2 + \sum_{d,j} N_{d,j} ( 2d - 2 ) + 2 N_L \ .
\end{equation}

Since in ChPT, the derivative expansion is performed in the low energy region around
the $\pi$ mass scale: The common energy scale is on the order of
the $\pi$ mass, $E \sim m_\pi$, and both $E$ and $m_\pi$ are much
smaller than the chiral symmetry breaking scale $\Lambda_\chi$,
i.e.,
$E$, $m_\pi \ll \Lambda_\chi$.
Then. the order of the matrix element $M$ in the derivative expansion,
denoted by $\bar{D}$,
is determined by counting the dimension
of $E$ and $m_\pi$ appearing in $M$:
\begin{equation}
\bar{D} = D + D_3 = 2 + \sum_{d,j} N_{d,j} ( 2d + 2 j - 2 )
+ 2 N_L \ .
\end{equation}
Note that $N_{1,0}$ and $N_{0,1}$ can be any number: these do not
contribute to $\bar{D}$ at all.

Based on the above discussions, we can classify the diagrams contributing to the matrix element $M$ according to the value of the above $\bar{D}$.
Let us list all the possible contributions for $\bar{D} = 2$ and $4$.
\begin{enumerate}
\item $\bar{D} = 2$ \\
This is the lowest order.
In this case, $N_L=0$: There are no loop contributions.
The leading order diagrams are tree diagrams in which the vertices are
described by the two types of terms: $(d,\,j)=(1,0)$ or
$(d,\,j)=(0,1)$.
Note that $(d,\,j)=(1,0)$ term includes $\pi$ kinetic term, and
$(d,\,j)=(0,1)$ term includes $\pi$ mass term.

\item $\bar{D} = 4$

\begin{enumerate}
\item $N_L=1$.

 In such case, $\sum_{d,j}N_{d,j}(2d + 2j - 2) = 0$. So that $N_{d,j}=0$ if $(d,j) \neq (1,0)$, $(0,1)$.  Then we conclude that these diagrams are one-loop diagrams in which all the vertices are of leading order.

\item $N_L=0$

In such case, $\sum_{d,j}N_{d,j}(2d + 2j - 2) = 2$. So that $(d,j) \neq (1,0)$, $(0,1)$.

\renewcommand{\theenumiii}{(\roman{enumiii})}
\begin{enumerate}
\item $N_{2,0}=1$, $(d,j) = (2,0)$;

\item $N_{1,1}=1$, $(d,j) \neq (1,1)$;

\item $N_{0,2}=1$, $(d,j) \neq (0,2)$.
\end{enumerate}

These diagrams are tree diagrams in which only one next order vertex is included. The next order vertices are described by
$(d,\,j)=(2,0)$, $(1,1)$ and $(0,2)$.

\end{enumerate}

\end{enumerate}

It should be noticed that we
included only logarithmic divergences in the above arguments. When we include quadratic divergences using some regularization scheme, loop integrals generate the terms proportional to the cutoff which are renormalized by the dimensional coupling constants.

\subsection{Topology of the nonlinear sigma model}

\label{ssecc:topology}

Here, we discuss the topology of the nonlinear sigma model which is essential for understanding the Skyrme model along the procedure of Ref.~\cite{Holzwarth:1985rb}. For this purpose, it convenient to consider the unitary field $U(x)$ defined in Eq.~(\ref{eq:defineUofM}). From (\ref{eq:currentlr}), the left- and right-handed currents are derived to be
\begin{eqnarray}
J_{\mu L} & = & J_{\mu L}^a T^a ={} - i f_\pi^2 U
\partial_\mu U^\dag \equiv {} - i f_\pi^2 L_\mu, \nonumber\\
J_{R\mu} & = &  = J_{R\mu}^aT^a = {} - i f_\pi^2 U^\dag
\partial_\mu U \equiv{} - i f_\pi^2 R_\mu,
\label{eq:lrcurrrents}
\end{eqnarray}
where $T^a = \tau^a/2$. One can show that under chiral transformation they transform in the
following way
\begin{eqnarray}
L_\mu & \rightarrow & g_L L_\mu g_L^\dagger , \;\;\;\; R_\mu
\rightarrow g_RR_\mu g_R^\dagger, \label{eq:lrcurrrentstrans}
\end{eqnarray}
which indicates that $R_\mu (L_\mu)$ is covariant under right (left)
chiral transformations.\footnote{Since $\det U = 1$, we have $\partial_\mu \det U = \partial_\mu \exp {\rm Tr}\ln U = {\rm Tr}\left[ L_\mu\right] = {\rm Tr}\left[ R_\mu \right] =
0.$}
And, for the weakly interacting pion fields, $L_\mu$ and $R_\mu$ reduce to
\begin{eqnarray}
L_\mu & = &{} -R_\mu \simeq \frac{i}{f_\pi} {\bm \tau}\cdot
\partial_\mu {\bm \pi}.\label{eq:chiralcurrentspi}
\end{eqnarray}

Because the matrix field $U(x)$ is unitary, at any fixed time, the matrix $U(\mathbf{x})$ defines a map from $R^3$ to the manifold $S^3$. Since at low energy limit, QCD goes to the vacuum accounted for by $\langle 0 | \sigma | 0 \rangle = f_\pi$,
\begin{eqnarray}
U(|\mathbf{x}| \to \infty)& = & \mathbf{1}.
\label{eq:configuxinfinite}
\end{eqnarray}
This limit tells us that all the points at $|\mathbf{x}| \to \infty$ are mapped onto the north pole of $S^3$ and energy of the system is finite. We then finally have the nontrivial map
\begin{eqnarray}
U(\mathbf{x}): R^3 \to S^3, \label{eq:mapU2S3}
\end{eqnarray}
for the static configuration $U(\mathbf{x})$. From mathematics we know that it is possible to categorize all the maps into homotopically distinct classes according to the times that the sphere $S^3$ is covered while $\mathbf{x}$ takes all values of coordinate space. In the language of topology, these maps constitute the third homotopy group $\pi_3(S^{\,3}) \sim Z$ with $Z$ being the additive group of integers which accounts for the times that $S^3$ is covered by the mapping $U(\mathbf{x})$, i.e., winding numbers. Because a change of the time coordinate can be regarded as homotopy transformation which cannot transit between the field configuration in homotopically distinct classes, the winding number is a conserved quantity in the homotopy transformation by the unitary condition of the field $U(x)$ and condition \eqref{eq:configuxinfinite}.

\begin{figure}[htbp]
\begin{center}
\includegraphics[scale=0.5]{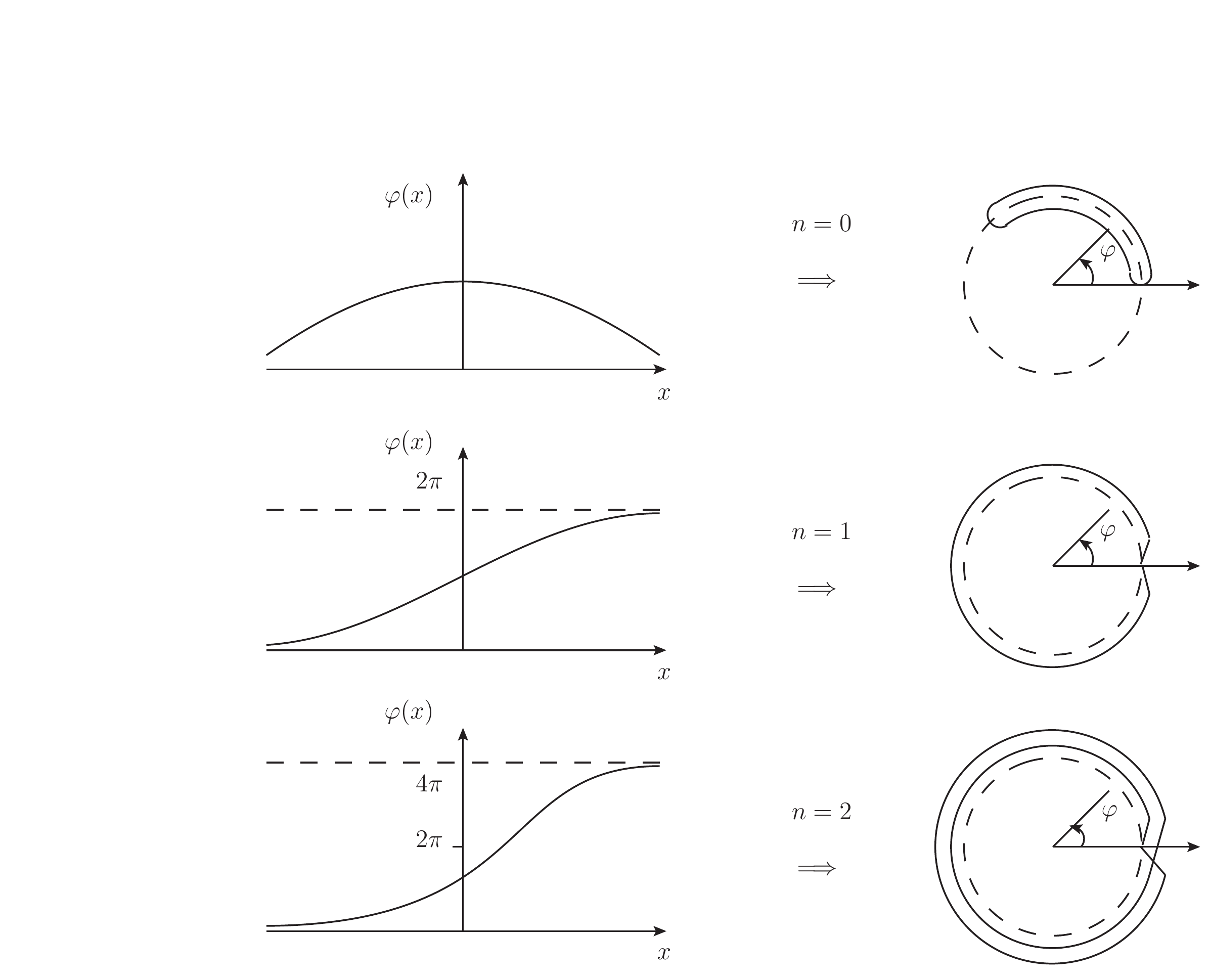}
\end{center}
\caption[]{%
Mappings of the $x$ axis on $S^1$ by $\varphi(x)$
with winding numbers $n = 0,1,2$.} \label{fig:winding1d}
\end{figure}

To illustrate the above discussion, we first consider a one-dimension example~\cite{Holzwarth:1985rb} where there is only one static field variable $\varphi(x)$. We assume a system with the energy given by
\begin{eqnarray}
E & = & \int_{-\infty}^{+\infty}\left[\left(\frac{\partial
\varphi}{\partial x}\right)^2 +
\sin^2\left(\frac{\varphi}{2}\right)\right]dx. \label{eq:energy1d}
\end{eqnarray}
In this expression, the two terms in the integrand are both non-negative. Therefore, if the system has finite energy, the static field $\varphi(x)$ should satisfy the boundary conditions
\begin{eqnarray}
\left\{
  \begin{array}{ll}
    \varphi \to 2\pi n_+,  & \;\;\;\;\;\;\; \hbox{for $x \to +\infty$;} \\
    \varphi \to 2\pi n_-,   & \;\;\;\;\;\;\; \hbox{for $x \to -\infty$,}
  \end{array}
\right.,  \label{eq:boundary1d}
\end{eqnarray}
with $n_\pm$ being integers and $ d\varphi/dx \to 0$ for $x \to \pm \infty$. Any function $\varphi(x)$ which is continuous, differentiable and satisfy boundary conditions (\ref{eq:boundary1d}) defines a map from the $x$ axis to a circle $S^1$ labelled by the angle
$\varphi(x)$. The mappings of $\varphi(x)$ can be illustrated in Fig.~\ref{fig:winding1d}.

From Fig.~\ref{fig:winding1d} one can easily arrive at the conclusions:
\begin{itemize}
\item $n_+ - n_- = 0$: The image of the $x$ axis on the circle $S^1$ may be contracted to a point, i.e., does not wind the circle.
\item $n_+ - n_- = 1$:  The image of the $x$ axis winds the circle $S^1$ and cannot be contracted on $S^1$ to a point.
\end{itemize}
Then, one can generalize these conclusions to say that the difference $n = n_+ - n_-$ counts the number of times of the image of the $x$ axis winds $S^1$, and therefore is called the winding number or the topological index.
Note that even we extend the fields $\varphi(x)$ as time dependent quantities $\varphi(x,t)$, the index $n$ is conserved because the continuous changes of the boundary conditions will involve infinite energy configurations for the non-integer $n_\pm$ and therefore are forbidden.

The winding number $n$ of the system can be determined by defining a
conserved current
\begin{eqnarray}
B^\mu & = & \frac{1}{2\pi}\epsilon^{\mu\nu}\partial_\nu
\varphi(x,t),
\label{eq:1dWindN}
\end{eqnarray}
where $\epsilon^{\mu\nu}$ is the antisymmetric tensor in two-dimension. Here we take the convention $\epsilon^{01} ={} - \epsilon^{10} = 1$ and $x_0 = t, x_1 = x$. From definition~\eqref{eq:1dWindN} one can conclude $\partial_\mu B^\mu = 0$ and compute the corresponding ``charge'' as,
\begin{eqnarray}
B & = & \int_{-\infty}^{+\infty} B^0 dx = \frac{1}{2\pi}\left(
\varphi(+\infty,t) - \varphi(-\infty,t)\right) = n_{+} - n_{-}  \, ,
\end{eqnarray}
which is the topological index $n$.

In topology, the above example in one spatial dimension could be stated that the homotopy group $\pi_1(S^1)$ is the group of integers under addition.

We next generalize the above discussion to the nonlinear sigma model in three spatial dimension. In such a case, we have to deal with the mapping $B^0$ which maps $R^3$ into $S^3$, i.e.,
\begin{eqnarray}
B^0: \pi_3(S^{\,3}) = Z.
\end{eqnarray}
For convenience, we introduce the notation
\begin{eqnarray}
\phi^0 = \frac{\sigma}{f_\pi},\;\;\;\;\; \phi^i =
\frac{\pi^i}{f_\pi}. \label{eq:topnonlinearphi}
\end{eqnarray}
which is the vector
representation of ${\rm SU}(2)_L \times {\rm SU}(2)_R$. In the nonlinear case, using the constraint (\ref{eq:constraintlsm}), the $\phi^i(x)$ can be regarded as an angular variable. Then the discussion in the one-dimension example can be easily extended to the nonlinear sigma model.

\begin{figure}[htbp]
\includegraphics[scale=0.7]{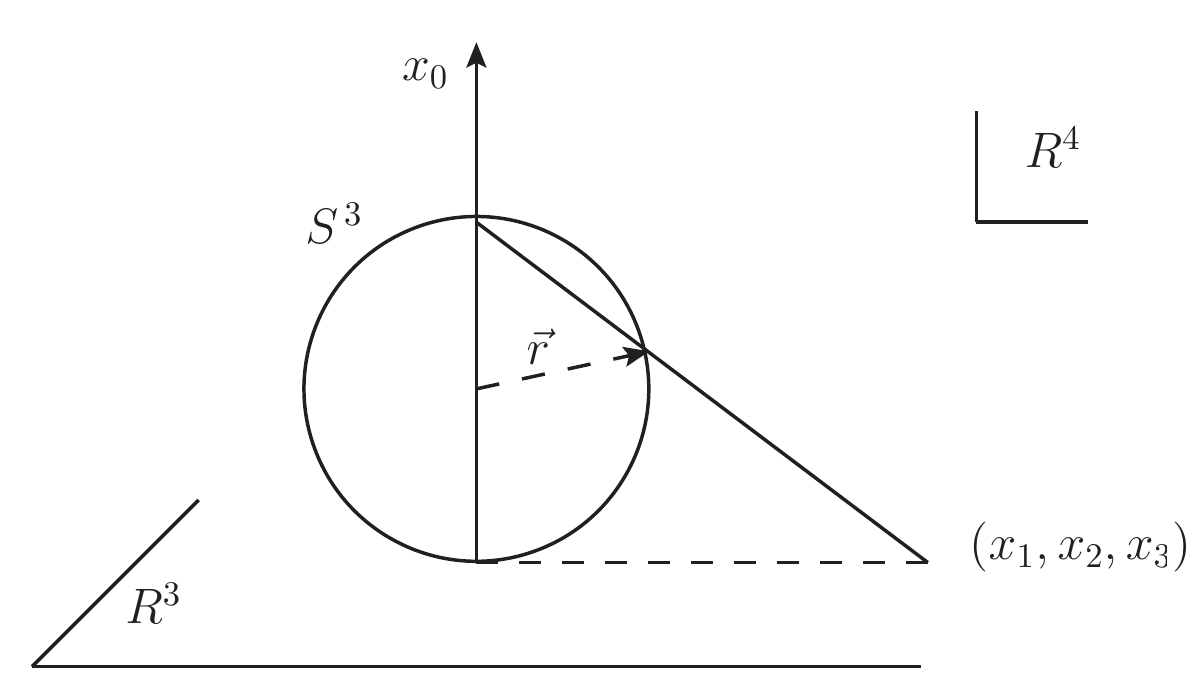}
\caption[Stereographic projection of a point on $S^3$ onto $R^3$ in $R^4$. The north pole is the center of the projection.]{%
Stereographic projection of a point on $S^3$ onto $R^3$ in $R^4$.}
\label{fig:stereoproj}
\end{figure}

In the group manifold, in terms of $\phi^i$, a fundamental surface element is characterized
by
\begin{eqnarray}
d^3\Sigma =
\epsilon^{ijkl}\phi^i\partial_1\phi^j\partial_2\phi^k\partial_3\phi^l
dx^1 dx^2 dx^3, \label{eq:surface}
\end{eqnarray}
where, as illustrated
in Fig.~\ref{fig:stereoproj}, $x^i$ are the corresponding coordinates on $R^3$
obtained by stereographic projection from $S^{\,3}$.
In affine geometry, \eqref{eq:surface} is the Jacobian associated to the transformation: $S^{\,3}\to R^3$. Hence, in analogy to \eqref{eq:1dWindN}, one concludes the normalized
topological density as
\begin{eqnarray}
B^0 = \frac{1}{12\pi^2}\epsilon^{ijkl}\epsilon^{\nu\alpha\beta}\phi^i
\partial_\nu \phi^j \partial_\alpha \phi^k \partial_\beta \phi^l.
\label{eq:normtopdensity}
\end{eqnarray}
By using (\ref{eq:topnonlinearphi}), one can write Eq.~(\ref{eq:normtopdensity}) in terms of the left-handed current
(\ref{eq:lrcurrrents}) as
\begin{eqnarray}
B^0 & = & \frac{(-i)^3}{24\pi^2}\epsilon^{\nu\alpha\beta}{\rm
Tr}\left[
\partial_\nu\left(\frac{i\bm{\tau}\cdot\bm{\pi}}{f_\pi}\right)\partial_\alpha\left(\frac{i\bm{\tau}\cdot\bm{\pi}}{f_\pi}\right)\partial_\beta\left(\frac{i\bm{\tau}\cdot\bm{\pi}}{f_\pi}\right)\right] +
\mathcal{O}\left(\frac{1}{f_\pi^4}\right) \nonumber\\
& = & \frac{i}{24\pi^2} \epsilon^{0\nu\alpha\beta}{\rm Tr}\left[
L_\nu L_\alpha L_\beta \right], \label{eq:normtopdensitycurrent}
\end{eqnarray}
where the last equation follows from (\ref{eq:chiralcurrentspi}). Notice
that $B^0$ does not vanish if and only if all the $3$ pion degrees of
freedom, $\pi^0$, $\pi^\pm$, are excited. The expressed \eqref{eq:normtopdensitycurrent} clearly shows that
\begin{eqnarray}
\int_{R^3}B^0 d^3x & = & \mbox{winding number}\, .
\label{eq:topcharge}
\end{eqnarray}

The topological charge (\ref{eq:normtopdensitycurrent}) can be easily written into a Lorentz covariant form
\begin{eqnarray}
B^\mu & = & \frac{i\epsilon^{\mu\nu\alpha\beta}}{24\pi^2}{\rm
Tr}\left[L_\nu L_\alpha L_\beta \right],\label{eq:topcurrentskyr}
\end{eqnarray}
which is conserved in $R^3$. Note that the conservation of the current \eqref{eq:topcurrentskyr} is independent of the equations of motion of pion.

\newpage

\section{The Skyrme model}

\label{sec:skyrmodel}

We have learned that the finite energy configuration of the nonlinear sigma model has an intrinsically non-trivial structure characterized by the homotopy group $\pi_3(S^3)$. Consequently, the nonlinear sigma model has static and finite configurations that are characterized by the conserved topological charges. And one can expect that these conserved topological charges can be explained as some conserved quantities of QCD, such as the baryon number~\footnote{Note that, in the nonlinear sigma model, baryons arise as topological charges while mesons are fluctuations with respect to the trivial QCD vacuum. The sources of these two kinds are different so that they both can accommodate in one model.}.

In the case that one of the conserved topological charges in the simplest nonlinear sigma model is identified as the baryon number, the model might be regarded as an effective model for baryons. However, the static energy corresponding to the field configuration $U(\mathbf{x})$ in the nonlinear sigma model is unstable against the rescaling of the space coordinate. So that, to have a stable energy configuration, the nonlinear sigma model should be stabilized by introducing other terms such as the Skyrme term due to the pioneer work by T.H.R.Skyrme~\cite{Skyrme:1961vq}. The nonlinear sigma model with the Skyrme term is called Skyrme model. To endow the definite quantum numbers to the static solution $U(\mathbf{x})$, the collective rotation should be made and the standard quantum mechanics quantization should be done. After quantization, the effective model for baryon is established. In this section, we will discuss all these points in turn and also some applications of the model.

\subsection{The model}

Let us start to discuss the Skyrme model from the nonlinear sigma model Lagrangian
(\ref{eq:lagrnls}) expressed in terms of the field $U(x)$ which satisfies the classical equation of motion. From (\ref{eq:lagrnls}), one can derive the canonical momentum conjugating to the field $U(x)$ as
\begin{eqnarray}
\Pi_{ij} & = & \frac{\partial {\cal L}_{\rm NL\sigma}}{\partial
\dot{U}_{ij}} = \frac{f_\pi^2}{4} \frac{\partial }{\partial
\dot{U}_{ij}}\left(\partial_\mu U_{lm}\partial_\mu
U^\dag_{ml}\right) = \frac{f_\pi^2}{4} \frac{\partial }{\partial
\dot{U}_{ij}}\left(\partial_0 U_{lm}\partial_0 U^\dag_{ml}\right).
\end{eqnarray}
Using identity
\begin{eqnarray}
\partial_\mu U^\dag & = & - U^\dag \partial_\mu U U^\dag,
\end{eqnarray}
one has
\begin{eqnarray}
\Pi_{ij} & = &{} -\frac{f_\pi^2}{4} \frac{\partial }{\partial
\dot{U}_{ij}}\left(\partial_0 U_{lm}U^\dag_{mk}\partial_0
U_{kn}U^\dag_{nl}\right) ={} - \frac{f_\pi^2}{2} \left(U^\dag_{jl}
\partial_0
U_{lm}U^\dag_{mi}\right) = \frac{f_\pi^2}{2} \partial_0 U^\dag_{ji}.
\end{eqnarray}
So that the Hamiltonian density of the system can be obtained as
\begin{eqnarray}
\mathcal{H}_{\rm NL\sigma} & = & \Pi_{ij}\dot{U}_{ij} - {\cal L}_{\rm NL\sigma} = \frac{f_\pi^2}{4}{\rm Tr}\left(
\partial_0 U^\dag\partial_0 U\right) + \frac{f_\pi^2}{4}{\rm Tr}\left(\partial_i U(x)
\partial_i U(x)^{\dag} \right).
\end{eqnarray}
After space integral one can express the energy of the system as
\begin{eqnarray}
E_{\rm NL\sigma} & = & \int d^3 x \mathcal{H}_{\rm NL\sigma} \equiv E_{\rm rotation}^{\rm NL\sigma} + E_{\rm static}^{\rm
NL\sigma},\label{eq:energynlsm}
\end{eqnarray}
where
\begin{eqnarray}
E_{\rm rotation}^{\rm NL\sigma} & = & \frac{f_\pi^2}{4}\int d^3
x{\rm Tr}\left(
\partial_0 U^\dag\partial_0 U\right),\nonumber\\
E_{\rm static}^{\rm NL\sigma} & = & \frac{f_\pi^2}{4}\int d^3 x{\rm
Tr}\left(\partial_i U(x)
\partial_i U(x)^{\dag} \right).
\end{eqnarray}
So that, in the case of the static solution $U(\mathbf{x})$, only $E_{\rm static}^{\rm NL\sigma}$ exists.

To illustrate the stability of the nonlinear sigma model with the static configuration $U(\mathbf{x})$, let us consider the rescaling
of the space coordinates in $U(\mathbf{x})$ through
\begin{eqnarray}
U(\mathbf{x})\rightarrow U(\lambda\mathbf{x}) ,
\label{eq:rescalingU}
\end{eqnarray}
and for generality write the dimension of the space as $D$. Then the
scaling behavior of the static energy is
\begin{eqnarray}
E_{\rm static}^{\rm NL\sigma}(\lambda) & = & \frac{f_\pi^2}{4}\int
d^D x{\rm Tr}\left(\partial_i U(\lambda\mathbf{x})
\partial_i U(\lambda\mathbf{x})^{\dag} \right) \nonumber\\
& = & \frac{f_\pi^2}{4}\int d^D (\lambda x) \lambda^{-D}\lambda^2 {\rm
Tr}\left(\partial^\lambda _i U(\lambda\mathbf{x})
\partial^{\lambda}_{i} U(\lambda\mathbf{x})^{\dag} \right)\nonumber\\
& = & \lambda^{2-D} E_{\rm static}^{\rm
NL\sigma}.\label{eq:scalenergynlsm}
\end{eqnarray}
In the case of $D = 3$, one has
\begin{eqnarray}
E_{\rm static}^{\rm NL\sigma}(\lambda) & = & \frac{1}{\lambda}
E_{\rm static},
\end{eqnarray}
which explicitly shows that the energy decreases with the increase of the space scale. So that in three-dimensional space the configuration $U(\mathbf{x})$ is not stable.

To avoid the stability problem of the static energy, Skyrme introduced a term, the so-called Skyrme term, to stabilize the static energy by extending the nonlinear sigma model Lagrangian as
\begin{eqnarray}
{\cal L}_{\rm Skyr} & = & \frac{f_\pi^2}{4}{\rm
Tr}\left[\partial_\mu U\partial^\mu U^\dag \right] +
\frac{1}{32e^2}{\rm Tr}\left\{\left[U^\dag\partial_\mu U,U^\dag
\partial_\nu U \right]\left[U^\dag\partial^\mu U,U^\dag
\partial^\nu U \right]\right\} , \label{eq:lagrskyrme}
\end{eqnarray}
with $e$ as a dimensionless parameter which indicates the magnitude of the soliton. Using the same method as that was used in the derivation of (\ref{eq:energynlsm}) one can get the energy of the Skyrme model as
\begin{eqnarray}
E^{\rm Skyr} & = & E_{\rm rotation}^{\rm Skyr} + E_{\rm static}^{\rm
Skyr},
\end{eqnarray}
where
\begin{eqnarray}
E_{\rm static}^{\rm Skyr} & = & - \int d^3 x{\rm
Tr}\left[\frac{f_\pi^2}{4}L_i L_i
+ \frac{1}{32e^2}[L_i,L_j]^2 \right] = E_{\rm static}^{\rm Skyr,(2)} + E_{\rm static}^{\rm Skyr,(4)}, \nonumber\\
E_{\rm rotation}^{\rm Skyr} & = & -\int d^3 x{\rm Tr}\left[
\frac{f_\pi^2}{4}L_0 L_0 + \frac{1}{16e^2}[L_0,L_i]^2 \right] =
E_{\rm rotation}^{\rm Skyr,(2)} + E_{\rm rotation}^{\rm Skyr,(4)},
\label{eq:energyskyr}
\end{eqnarray}
with the subscript $(2)$ standing for the contribution from the nonlinear sigma model while the subscript $(4)$ representing the effect of the Skyrme term by regarding them as the $\mathcal{O}(p^2)$ and $\mathcal{O}(p^4)$ terms in the ChPT, respectively. And for simplicity, we have defined $L_\mu = U^\dag \partial_\mu U$.

Using the identity $\epsilon_{ijk}\epsilon_{lmk} = \delta_{il}\delta_{jm} - \delta_{im}\delta_{jl}$ one can prove the following relation
\begin{eqnarray}
{\rm Tr}\left(\epsilon_{ijk}L_iL_j \right)^2 & = & {\rm
Tr}\left[\epsilon_{ijk}\epsilon_{lmk}L_iL_jL_lL_m\right] = {\rm
Tr}\left[ L_lL_mL_lL_m - L_mL_mL_lL_l\right] \nonumber\\
& = & \frac{1}{2}{\rm
Tr}\left[ L_i, L_j\right]^2.
\end{eqnarray}
So that, with respect to the Cauchy-Schwartz inequality $A^2 + B^2 \geq 2 AB$ we have the following inequality for the static energy
\begin{eqnarray}
E_{\rm static}^{\rm Skyr} & = & - \int d^3 x{\rm
Tr}\left[\frac{f_\pi^2}{4}L_i L_i
+ \frac{1}{16e^2}(\epsilon_{ijk}L_iL_j)^2 \right]\nonumber\\
& = & - \frac{f_\pi^2}{4}\int d^3 x{\rm Tr}\left[L_i L_i
+ \frac{1}{8e^2f_\pi^2}(\sqrt{2}\epsilon_{ijk}L_iL_j)^2 \right]\nonumber\\
& = & \frac{f_\pi^2}{4}\int d^3 x{\rm Tr}\left[L_i L_i^\dag
+ \frac{1}{4e^2f_\pi^2}(\epsilon_{ijk}L_iL_j)(\epsilon_{lmk}L_l^\dag L_m^\dag) \right]\nonumber\\
& \geq & \frac{f_\pi^2}{4}\int d^3 x\left|{\rm Tr}\left( \frac{1}{ef_\pi}\epsilon_{ijk}L_iL_jL_k \right)\right|\nonumber\\
& \geq & 0 \, , \label{eq:lowboundskyreneg}
\end{eqnarray}
which means that the static energy is bounded from below and that for the Skyrme model should be larger than or equal to zero. In terms of the topological charge $B$ from Eq.~\eqref{eq:topcurrentskyr}, the above relation becomes
\begin{eqnarray}
E_{\rm static}^{\rm Skyr} & \geq & 6 \pi^2 \frac{f_\pi}{e} \left|B\right| \, ,
\label{eq:lowboundskyrenegBog}
\end{eqnarray}
which is the Bogomol'ny bound. The lower limit is saturated in case of $L_i$ is a self-dual field, i.e.,
\begin{eqnarray}
L_i & = &  \frac{1}{2ef_\pi} \epsilon_{ijk}L_iL_j \, ,
\end{eqnarray}
which is incompatible with the Maurer-Cartan equation \eqref{eq:maurercartaneq} given in the following. This means that the skyrmion energy should be larger than the Bogomol'ny bound.

Now, let us prove that, in the Skyrme model, the soliton is stable in three-dimension space. Considering the rescaling of the field $U(\mathbf{x})$ given by Eq.~(\ref{eq:rescalingU}) and using the same method as that was used in the derivation of Eq.~(\ref{eq:scalenergynlsm}) one has the following scaling behavior of the static energy
\begin{eqnarray}
E_{{\rm static}}^{\rm Skyr}(\lambda) & = & \lambda^{2-D} E_{\rm
static}^{\rm Skyr,(2)} + \lambda^{4-D} E_{\rm static}^{\rm
Skyr,(4)}.
\end{eqnarray}
So that, for three-dimension space, i.e., $D = 3$, one has
\begin{eqnarray}
\frac{d E_{{\rm static}}^{\rm
Skyr}(\lambda)}{d\lambda}\Big|_{\lambda = 1} & = &
\left[(2-D)\lambda^{1-D}E_{\rm static}^{\rm Skyr,(2)} +
(4-D)\lambda^{3-D}E_{\rm
static}^{\rm Skyr,(4)}\right]_{\lambda=1,D=3}\nonumber\\
& = & - E_{\rm static}^{\rm Skyr,(2)} + E_{\rm
static}^{\rm Skyr,(4)},\label{eq:1stdetiveskyrE}\\
\frac{d^2 E_{{\rm
static}}^{Skyr}(\lambda)}{d\lambda^2}\Big|_{\lambda = 1} & = &
\left[(2-D)(1-D)\lambda^{-D}E_{\rm static}^{\rm Skyr,(2)} +
(4-D)(3-D)\lambda^{2-D}E_{\rm
static}^{\rm Skyr,(4)}\right]_{\lambda=1,D=3}\nonumber\\
& = & 2 E_{\rm static}^{\rm Skyr,(2)}. \label{eq:2nddetiveskyrE}
\end{eqnarray}
The requirement for the extremum stable condition
\begin{eqnarray}
\frac{d E_{{\rm static}}^{\rm Skyr}(\lambda
)}{d\lambda}\Big|_{\lambda = 1} & = & 0
\, ,
\end{eqnarray}
leads to
\begin{eqnarray}
E_{\rm static}^{\rm Skyr,(2)} & = & E_{\rm static}^{\rm Skyr,(4)} =
\frac{1}{2} E_{\rm static}^{\rm Skyr},
\end{eqnarray}
which, using (\ref{eq:lowboundskyreneg}), shows that $E_{\rm
static}^{\rm Skyr,(2)} \geq 0$. Then, from the
Eq.~(\ref{eq:2nddetiveskyrE}), one has
\begin{eqnarray}
\frac{d^2 E_{\lambda {\rm static}}^{Skyr}}{d\lambda^2}\Big|_{\lambda
= 1} & = & 2 E_{\rm static}^{\rm Skyr,(2)} \geq 0.
\end{eqnarray}
This equation is the minimum stable condition which implies that the static energy (\ref{eq:energyskyr}) is indeed stable against the
space scaling.

After adding the Skyrme term, certain solutions of the equation of motion in the nonlinear sigma model becomes stable. The stabilized solutions in the Skyrme model are called Skyrme solitons or skyrmions. Here, soliton is the classical, stable structure with
finite energy in the nonlinear field theory. Skyrme believed that, in his theory, the solution with winding number $1 (B=1)$ is
a fermion, and, he also guessed that the skyrmion is a classical baryon~\footnote{In Ref.~\cite{Witten:1983tx}, Witten showed that when the color number of the underlying strong dynamics is odd, the soliton must be a fermion while when the color number of the underlying strong dynamics is even, for example in the technicolor theory which trigures the breaking of electroweak symmetry (for a review, see, e.g., Ref.~\cite{Hill:2002ap}), the sliton can be a boson. So that in QCD in which the color number $N_c = 3$, soliton must be a fermion.}.

Note that the Skyrme term can be interpreted as the higher order correction to the nonlinear sigma model, so that it is not the only term stabling the skyrmion as was shown in Eq.~(\ref{eq:LP4}). Here we will not consider other possibilities but only discuss the physics of the Skyrme model.

\subsection{Equation of motion of the skyrmion}

The Euler-Lagrange equation of the skyrmion can be derived from the least action principle~\footnote{Here we derive the EoM in terms of field $U(\mathbf{x})$. In the next part, the EoM of skyrmion can be derived in terms of the hedgehog ansatz field $F(r)$ in a compact way.}
\begin{eqnarray}
\delta S & = & \int d^4x \delta{\cal L}_{\rm Skyr} = 0.
\end{eqnarray}
From Eq.~(\ref{eq:lagrskyrme}) one has
\begin{eqnarray}
\delta {\cal L}_{\rm Skyr} & = & -\frac{f_\pi^2}{2} {\rm Tr}\left[\left(\delta L_\mu
\right)L^\mu \right] + \frac{1}{16e^2} {\rm
Tr}\left\{\delta\left(\left[L_\mu ,L_\nu \right]\right)\left[L^\mu ,
L^\nu \right]\right\},\label{eq:deltalagr}
\end{eqnarray}
where $L_\mu$ was defined as $L_\mu = U^\dag \partial_\mu U$.

By using the unitary condition $UU^\dag = 1$ one can easily obtain
\begin{eqnarray}
\delta U^\dag & = & - U^\dag \left(\delta U\right) U^\dag,
\end{eqnarray}
which leads to the following relation
for $\delta L_\mu$
\begin{eqnarray}
\delta L_\mu & = & \delta (U^\dag \partial_\mu U) = - U^\dag \left(\delta U\right) L_\mu + L_\mu U^\dag (\delta U)
+ \partial_\mu( U^\dag \delta U). \label{eq:deltalmu}
\end{eqnarray}
Consequently, the first trace term in Eq.~(\ref{eq:deltalagr}) is reduced to
\begin{eqnarray}
{\rm Tr}\left[\left(\delta L_\mu \right)L^\mu \right] & = & {\rm
Tr}\left[ - U^\dag \left(\delta U\right) L_\mu L^\mu + L_\mu U^\dag
(\delta U)L^\mu + \partial_\mu( U^\dag \delta U)L^\mu \right] = {\rm
Tr}\left[ \partial_\mu( U^\dag \delta U)L^\mu \right] \nonumber\\
& = & -{\rm Tr}\left[ (\partial_\mu L^\mu) U^\dag \delta U \right] +
{\rm Tr}\left[ \partial_\mu( U^\dag \delta U L^\mu)
\right].\label{eq:deltalmutr}
\end{eqnarray}

To obtain the contribution to the equation of motion from the second
term of the Skyrme model, one should resort to the Maurer-Cartan
equation of $L_\mu$\footnote{Using the definition of $L_\mu$ we
have
\begin{eqnarray}
\partial_\mu L_\nu & = & \partial_\mu \left(U^\dag \partial_\nu
U \right) = - U^\dag \partial_\mu U U^\dag \partial_\nu U + U^\dag
\partial_\mu\partial_\nu U.
\end{eqnarray}
So that
\begin{eqnarray}
\partial_\mu L_\nu - \partial_\nu L_\mu & = & - U^\dag \partial_\mu U U^\dag \partial_\nu U + U^\dag
\partial_\mu\partial_\nu U + U^\dag \partial_\nu U U^\dag \partial_\mu U - U^\dag
\partial_\nu\partial_\mu U \nonumber\\
& = & - U^\dag \partial_\mu U U^\dag \partial_\nu U + U^\dag
\partial_\nu U U^\dag \partial_\mu U \nonumber\\
& = & - \left[ L_\mu, L_\nu \right],
\end{eqnarray}
i.e,
\begin{eqnarray}
\partial_\mu L_\nu - \partial_\nu L_\mu + \left[ L_\mu, L_\nu
\right] = 0 \, .
\end{eqnarray}
}
\begin{eqnarray}
\partial_\mu L_\nu - \partial_\nu L_\mu & = & - \left[ L_\mu, L_\nu
\right]. \label{eq:maurercartaneq}
\end{eqnarray}
From this equation we have
\begin{eqnarray}
\delta\left[ L_\mu, L_\nu \right] & = & -\partial_\mu \delta L_\nu +
\partial_\nu \delta L_\mu \nonumber\\
& = & -\partial_\mu \left[ - U^\dag \left(\delta U\right) L_\nu +
L_\nu U^\dag (\delta U) \right] +
\partial_\nu \left[ - U^\dag \left(\delta U\right) L_\mu +
L_\mu U^\dag (\delta U)  \right],
\end{eqnarray}
where use has been made to Eq.~(\ref{eq:deltalmu}). So that we
obtain the second trace term in Eq.~(\ref{eq:deltalagr}) as
\begin{eqnarray}
 {\rm
Tr}\left\{\delta\left(\left[L_\mu ,L_\nu \right]\right)\left[L^\mu ,
L^\nu \right]\right\} & = &  2{\rm Tr}\left\{\partial_\mu \left[ U^\dag \left(\delta
U\right) \left[L_\nu, \left[L^\mu , L^\nu \right]\right]\right] -
U^\dag \left(\delta U\right) \left[L_\nu,\partial_\mu\left[L^\mu ,
L^\nu \right]\right] \right\} . \nonumber\\
\label{eq:deltalmulnu}
\end{eqnarray}

Combining Eqs.~(\ref{eq:deltalagr},\ref{eq:deltalmutr},\ref{eq:deltalmulnu}) we obtain
\begin{eqnarray}
\delta S & = & \int d^4x \left\{\frac{1}{8e^2}{\rm Tr}\left\{\partial_\mu \left[
U^\dag \left(\delta U\right) \left[L_\nu, \left[L^\mu , L^\nu
\right]\right]\right]\right\}  - \frac{f_\pi^2}{2} {\rm Tr}\left[
\partial_\mu( U^\dag \delta U L^\mu) \right] \right. \nonumber\\
& & \left. \qquad\qquad + \frac{f_\pi^2}{2}{\rm Tr}\left[ (\partial_\mu L^\mu)
U^\dag \delta U \right] - \frac{1}{8e^2} {\rm Tr}\left\{ U^\dag
\left(\delta U\right) \left[L_\nu,\partial_\mu\left[L^\mu , L^\nu
\right]\right] \right\}\right\} .
\end{eqnarray}
Omitting the surface term, we arrive at the equation of motion of
the field $U(x)$ as
\begin{eqnarray}
\frac{f_\pi^2}{2} (\partial_\mu L^\mu) - \frac{1}{8e^2}
\left[L_\nu,\partial_\mu\left[L^\mu , L^\nu \right]\right] =
0.\label{eq:eomu1}
\end{eqnarray}

By using the relation
\begin{eqnarray}
\partial_\mu L_\nu & = & \left(\partial_\mu U^\dag\right) \partial_\nu U + U^\dag
\left(\partial_\mu\partial_\nu U\right) =  -L_\mu L_\nu + U^\dag
\partial_\mu\partial_\nu U,
\end{eqnarray}
we obtain the following relation
\begin{eqnarray}
\left[L_\nu,\partial_\mu\left[L^\mu , L^\nu \right]\right] & = &
\partial_\mu\left[L_\nu,\left[L^\mu,L^\nu\right]\right]- \partial_\mu L_\nu\left[L^\mu ,
L^\nu \right] +
\left[L^\mu , L^\nu \right]\partial_\mu L_\nu \nonumber\\
& = &
\partial_\mu\left[L_\nu,\left[L^\mu,L^\nu\right]\right] + L_\mu L_\nu \left[L^\mu , L^\nu \right] -
\left[L^\mu , L^\nu \right]L_\mu L_\nu \nonumber\\
& = &
\partial_\mu\left[L_\nu,\left[L^\mu,L^\nu\right]\right].
\end{eqnarray}
With respect to this relation we finally obtain the equation of
motion of $U(\mathbf{x})$ as
\begin{eqnarray}
\frac{f_\pi^2}{2} (\partial_\mu L^\mu) - \frac{1}{8e^2}
\partial_\mu\left[L_\nu,\left[L^\mu , L^\nu \right]\right] = 0.\label{eq:eomu}
\end{eqnarray}

\subsection{The hedgehog ans$\ddot{a}$tz of the skyrmion}

Eq.~(\ref{eq:eomu}) is a highly nonlinear equation which, therefore, can only be handled in some special cases. Under the assumption of
maximal symmetry, Skyrme proposed a so-called hedgehog
ansatz of the solution of Eq.~(\ref{eq:eomu})
\begin{eqnarray}
U(\mathbf{x}) & = & \exp\left(i \bm{\tau}\cdot
\hat{\mathbf{x}}F(r)\right) = \cos F(r) + i \bm{\tau}\cdot
\hat{\mathbf{x}} \sin F(r). \label{eq:hedgehog}
\end{eqnarray}
Ansatz (\ref{eq:hedgehog}) is based on the following
considerations~\cite{Holzwarth:1985rb}: To have a nonvanishing topological charge
(\ref{eq:topcharge}), the mapping $U(\mathbf{x})$ should cover the $3$-sphere at least once in a non-contractible way. This means, in the general parametrization $U = \exp(i{\bm\tau}\cdot{\bm\phi}) = \cos\phi + i {\bm\tau}\cdot
\hat{{\bm\phi}}\sin \phi$, for every value of $\phi$, $\hat{{\bm\phi}}$ must
cover the unit sphere $S^2$ in the isospace in a non-contractible way. In other words, for any constant $\phi$, if $\mathbf{x}$ takes all values in three-dimensional space, the unit isovector
$\hat{{\bm\phi}}(\mathbf{x})$ must cover the full
solid angle $4\pi$ in isospace. Then, the simplest choice of $\hat{{\bm\phi}}(\mathbf{x})$ is
\begin{eqnarray}
\hat{{\bm\phi}}(\mathbf{x}) & = & \hat{\mathbf{x}}.
\end{eqnarray}
With respect to the fact that the static energy \eqref{eq:energyskyr} involves only squares of derivatives of $U$, it is reasonable to expect that the minimal energy of the system can be obtained from a purely radial dependent chiral angle
$\phi(\mathbf{x}) = F(r)$.

The boundary conditions of $F(r)$ can be established in the following way:
\begin{itemize}
\item To keep the total energy of the system finite, $U(\mathbf{x})$ must smoothly
approach to a real constant for $r \to \infty$. Therefore, at $r = \infty$, one can choose $F(\infty) = n_{\infty} \pi$ with $n_\infty$ being integers.

\item Since the origin of the three-dimensional space must be mapped onto a
single point on $S^3$, one has to require $F(0) =
n_0\pi$. And, to have a nonzero winding number, one should have $n = n_{\infty} - n_0 \neq 0$. Without loss of generality, we choose $n_{\infty} = 0$. Then, all functions satisfying $F(0) = n\pi$ lead to $n$-fold non-contractible covering of $S^3$.
\end{itemize}

From
the ansatz (\ref{eq:hedgehog}), one can easily check that neither isospin $(I)$ nor spin
$(J)$ is a good quantum number but their sum
\begin{eqnarray}
\mathbf{K} & = & \mathbf{J} + \mathbf{I}
\end{eqnarray}
is a good quantum number. It is easy to check that $U(\mathbf{x})$ is invariant under
rotations in $K$-space
\begin{eqnarray}
[\mathbf{K}, U(\mathbf{x})] & = & i \sin
F\left\{\left[\left(\mathbf{x} \times
\frac{\bm{\nabla}}{i}\right),\bm{\tau}\cdot \hat{\mathbf{x}}\right]
+ \left[ \frac{\bm{\tau}}{2}, \bm{\tau}\cdot
\hat{\mathbf{x}}\right]\right\} \nonumber\\
& = & i\sin F\left\{-i\left( \bm{\tau}\times \hat{\mathbf{x}}\right)
- i\left( \hat{\mathbf{x}}\times\bm{\tau} \right)\right\} = 0 .
\label{eq:kspinskyr}
\end{eqnarray}
By using~\eqref{eq:defineUP} one concludes that that the ansatz (\ref{eq:hedgehog}) is invariant under parity transformation. Therefore, in the hedgehog ansatz, skyrmions have quantum numbers $K^P = 0^+$ and can be regarded as an admixture of states with $I = J$.

Substituting the ansatz (\ref{eq:hedgehog}) into Eq.~(\ref{eq:eomu}) and using identities
\begin{eqnarray}
r^2 & = & \mathbf{x}^2 = x^i x^i, \;\;\;\;\;\;\; {\bm\tau}\cdot \mathbf{x} =
\tau^i x^i, \nonumber\\
\partial_i r & = & \frac{\partial}{\partial x^i}\sqrt{ x^j x^j} =
\frac{x^j}{r} = \hat{x}^j , \quad \partial_i({\bm\tau} \cdot \mathbf{x}) =
\frac{\partial}{\partial x^i}({\bm\tau} \cdot
\mathbf{x}) = \tau^i,
\end{eqnarray}
one can easily obtain the soliton mass as
\begin{eqnarray}
M_{\rm Skyr} & = & 4\pi \int d r r^2\left[\frac{f_\pi^2}{2}\left( 2 \frac{\sin^2F}{r^2} + F^{\prime 2}\right) + \frac{1}{2 e^2} \frac{\sin^2F}{r^2}\left( \frac{\sin^2F}{r^2} + 2 F^{\prime 2}\right) \right]
,\label{eq:skyrmass}
\end{eqnarray}
which, as expected in the Large $N_c$ expansion, is of
$\mathcal{O}(N_c).$

Using the soliton mass \eqref{eq:skyrmass} one can obtain the equation of motion for the profile function $F(r)$ though minimizing the $M_{\rm Skyr}$. Straightforward derivation yields
\begin{eqnarray}
\left( r^2 + \frac{2}{e^2 f_\pi^2} \sin^2 F\right) F^{\prime\prime} + 2r F^{\prime} + \frac{1}{ e^2 f_\pi^2} \sin 2F \, F^{\prime2} - \sin 2F - \frac{\sin^2F \sin 2F}{e^2 f_\pi^2 r^2 }& = & 0,
\label{eq:eomprofileF}
\end{eqnarray}
and, to describe baryon number-one baryons, the solution of this equation should satisfies the boundary
conditions
\begin{eqnarray}
F(r=0) & = & \pi,\;\;\;\;\; F(r \to \infty) = 0 .
\label{eq:boundaryF}
\end{eqnarray}

\begin{figure}[htbp]
\includegraphics[scale=0.5]{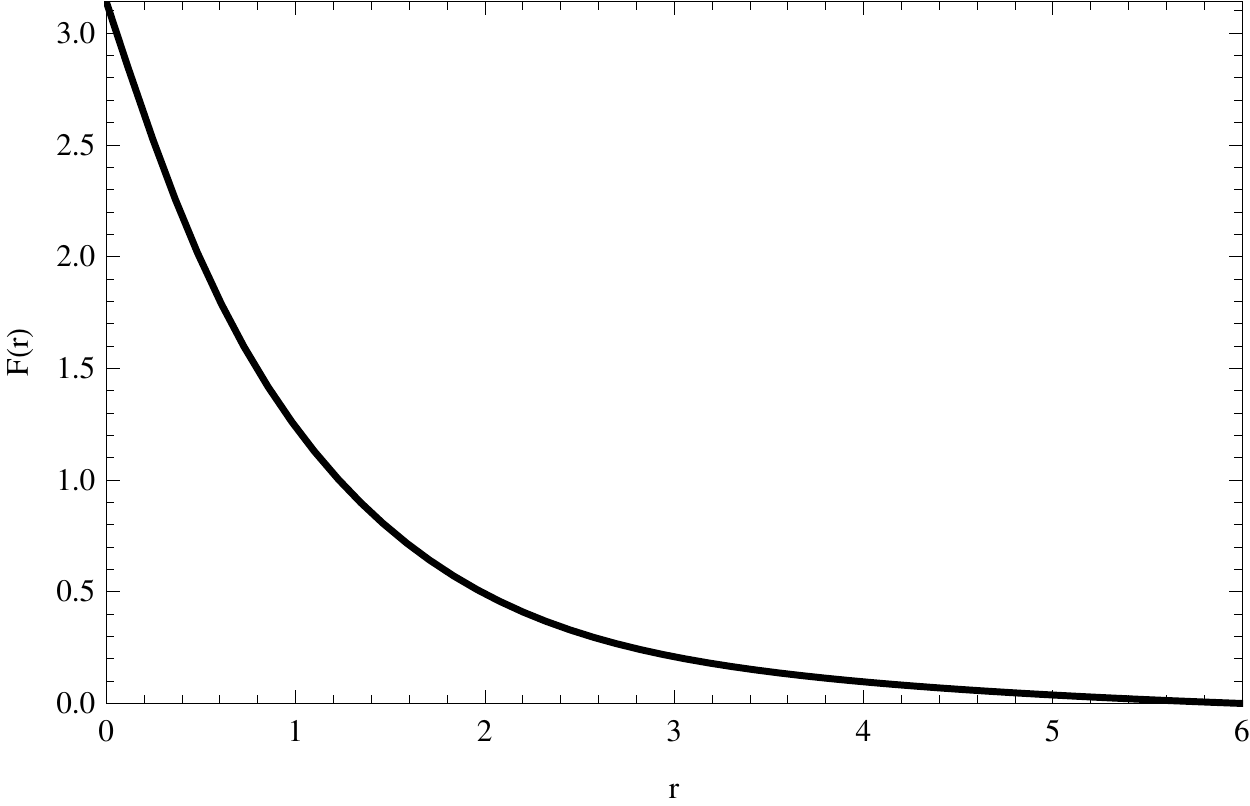}
\caption{Solution of the EoM~\eqref{eq:eomprofileFrs} in terms of dimensionless coordinate through rescaling~\eqref{eq:rescale}.}
\label{fig:numericalF}
\end{figure}

Next we make the coordinate transformation
\begin{eqnarray}
r \to r^\prime /(ef_\pi) \label{eq:rescale}
\end{eqnarray}
 then the skyrmion mass and equation of motion of profile $F(r)$ are reexpressed as
\begin{eqnarray}
& & M_{\rm Skyr} = 4\pi \frac{f_\pi}{e}\int d r r^2\left[\frac{1}{2}\left( 2 \frac{\sin^2F}{r^2} + F^{\prime 2}\right) + \frac{1}{2} \frac{\sin^2F}{r^2}\left( \frac{\sin^2F}{r^2} + 2 F^{\prime 2}\right) \right], \label{eq:skyrmassrs}\\
& & \left( r^2 + 2 \sin^2 F\right) F^{\prime\prime} + 2r F^{\prime} + \sin 2F \, F^{\prime2} - \sin 2F - \sin^2F \sin 2F  = 0,\label{eq:eomprofileFrs}
\end{eqnarray}
where we have written the dimensionless coordinate $r^\prime$ as $r$. EoM~\eqref{eq:eomprofileFrs} tells us that, in the Skyrme model, in terms of the dimensionless coordinate, the solution of the profile function $F(r)$ is independent of the parameter $f_\pi$ and $e$. Moreover, in terms of the dimensionless coordinate, the skyrmion mass can be calculated as $M_{\rm Skyr} = C  \times f_\pi/e$ with $C$ as a dimensionless quantity independent of $f_\pi/e$ which indicates that $M_{\rm Skyr} \sim O(N_c)$ since $f_\pi \sim O(\sqrt{N_c})$ and $e \sim O(1/\sqrt{N_c})$. The solution of the EoM~\eqref{eq:eomprofileFrs} can be plotted as Fig.~\ref{fig:numericalF} and the skyrmion mass is obtained as
\begin{eqnarray}
M_{\rm Skyr} = 74.58 \times \left(\frac{f_\pi}{e}\right). \label{eq:massSkyrdimless}
\end{eqnarray}

By using the profile function \eqref{eq:hedgehog}, one can obtain the
topological charge density (\ref{eq:normtopdensitycurrent}) as
\begin{eqnarray}
B^0(r) & = &
\frac{1}{2\pi^2}\sin^2F\frac{F^\prime}{r^2},\label{eq:baryondensityF}
\end{eqnarray}
which leads to the topological charge
\begin{eqnarray}
B & = & \frac{1}{\pi}\left(F(0) - F(\infty) \right) +
\frac{1}{2\pi}\left(\sin 2F(\infty) - \sin2F(0)\right).
\end{eqnarray}
Threfore, with the boundary conditions given by Eq.~\eqref{eq:boundaryF}, Skyrme model indeed describes the $B = 1$ classical baryon physics.

Although the boundary conditions $F(r=0) = n\pi,  F(r \to \infty) = 0 ~(n=2,3,\cdots)$ yield the topological charge $n$ baryon, this does not mean that one can safely choose this boundary conditions in the Skyrme model to describe the nuclei with baryon number $n$. This is because, using these boundary conditions, the obtained mass of the nuclei is larger than the total mass of the constituents therefore unstable. For example, when we use $F(r=0) = 2\pi,  F(r \to \infty) = 0$, the nuclei (Deutron) mass is three times of the mass of the nucleon obtained by taking $n=1$~\cite{Weigel:1986zc}.
For discribing nuclei by using Skyrme model, other ansatz than the hedgehog, such as the rational map ansatz~\cite{Houghton:1997kg}, should be used.

\subsection{The collective rotation}

In the previous discussion, since skyrmion is still a classical object, it does not have any quantum numbers. To endow skyrmion with definite spin and isospin quantum numbers, the system should be quantized. This quantization could be done by a collective rotation of the skyrmion which will be discussed in this part.

One can easily check that the Skyrme Lagrangian \eqref{eq:lagrskyrme} is invariant under the following rotations:
\begin{eqnarray}
U(\mathbf{x}) & \to & U(R\mathbf{x}); \;\;\;\; U(\mathbf{x}) \to C
U(\mathbf{x})C^\dag,
\end{eqnarray}
where $R$ is the spacial rotation matrix and $C$ is the isorotation matrix. Since the hedgehog profile function correlates the space rotation and the isorotation, it can be regarded as a superposition of states with all possible values of $C$ with $C$ as a time dependent ${\rm SU}(2)$-valued matrix.

Specifically, we introduce the ${\rm SU}(2)$ collective variables by
\begin{eqnarray}
U_c(\mathbf{x}) & \rightarrow & U(\mathbf{x},t) = C(t)
U_c(\mathbf{x}) C^\dag(t),\label{eq:mesoncollectiveskyr}
\end{eqnarray}
where $C(t)$ is a unitary matrix satisfying $C(t)C(t)^\dag = C(t)^\dag C(t) = 1$ and the subscript $c$ indicates that field $U$ is independent of time. Substituting (\ref{eq:mesoncollectiveskyr}) into the Skyrme model Lagrangian (\ref{eq:lagrskyrme}) one obtains the energy induced by the collective rotation as ($L_0$ depends on the time derivative of $C(t)$) Eq.~(\ref{eq:energyskyr}).

Defining the angular velocity $\Omega_i$ corresponding to the collective coordinate rotation by
\begin{eqnarray}
\frac{i}{2} \bm{\tau} \cdot \bm{\Omega} & \equiv & C^\dag
\partial_0 C,\label{eq:angularvelocity}
\end{eqnarray}
one can express the rotation energy $E_{\rm rotation}^{\rm Skyr}$ given in Eq.~(\ref{eq:energyskyr}) in terms of the angular velocity as
\begin{eqnarray}
E_{\rm rotation}^{\rm Skyr} & = & \mathcal{I}_{\rm Skyr} {\rm
Tr}\left[\partial_0 C^\dag(t)\partial_0C(t)\right] = \frac{1}{2}
\mathcal{I}_{\rm Skyr} \Omega^2, \label{eq:energyrotomega}
\end{eqnarray}
with $\mathcal{I}_{\rm Skyr}$ being the moment of inertia of the soliton configuration with respect to the rotation (\ref{eq:mesoncollectiveskyr}) which can be expressed in terms of $F(r)$ by using the same trick applied in the above calculations.

Explicit derivation of the moment of inertia $\mathcal{I}_{\rm Skyr}$ in terms of profile function $F(r)$ is the following: Under the rotation (\ref{eq:mesoncollectiveskyr}), we have
\begin{eqnarray}
L_\mu  & = & U^\dag \partial_\mu U = C(t)U_c^\dag C^\dag(t)
\partial_\mu \left( C(t)U_c C^\dag(t)\right) \nonumber\\
& = & C(t)U_c^\dag C^\dag(t)
\partial_\mu C(t)U_c C^\dag(t) + C(t)U_c^\dag \partial_\mu U_c C^\dag(t) + C(t)\partial_\mu
C^\dag(t),
\end{eqnarray}
which gives
\begin{eqnarray}
L_0  & = & \frac{i}{2}C(t)\left[ - 2 \sin^2 F {\bm \tau}\cdot {\bm \Omega} + 2 \sin^2F \mathbf{x}\cdot{\bm \Omega}\mathbf{x}\cdot{\bm \tau} - \sin 2F \epsilon_{ijk}\Omega_i x_j \tau_k \right] C^\dag(t) ,\nonumber\\
L_i & = & C(t)U_c^\dag \partial_i U_c C^\dag(t).
\label{eq:l0licollectskyr}
\end{eqnarray}
So that we have
\begin{eqnarray}
{\rm Tr}\left[L_0L_0\right] = - 2\sin^2F \left[\bm{\Omega}^2 - \left(\bm{\Omega}\cdot
\hat{\mathbf{x}}\right)^2\right].
\end{eqnarray}
Using
\begin{eqnarray}
\int \sin\theta d\theta d\phi & = &  4\pi; \;\;\;\;\; \int \hat{x}_i
\hat{x}_j \sin\theta d\theta d\phi = \frac{4\pi}{3}\delta_{ij},
\end{eqnarray}
one obtains
\begin{eqnarray}
-\frac{f_\pi^2}{4}\int d^3x {\rm Tr}\left[L_0L_0\right] & = & \frac{4\pi}{3} f_\pi^2 \int r^2 dr \sin^2F \bm{\Omega}^2.
\end{eqnarray}
Similar calculation leads to
\begin{eqnarray}
\int d^3 x {\rm Tr}\left[ L_0, L_i\right]^2 & = & -
\frac{64\pi}{3}\int r^2 dr \sin^2F \left( F^{\prime~2}+ \frac{\sin^2
F}{r^2}\right) \bm{\Omega}^2.
\end{eqnarray}
We then finally obtain the moment of inertia of the Skyrme model as
\begin{eqnarray}
\mathcal{I}_{\rm Skyr} & = & \frac{8\pi}{3} \int_0^\infty r^2 dr
\sin^2F \Big\{f_\pi^2 + \frac{1}{e^2}(F^{\prime \, 2} +
\frac{\sin^2F}{r^2})\Big\}.\label{eq:momentinatiaskyr}
\end{eqnarray}
Then, after scaling \eqref{eq:rescale} we express the moment of inertia as
\begin{eqnarray}
\mathcal{I}_{\rm Skyr} & = & \left(\frac{1}{e^3f_\pi}\right)\frac{8\pi}{3} \int_0^\infty r^2 dr
\sin^2F \Big\{1 + F^{\prime \, 2} +
\frac{\sin^2F}{r^2}\Big\},
\end{eqnarray}
which, by using solution of EoM $F(r)$, yields
\begin{eqnarray}
\mathcal{I}_{\rm Skyr} & = & 38.11 \times \left(\frac{1}{e^3f_\pi}\right).\label{eq:inertiaSkyrdimless}
\end{eqnarray}
This expression shows that $\mathcal{I}_{\rm Skyr} \sim O(N_c)$.

Following the classical mechanics, angular momentum of skyrmion can be stated as
\begin{eqnarray}
\bm{J} & = & \frac{\partial E_{\rm rotation}^{\rm Skyr}}{\partial
\bm{\Omega}} = \mathcal{I}_{\rm Skyr} \bm{\Omega},
\end{eqnarray}
so that rotation energy of skyrmion is
\begin{eqnarray}
E_{\rm rotation}^{\rm Skyr} & = & \frac{\bm{J}^2}{2\mathcal{I}_{\rm
Skyr} } .
\end{eqnarray}
Following the standard quantum mechanics, angular momentum is given by
\begin{eqnarray}
\bm{J}^2 & = & j(j+1)\hbar^2,
\end{eqnarray}
where $j = 0, 1/2 , 1, 3/2 , 2, \cdots $ and $\hbar = h/(2\pi)$ is the
Planck constant which could be conveniently taken as $\hbar = 1$. Then, after
this standard quantization procedure, the baryon masses can be
expressed as
\begin{eqnarray}
m_B & = & M_{\rm static}^{\rm Skyr} + \frac{j(j+1)}{2
\mathcal{I}_{\rm Skyr}}. \label{eq:massbaryonqunti1}
\end{eqnarray}
Specifically, the nucleon $J=1/2$ and $\Delta$ resonance $J=3/2$ masses could be ontained as
\begin{eqnarray}
M_N & = & M_{\rm Skyr} + \frac{3}{8\mathcal{I}_{\rm Skyr}}, \;\;\;\;
M_\Delta = M_{\rm Skyr} + \frac{15}{8\mathcal{I}_{\rm Skyr}},
\label{eq:massNDelta}
\end{eqnarray}
which gives the $N$-$\Delta$ mass splitting
\begin{eqnarray}
M_\Delta - M_N & = & \frac{3}{2\mathcal{I}_{\rm Skyr}}.
\label{eq:massNDeltasplit}
\end{eqnarray}

The numerical values of $ M_{\rm static}^{\rm Skyr}$ and $\mathcal{I}_{\rm Skyr}$ cannot be obtained before fixing the values of $f_\pi$ and $e$. One way to overcome this obstacle is to resort to the baryon spectrum, for example, take the masses of $\Delta$ and $N$ as input values as done in Ref.~\cite{Adkins:1983ya}~\footnote{An alternative way to determine these parameters is to use the meson dynamics from which $f_\pi$ is taken as the pion decay constant and $e$ is fixed from $\pi$-$\pi$ scattering. This will be discusses in the next chapter.}. In such a way, using \eqref{eq:massSkyrdimless}, \eqref{eq:inertiaSkyrdimless} and \eqref{eq:massNDelta}, we obtain the following values of $f_\pi$ and $e$
\begin{eqnarray}
f_\pi & = & 58.45~{\rm MeV}, \qquad e = 5.03 .\label{eq:paraskyr}
\end{eqnarray}

The quantization procedure can be achieved alternatively by using collective coordinates in terms of which both the physical operators and baryon states can be explicitly constructed. Since $C(t) \in {\rm SU}(2)$ and $C^\dag(t)C(t) = C(t)C^\dag(t) = 1$, it can be locally parametrized as
\begin{eqnarray}
C(t) = a_0(t) + i\bm{\tau}\cdot \mathbf{a}(t),\label{eq:paracollec}
\end{eqnarray}
with constraint
\begin{eqnarray}
a_0^2 + \mathbf{a}^2 = 1. \label{eq:constraintparacollec}
\end{eqnarray}
Thus $C(t)$ can be regarded as the collection of the time dependent canonical coordinates with the conjugate momentum to $a_k$ as
\begin{eqnarray}
\pi_k & = & \frac{\partial {\cal L}_{\rm Skyr}}{\partial \dot{a}_k}
= 4 \mathcal{I}_{\rm Skyr} \dot{a}_k;\;\;\;\;\; k = 0,1,2,3,
\label{eq:conjumomenta}
\end{eqnarray}
where in the last step, we have substituted the rotation (\ref{eq:mesoncollectiveskyr}) into Eq.~(\ref{eq:energyskyr}). Then the Hamiltonian associated to the collective rotation reads
\begin{eqnarray}
H & = & \int d^3 x \left[\sum_{k=0}^3\pi_i\dot{a}_i - {\cal L}_{\rm
Skyr}\right] = M_{\rm Skyr} + \frac{1}{8\mathcal{I}_{\rm
Skyr}}\sum_{k=0}^3\pi_k^2.
\end{eqnarray}
By using the operator form of the conjugate momentum, i.e. $\pi_k = -i\partial/\partial a_k$, one obtains the quantized Hamiltonian as
\begin{eqnarray}
H & = & M_{\rm Skyr} + \frac{1}{8\mathcal{I}_{\rm
Skyr}}\sum_{k=0}^3\left(\frac{1}{i}\frac{\partial}{\partial
a_k}\right)^2.\label{eq:hamiltcollec}
\end{eqnarray}

We next, construct the spin ($\mathbf{J}$) and isospin ($\mathbf{I}$) operators in terms of the collective coordinates~\cite{Zahed:1986qz}. Since in the parameterization of the collective rotation (\ref{eq:paracollec}) $a_i$ satisfies the constraint (\ref{eq:constraintparacollec}), one can parameterize $a_i$ in terms of the three independent variables $\xi$'s $(\xi = \theta, \phi,\psi)$ on $S^3$, e.g.,
\begin{eqnarray}
a_0 & = & \cos\theta, \;\;\;\; a_1 = \sin\theta \cos\phi, \;\;\; a_2
= \sin\theta \sin\phi \cos\psi, \;\;\; a_3 = \sin\theta \sin\phi
\sin\psi .\label{eq:paraais3angle}
\end{eqnarray}
In addition, the rotation induced energy (\ref{eq:energyrotomega}) is invariant under, respectively, ``rotations'' and ``isorotations''
\begin{eqnarray}
C(t) & \to & C(t)f\;\;\;\;\;\; \mbox{and}\;\;\;\;\;\; C(t) \to f
C(t), \;\;\;\;\;\;\; \mbox{with $f \in {\rm SU}(2)$},
\end{eqnarray}
and also a discrete $Z_2$-symmetry, $C(t) \to \pm C(t)$. Since $\det
C(t) = e^{\ln\det C(t)} = e^{{\rm Tr}\ln C(t)} = 1$, one has
\begin{eqnarray}
0 & = & {\rm Tr}\left[\frac{\partial \ln C(t)}{\partial \xi^a}\right] = {\rm Tr}\left[ C^{\dag}(t)\frac{\partial  C(t)}{\partial \xi^a}\right] = - {\rm Tr}\left[ C(t)\frac{\partial  C^{\dag}(t)}{\partial \xi^a}\right] .
\end{eqnarray}
Then, in terms of $\xi$'s one has the rotation induced Lagrangian \eqref{eq:energyrotomega} as
\begin{eqnarray}
{\cal L}_C = \frac{1}{2}\mathcal{I}_{\rm Skyr}\dot{\xi}_a
h_{ac}h_{bc}\dot{\xi}_b = \frac{1}{2}\mathcal{I}_{\rm
Skyr}\dot{\xi}^{\rm T} hh^{\rm T}\dot{\xi},
\label{eq:LagC}
\end{eqnarray}
where the index ${\rm T}$ stands for transposition and $h$ is defined though
\begin{eqnarray}
C^\dag(t) \frac{\partial C(t)}{\partial \xi^a} = i h_{ab}T_b,
\label{eq:derivecollectsphe}
\end{eqnarray}
with $T^a = \tau^a/2$.

We write the canonical momentum conjugating to $\xi$ as $\tilde{\pi}$, i.e.,
\begin{eqnarray}
\tilde{\pi}^{\rm T} = \frac{\partial {\cal L}}{\partial \dot{\xi}} =
\mathcal{I}_{\rm Skyr} \dot{\xi}^{\rm T} hh^{\rm T}.
\label{eq:conjugatemomentumxi}
\end{eqnarray}
Then, from ${\cal L}_C$, one can derive the corresponding hamiltonian density ${\cal H}$ as
\begin{eqnarray}
{\cal H} & = & \tilde{\pi}^{\rm T} \dot{\xi} - {\cal L}_C =
\frac{1}{2 \mathcal{I}_{\rm Skyr}}\tilde{\pi}^{\rm T}\left(hh^{\rm
T}\right)^{-1}\tilde{\pi}, \label{eq:hamiltonianxi}
\end{eqnarray}
where $\xi$ and $\tilde{\pi}$ satisfy the following Poisson brackets
\begin{eqnarray}
\{ \xi_a,\xi_b\}_{\rm P.B} & = & \{
\tilde{\pi}_a,\tilde{\pi}_b\}_{\rm P.B} = 0, \;\;\;\;\; \{
\xi_a,\tilde{\pi}_b\}_{\rm P.B} = \delta_{ab}.
\end{eqnarray}
Canonical quantization consists in postulating
\begin{eqnarray}
[ \xi_a,\xi_b ] & = & [ \tilde{\pi}_a,\tilde{\pi}_b ] = 0,
\;\;\;\;\; [ \xi_a,\tilde{\pi}_b ] = i\delta_{ab}.
\end{eqnarray}
By using Lagrangian \eqref{eq:LagC}, one can derive the the classical spin and isospin charges as
\begin{eqnarray}
J_a & = & i \mathcal{I}_{\rm Skyr}{\rm Tr}\left[\tau_a
\dot{C}^\dag(t) C(t)\right], \;\;\;\;\; I_a = i \mathcal{I}_{\rm
Skyr}{\rm Tr}\left[\tau_a \dot{C}(t) C(t)^\dag\right].
\label{eq:defineJI}
\end{eqnarray}
From these expressions, one can easily see that $J_a$ and $I_a$ link to each other though the substitution $C \leftrightarrow C^\dag$ which reflects the fact of the spin-isospin correlation in the skyrmion approach. From \eqref{eq:defineJI} we have
\begin{eqnarray}
J_k & = &
{} - 2 \mathcal{I}_{\rm
Skyr}\left(\dot{a}_0 a_k - \dot{a}_k a_0 + \epsilon_{klm}\dot{a}_l a_m\right).
\end{eqnarray}
So that, by using \eqref{eq:conjugatemomentumxi} we obtain
\begin{eqnarray}
J_k & = &
{} - \frac{1}{2} \left(\pi_0 a_k - \pi_k a_0 + \epsilon_{klm}{\pi}_l a_m \right) \nonumber\\
& = & \frac{i}{2}\left(a_k\frac{\partial}{\partial a_0} -
a_0\frac{\partial}{\partial a_k} - \epsilon_{klm}a_l
\frac{\partial}{\partial a_m}\right),
\label{eq:operspin}
\end{eqnarray}
where in the last equation we have substituted the moment $\pi_i$ with its operator expression. Similarly we obtain
\begin{eqnarray}
I_k & = & \frac{i}{2}\left(a_0\frac{\partial}{\partial a_k} -
a_k\frac{\partial}{\partial a_0} - \epsilon_{klm}a_l
\frac{\partial}{\partial a_m}\right).\label{eq:operisospin}
\end{eqnarray}
From the operator expressions \eqref{eq:operspin} and \eqref{eq:operisospin} we obtain
\begin{eqnarray}
\mathbf{J}^2 & = &
\frac{1}{4}\sum_{k=0}^3\left(-\frac{\partial^2}{\partial
a_k^2}\right).
\end{eqnarray}
Moreover, making use of Eqs.~(\ref{eq:derivecollectsphe},\ref{eq:conjugatemomentumxi}), we have the corresponding generators in (${\rm SU}(2); C$) as
\begin{eqnarray}
J_a & = & h_{ab}^{-1}\tilde{\pi}_b, \;\;\;\;\; I_a =
k^{-1}_{ab}\tilde{\pi}_b.\label{eq:operatorxi}
\end{eqnarray}
which fulfill the standard ${\rm SU}(2)$ algebra ( for a proof, see Ref.~\cite{Zahed:1986qz}). Similar relations hold for $I_a$'s. By using \eqref{eq:operatorxi} we obtain
\begin{eqnarray}
[J_a, C(t)] & = & h_{ab}^{-1}[\tilde{\pi}_b,C(t)] =
-ih^{-1}_{ab}\frac{\partial C(t)}{\partial \xi_b} = C(t)T_a,
\end{eqnarray}
where using has been made to Eq.~(\ref{eq:derivecollectsphe}). Similarly, we have
\begin{eqnarray}
[I_a, C(t)] & = & - T_a C(t).
\end{eqnarray}
The transformations of $C(t)$ under $\mathbf{J}$ and $\mathbf{I}$ state that $J_a$ and $I_a$ trigger ${\rm SU}(2)_R$ and ${\rm SU}(2)_L$ rotations, respectively. We summarize the matrix elements of $C(t)$ corresponding to the fundamental representation of spinor in Table.~\ref{table:spinreprect} and from the polynomials in $C(t)$ one can construct higher representations.
\begin{table}%
\caption{The matrix elements of $C(t)$ corresponding to the
fundamental spinor representation. \label{table:spinreprect}}
\begin{tabular}{llllllll}
\hline \hline  ~\, & \, $C_{11}$ \, & \, $C_{12}$ \, & \, $C_{21}$  \, & \, $C_{22}$  \\
\hline
 $J_3$ & \, $1/2$ \, & \, $-1/2$ \, & \, $1/2$ \, & \, $-1/2$ \\
 $I_3$ & \, $-1/2$ \, & \, $-1/2$ \, & \, $1/2$ \, & \, $1/2$ \\
 ~ & \, $| n\uparrow\rangle$ \, & \, $|n\downarrow\rangle$ \, & \, $|p\uparrow\rangle$ \, & \, $|p\downarrow\rangle$ \\
\hline \hline
\end{tabular} 
\end{table}
From Table.~\ref{table:spinreprect}, one can construct the wave functions of proton and neutron as
\begin{eqnarray}
| p\uparrow\rangle & = & \frac{1}{\pi}\left( a_1+ia_2\right)
;\;\;\;\;\; | p\downarrow\rangle = - \frac{i}{\pi}\left( a_0 -
ia_3\right) \nonumber\\
| n\uparrow\rangle & = & \frac{i}{\pi}\left( a_0 + ia_3\right)
;\;\;\;\;\; | n\downarrow\rangle = - \frac{1}{\pi}\left( a_1 -
ia_2\right), \label{eq:wavepn}
\end{eqnarray}
where the coefficients are normalization factors on $S^3$.
By using the polynomials in $C(t)$ one can show that the Skyrme model generates a tower of states with $I = J = 1/2, 3/2, 5/2, \cdots$. However, in a hedgehog configuration, by using a non-relativistic quark model, it has been argued that $I_{\rm max} = J_{\rm
max} = N_c/2$ which suggests that for $N_c= 3$ only $I = J = 1/2, 3/2$ are relevant and the rest are spurious~\cite{Zahed:1986qz}.

In terms of the generators (\ref{eq:operatorxi}), the Hamiltonian
(\ref{eq:hamiltonianxi}) reads
\begin{eqnarray}
H & = & \frac{J^2}{2\mathcal{I}_{\rm Skyr}} =
\frac{I^2}{2\mathcal{I}_{\rm Skyr}},
\end{eqnarray}
which indicates that the left and right Casimir operators of ${\rm
SU}(2)$ are identical which is consistent with
Eq.~(\ref{eq:hamiltcollec}). 

The nucleon and $\Delta$ resonance masses and their splitting can also be derived and all of them are consistent with Eqs.~(\ref{eq:massNDelta}, \ref{eq:massNDeltasplit}).

\subsection{Applications of the Skyrme model}


In this part, following Ref.~\cite{Adkins:1983ya}, we will make some applications of the Skyrme model to study some quantities of nucleons, such as the axial coupling $g_A$, the charge radii and magnetic moments of baryon.

\subsubsection{The axial coupling $g_A$}

The axial coupling $g_A$ is a quantity which measures the spin-isospin correlation in the nucleon. It is defined through the expectation value of the axial-vector current $J_{\mu5}^a(x)$ in a nucleon state at the limit of zero momentum transfer. From Lorentz structure and also the $C, P, T$ invariances, the matrix element is decomposed as
\footnote{The term of the form $(p_1+p_2)_\mu\bar{u}(p_2)\frac{\tau^a}{2} g_2(q^2)\gamma_5 u(p_1)$
is excluded by the CP invariance together with the hermiticity of
the axial-vector current $J_{\mu5}^a$.}
\begin{eqnarray}
\langle N(p_2)| J_{\mu5}^a| N(p_1)\rangle & = &
\bar{u}(p_2)\frac{\tau^a}{2}\left[ g_A(q^2)\gamma_\mu \gamma_5 + h_A
(q^2)q_\mu \gamma_5\right]u(p_1). \label{eq:defga}
\end{eqnarray}
with $(q = p_2 - p_1)$ being the momentum transferred to the axial-vector current.
With respect to the axial-current conservation in the chiral limit, one has
\begin{eqnarray}
2m_N g_A(q^2) + q^2h_A(q^2) = 0,
\end{eqnarray}
where the equation of the nucleon is used. Using this relation one can rewrite the matrix element (\ref{eq:defga}) in terms of $g_A$ as
\begin{eqnarray}
\langle N(p_2)| J_{\mu5}^a| N(p_1)\rangle & = &
\bar{u}(p_2)\frac{\tau^a}{2}\left[ g_A(q^2)\gamma_\mu \gamma_5
-q_\mu \gamma_5 \frac{2m_N}{q^2}g_A(q^2)\right]u(p_1) \nonumber\\
& = & \bar{u}(p_2)\frac{\tau^a}{2}\left[ g_A(q^2)\gamma_\mu \gamma_5
-q_\mu q_\nu \gamma_\nu \gamma_5 \frac{1}{q^2}g_A(q^2)\right]u(p_1),
\end{eqnarray}
where the EoM of the initial and final nucleon states have been used. Then, in the non-relativistic limit and soft pion limit, i.e., $q_0 = 0$ and $\mathbf{q} = 0$ we have
\begin{eqnarray}
\lim_{q\to 0}\langle N(p_2)| J_{i5}^a| N(p_1)\rangle & = &
\lim_{q\to 0}\bar{u}(p_2)\frac{\tau^a}{2}\left[g_A(q^2)g^\nu_i - \frac{q_i q^\nu}{q^2}g_A(q^2)\right]\gamma_\nu \gamma_5u(p_1) \nonumber\\
& = & \lim_{q\to 0} g_A(0) \left(\delta_{ij} -
\frac{q_iq_j}{\mathbf{q}^2}\right)\langle N|\sigma_j
\frac{\tau^a}{2}|N\rangle,
\end{eqnarray}
where we have used
\begin{eqnarray}
\gamma_j\gamma_5 & = & \left(
                         \begin{array}{cc}
                           \sigma_j & 0 \\
                           0 & -\sigma_j \\
                         \end{array}
                       \right),
\end{eqnarray}
and $| N\rangle$ corresponds to the large component of positive energy solution of the Dirac equation. Once the limit is taken in the symmetric way, i.e., $q_iq_j \to \delta_{ij}\mathbf{q}^2/3$, one has
\begin{eqnarray}
\lim_{q\to 0}\langle N(p_2)| J_{i5}^a| N(p_1)\rangle & = &
\frac{2}{3}g_A(0) \langle N|\sigma_i
\frac{\tau^a}{2}|N\rangle.\label{eq:gacurrent}
\end{eqnarray}
All the above derivations are based on the current algebra for a given value of $g_A(0)$.

Here, we determine the value of $g_A$ from the Skyrme model. In the Skyrme model, the space
part of the axial-vector current can be derived using the Noether's construction. Explicit calculation yields
\begin{eqnarray}
\int d^3x J_{i5}^a(\mathbf{x}) & = & - \frac{1}{2}g_{\rm Skyr}{\rm
Tr}\left(\tau_i C^\dag \tau^a C\right) \, ,
\label{eq:axialcurrentshyr}
\end{eqnarray}
where $g_{\rm Skyr}$ is expressed in terms of the profile function $F$ as
\begin{eqnarray}
g_{\rm Skyr} & = & \frac{4\pi}{3} \int_0^\infty r^2 dr
\left\{f_\pi^2\left(F^\prime + \frac{\sin2F}{r}\right) +
\frac{1}{e^2}\left(F^{\prime2}\frac{\sin2F}{r} +
2F^\prime\frac{\sin^2F}{r^2} +
\frac{\sin^2F}{r^3}\sin2F\right)\right\}. \nonumber\\
\label{eq:gaskyr}
\end{eqnarray}
After rescaling \eqref{eq:rescale} we obtain
\begin{eqnarray}
g_{\rm Skyr} & = & \frac{4\pi}{3} \frac{1}{e^2}\int_0^\infty r^2 dr
\left\{\left(F^\prime + \frac{\sin2F}{r}\right) +
\left(F^{\prime2}\frac{\sin2F}{r} +
2F^\prime\frac{\sin^2F}{r^2} +
\frac{\sin^2F}{r^3}\sin2F\right)\right\}, \nonumber\\
\end{eqnarray}
which indicates that $g_{\rm Skyr} \sim \mathcal{O}(N_c)$ since $e \sim \mathcal{O}(1/\sqrt{N_c})$.

Sandwiching the current (\ref{eq:axialcurrentshyr}) between the
nucleon states we have
\begin{eqnarray}
\lim_{q\to 0}\int d^3x\exp^{i\mathbf{q}\cdot \mathbf{x}}\langle N|
J_{i5}^a(\mathbf{x})| N\rangle & = & - \frac{1}{2}g_{\rm Skyr}
\langle N| {\rm
Tr}\left(\tau_i C^\dag \tau^a C\right)| N\rangle \nonumber\\
& = & \frac{2}{3}g_{\rm Skyr}  \langle N|\sigma_i
\frac{\tau^a}{2}|N\rangle,
\end{eqnarray}
where we have used the relation $\langle N^\prime| {\rm Tr}(\tau_i
C^\dag \tau^a C)| N\rangle = \frac{2}{3}\langle N^\prime| \sigma_i
\tau^i| N\rangle$, which can be proved using the nucleon wave
function (\ref{eq:wavepn}), satisfied for any nucleon states $N$ and
$N^\prime$. Identifying this expression with (\ref{eq:gacurrent})
one gets $g_A(0) = g_{\rm Skyr}$.  Therefore the axial coupling $g_A$ is $\mathcal{O}(N_c)$.

Using the numerical solution of the EoM of $F(r)$ one can obtain the following value of $g_A$
\begin{eqnarray}
g_A & = & 24.72 \times \left(\frac{1}{e^2}\right)  = 0.98,
\end{eqnarray}
where the value of $e$ determined from baryon spectrum is used. This value is deviated by about 30\% from the experimental value of $g_A = 1.27$~\cite{Beringer:1900zz} which is acceptable since the Skyrme model calculation only takes into account the leading $\mathcal{O}(N_c)$ effect.

\subsubsection{The charge radii and magnetic moments of baryons}

The isoscalar charge radius of a nucleon which accounts for the distribution of matter in it is expressed as
\begin{eqnarray}
r_0 & = & \langle r^2 \rangle_{I=0}^{1/2} = \left[ \int d^3x \, r^2
\, \rho_0\right]^{1/2},
\end{eqnarray}
with $\rho_0$ being the normalized baryon density.

In the Skyrme model, the normalized baryon density is the topological charge density $B^0(r)$, which with the hedgehog ansatz, is given by Eq.~(\ref{eq:baryondensityF}), i.e., $\rho_0 = B^0(r)$. So that we have
\begin{eqnarray}
r_0 & = & \left[\frac{2}{\pi}\int dr \, r^2 \sin^2F \,
F^\prime\right]^{1/2}.
\end{eqnarray}
And after rescaling \eqref{eq:rescale} we have
\begin{eqnarray}
r_0 & = & \frac{1}{e f_\pi}\left[\frac{2}{\pi}\int dr \, r^2 \sin^2F \,
F^\prime\right]^{1/2},
\end{eqnarray}
which means that $r_0$ is $\mathcal{O}(N_c^0)$. This $N_c$ order agrees with the large $N_c$ argument of the nucleon properties discussed at the end of subsection~\ref{subsec:largeNc}. By using the solution of the profile function $F(r)$, the numerical value of $r_0$ can be obtained as
\begin{eqnarray}
r_0 & = & 0.96\times \left(\frac{1}{e f_\pi}\right) = 0.66~{\rm fm}.\label{eq:r0skyr}
\end{eqnarray}
This result is about $25\%$ smaller than the
data $0.877 \pm 0.005$~fm~\cite{Beringer:1900zz} which is also acceptable at the leading $\mathcal{O}(N_c)$.

The isoscalar moment $\bm{\mu} _{I=0}$ and isovector magnetic moment $\bm{\mu} _{I=1}$ in the nucleon are defined in the rest frame as
\begin{eqnarray}
\bm{\mu} _{I=0} & = & \frac{1}{2}\int d^3x \, \mathbf{x}\times
\mathbf{B}, \nonumber\\
\bm{\mu} _{I=1} & = & \frac{1}{2}\int d^3x \, \mathbf{x}\times
\mathbf{V}^3, \label{eq:defmagneticmoment}
\end{eqnarray}
where $\mathbf{B}$ is the space component of the baryon current (\ref{eq:topcurrentskyr}), and $\mathbf{V}^3$ is the third component of the isovector current. It can be calculated that, for an adiabatically rotating skyrmion
\begin{eqnarray}
B^i & = & \frac{i}{2\pi^2}\epsilon^{ijk}F^\prime \hat{x}^j \ {\rm
Tr}\left[ \tau^k \dot{C}^\dag(t) C(t)\right].
\end{eqnarray}
Substituting this into (\ref{eq:defmagneticmoment}) and using the proton wave function (\ref{eq:wavepn}) one has the third component of the proton isoscalar magnetic moment as
\begin{eqnarray}
\left(\bm{\mu} _{I=0}\right)_3 & = & - \frac{2i}{3\pi}\int_0^\infty
dr \, r^2 \, F^\prime\sin^2F \langle p\uparrow|{\rm Tr}\left[\tau_3
\dot{C}^\dag
C\right]|p\uparrow\rangle \nonumber\\
& = & \frac{i}{3}\langle r^2\rangle_{I=0} \langle p\uparrow|{\rm
Tr}\left[\tau_3 \dot{C}^\dag C\right]|p\uparrow\rangle.
\end{eqnarray}
Using the canonical prescription (\ref{eq:conjumomenta}) one can write the matrix as
\begin{eqnarray}
\langle p\uparrow|{\rm Tr}\left[\tau_3 \dot{C}^\dag
C\right]|p\uparrow\rangle & = & 2i\langle p\uparrow|\left(
\dot{\mathbf{a}} \times \mathbf{a}\right)_3|p\uparrow\rangle
\nonumber\\
& = & - \frac{1}{2 \mathcal{I}_{\rm Skyr}}\epsilon_{3kl} \langle
p\uparrow|a_k \frac{\partial}{\partial a_l}|p\uparrow\rangle
\nonumber\\
& = & - \frac{i}{2 \mathcal{I}_{\rm Skyr}}.
\end{eqnarray}
So that we have
\begin{eqnarray}
\bm{\mu} _{p}^{I=0} & = & \frac{i}{3}\langle r^2\rangle_{I=0} \left[
- \frac{i}{2 \mathcal{I}_{\rm Skyr}}\right] =  \frac{\langle
r^2\rangle_{I=0}}{6 \mathcal{I}_{\rm Skyr}} \nonumber\\
& = & \frac{m_\Delta - m_N}{9} \langle r^2\rangle_{I=0},
\end{eqnarray}
which is $\mathcal{O}(1/N_c)$. The equality of the proton and neutron isoscalar magnetic moments implies
\begin{eqnarray}
\bm{\mu} _{p}^{I=0} + \bm{\mu} _{n}^{I=0} & = &
\left[\frac{4}{9}\left(m_\Delta - m_N \right)m_N \langle
r^2\rangle_{I=0}\right]\mu_N \, ,
\end{eqnarray}
where $\mu_N = 1/(2m_N)$ is the nuclear magneton. By using the numerical result of $r_0$ given in Eq.~\eqref{eq:r0skyr} we obtain $\bm{\mu} _{p}^{I=0} + \bm{\mu} _{n}^{I=0} = 1.36 $ which is about $22\%$ less than data $1.76$~\cite{Beringer:1900zz}.

The isovector magnetic moment can be calculated in the same way. Explicit derivation leads to
\begin{eqnarray}
\left(\bm{\mu} _{I=1}\right)_3 & = &
-\frac{4\pi}{3}F_\pi^2\int_0^\infty dr \, r^2 \, \sin^2F\left[ 1 +
\frac{8\varepsilon^2}{F_\pi^2}\left( F^{\prime 2} +
\frac{\sin^2F}{r^2}\right)\right]\langle p\uparrow|\left(\tau_3
C^\dag \tau_3C\right)|p\uparrow\rangle.
\end{eqnarray}
The matrix element in the right hand of the above equation can be rewritten as
\begin{eqnarray}
\langle p\uparrow|\left(\tau_3 C^\dag
\tau_3C\right)|p\uparrow\rangle & = & 2\langle
p\uparrow|\left[1-2\left(a_1^2 +
a_2^2\right)\right]|p\uparrow\rangle = 2-
\frac{4}{\pi^2}\int_{S^3}d\mu(a)\left(a_1^2 + a_2^2\right)^2 . \nonumber
\end{eqnarray}
The integral in the second term can be calculated by using the polar parametrization of $S^3$ as
\begin{eqnarray}
\int_{S^3}d\mu(a)\left(a_1^2 + a_2^2\right)^2 & = & \int_0^\pi
d\alpha \sin^2\alpha \int_0^\pi d\beta \sin\beta
\int_0^{2\pi}d\gamma \sin^4\alpha \sin^4\beta = \frac{2\pi^2}{3},
\end{eqnarray}
so that
\begin{eqnarray}
\langle p\uparrow|\left(\tau_3 C^\dag
\tau_3C\right)|p\uparrow\rangle & = & - \frac{2}{3}.
\end{eqnarray}
Then we get the isovector magnetic moment in a proton state as
\begin{eqnarray}
\bm{\mu} _p^{I=1} & = & \frac{1}{3}\mathcal{I}_{\rm Skyr} =
\frac{1}{2}\left(m_\Delta - m_N\right),
\end{eqnarray}
which is $\mathcal{O}(1/N_c)$. Since $\bm{\mu} _n^{I=1}$ has an opposite sign, we deduce, in terms
of the nuclear magneton,
\begin{eqnarray}
\bm{\mu} _p^{I=1} - \bm{\mu} _n^{I=1} & = &
\left(\frac{2m_N}{m_\Delta - m_N}\right)\mu_N.
\end{eqnarray}
The numerical result $\bm{\mu} _{p}^{I=1} - \bm{\mu} _{n}^{I=1} = 6.44$ calculated from the Skyrme model is also
about $30\%$ deviation from the empirical value of $\bm{\mu} _{p}^{I=1} - \bm{\mu} _{n}^{I=1} = 9.41$~\cite{Beringer:1900zz}.

\newpage

\section{Many-body system and nuclear matter}

\label{sec:matter}

The Skyrme model, as the nonlinear sigma model stabilized by the Skyrme term which is one of the next to leading order terms of chiral perturbation theory, has great advantages in describing hadron physics. We have learned in the previous chapters that both baryon and meson physics in free space can be studied by using the Skyrme model. In this chapter we will learn that the Skyrme model can also be used to study nuclear matter and the medium modified hadron properties. We will first discuss the two-body nucleon-nucleon interaction from the Skyrme model. Then we discuss the crystal structures used so far in the exploration of the nuclear matter onto which skyrmions are put and give an explicit computation of the nuclear matter properties based on the face-centered cubic crystal.  We finally explore the medium modified hadron (here pion) properties by regarding the skyrmion matter as nuclear matter.

\subsection{The skyrmion-skyrmion interaction}

Here, for discussing the arrangement of skyrmions on the crystal lattice, we study the simplest case, the skyrmion-skyrmion interaction based on the product ansatz which was originally proposed by T.~H.~R.~Skyrme~\cite{Skyrme:1961vq}~\footnote{If one substitutes the boundary condition \eqref{eq:paraskyr} with $F(r=0) = 2\pi$, $F(r \to \infty)=0$ and uses the values \eqref{eq:boundaryF}, a state with baryon number two can be yielded. However, the mass of such computed baryon number-two state is larger than three proton mass so it can easily decay into two baryons, i.e., it is not a stable state~\cite{Weigel:1986zc}. }.

Supposing the two skyrmions we are considering are far away from each other, then it is reasonable to parameterize the field configurations by producting two hedgehog skyrmions with a relative rotation in spin-isospin space
\begin{eqnarray}
U_{cc}(\mathbf{x},\mathbf{x}_1,\mathbf{x}_2) & = &
U_{c}(\mathbf{x}-\mathbf{x}_1)C(\bm{\alpha})U_{c}(\mathbf{x}-\mathbf{x}_2)C^\dag(\bm{\alpha}),\label{eq:productansatzskyr}
\end{eqnarray}
where $C(\bm{\alpha}) = \exp(i\bm{\tau}\cdot\bm{\alpha}/2)$ is the rotation in the isospin space with rotation angle $\alpha$, $U_c(\mathbf{x})$ is the hedgehog ansatz satisfying the classical equation of motion of soliton, $\mathbf{x}_1$ and $\mathbf{x}_2$ are the centers of the two skyrmions. Substituting \eqref{eq:productansatzskyr} into Eq.~\eqref{eq:energyskyr} one can express the potential energy as
\begin{eqnarray}
V(\mathbf{x}_1,\mathbf{x}_2) & = & \int d^3x
\left\{\frac{1}{4}f_\pi^2{\rm Tr}\left[L_i(1,2)L_i(1,2)\right] +
\frac{1}{32e^2}{\rm Tr}\left[L_i(1,2),
L_j(1,2)\right]^2\right\} - E_1 -E_2,\label{eq:potentialproduct}
\end{eqnarray}
where the expression of $L_\mu(1,2)$ is
\begin{eqnarray}
L_\mu(1, 2) & = & U_{cc}^\dag(\mathbf{x},\mathbf{x}_1,\mathbf{x}_2) \partial_\mu U_{cc}(\mathbf{x},\mathbf{x}_1,\mathbf{x}_2) = C({\bm \alpha})\left(L_\mu(2) + U_2^\dag C^\dag
L_\mu(1)CU_2\right)C^\dag({\bm \alpha}).
\end{eqnarray}
Substituting this expression to Eq.~\eqref{eq:potentialproduct} one has
\begin{eqnarray}
V(\mathbf{x}_1,\mathbf{x}_2) & = & \frac{1}{2}F_\pi^2\int d^3x {\rm
Tr}\left\{ - C^\dag L_i(1)CR_i(2)\right\} + \cdots ,\label{eq:nnpotential}
\end{eqnarray}
where ``$\cdots$'' stands for the contribution from quartic terms. Using the asymptotic form in the field gradients in (\ref{eq:nnpotential}), one obtains
\begin{eqnarray}
V(\mathbf{x}_1,\mathbf{x}_2) & \sim & - \frac{1}{2}{\rm
Tr}\left[C^\dag \tau^a C \tau^b\right]\int d^3x
\partial_i^2\pi^a(1)\pi^a(2),
\label{eq:expandpi}
\end{eqnarray}
where the approximation Eq.~\eqref{eq:chiralcurrentspi} for the weakly interacting pion field is used and partial integral is done. Since $\partial_i^2\pi^a(1)$ survives only in the small area around $\mathbf{x}_1$, we can expand $\pi^a(2)$ in Taylor's series about
$\mathbf{x}_1$ as
\begin{eqnarray}
\pi^b(2) & = & f_\pi (\widehat{x-r})\sin F(|\mathbf{x}-\mathbf{r}|)
\sim f_\pi K^2
\partial_r\left(\frac{1}{|\mathbf{x}-\mathbf{r}|}\right),\nonumber\\
\partial_i^2\pi^a(1) & = & f_\pi \partial_i^2(\hat{x}^a\sin F) =
f_\pi \hat{x}^a \phi(x),
\end{eqnarray}
where $\widehat{x-r}$ is a unit vector parallel to $\mathbf{x}-\mathbf{r}$ with $\mathbf{r} = \mathbf{x}_1 - \mathbf{x}_2$ and $\phi(x)$ is a smooth function of $x$. Then, substituting \eqref{eq:expandpi} into \eqref{eq:nnpotential}, we asymptotically have
\begin{eqnarray}
V(r) & \sim & \frac{1}{6}K^2 F_\pi^2{\rm Tr}\left[ C^\dag \tau^a C
\tau^b\right]\partial_a\partial_b\left(\frac{1}{r}\right),
\end{eqnarray}
which is a reminiscent of one-pion exchange.

By using \eqref{eq:paracollec} we can express
\begin{eqnarray}
{\rm Tr}\left[ C^\dag \tau^a C
\tau^b\right] & = & {\rm Tr}\left[\left(a_0 - i {\bm\tau}\cdot\mathbf{a}\right) \tau_a \left(a_0 + i {\bm\tau}\cdot\mathbf{a}\right) \tau_b \right] \nonumber\\
& = & 2 \left[ \left(a_0^2 - \mathbf{a}^2\right)\delta_{ab} + 2a_a a_b - 2 a_0 a_c \epsilon_{acb}\right].
\label{A and a}
\end{eqnarray}
Using identity
\begin{eqnarray}
\partial_a \partial_b \left( \frac{1}{r}\right) & = & \frac{3 x_a x_b - r^2\delta_{ab}}{r^5},
\end{eqnarray}
we finally obtain
\begin{eqnarray}
V(r) & \sim & \frac{2}{3}K^2 F_\pi^2  \frac{3 \mathbf{x}\cdot \mathbf{a}\mathbf{x}\cdot \mathbf{a} - r^2 \mathbf{a}^2}{r^5}. \label{eq:potential2s}
\end{eqnarray}

From Eq.~\eqref{eq:potential2s} one concludes that the most attractive potential is given by $\mathbf{a}\cdot \mathbf{x} = 0, \mathbf{a}^2 = 1$. Considering the rotation $C({\bm \alpha}) = \exp \left(i {\bm \tau}\cdot {\bm \alpha}/2\right)$ we have
\begin{eqnarray}
\mathbf{a} & = & \hat{{\bm \alpha}} \sin \frac{\alpha}{2}.
\end{eqnarray}
So that $\mathbf{a}^2 = 1$ leads to $\alpha = \pi$. In addition, the condition $\mathbf{a}\cdot \mathbf{x} = 0$ implies that the rotation axis $\mathbf{a}$ should be perpendicular to $\mathbf{x}$. These discussions yield a conclusion: {\it To get the most attractive potential, the pair skyrmions should be arranged in such a way that they should mutually rotate in the isospin space by angle $\pi$ about the axis perpendicular to the line joining them.}  Potential Eq.~\eqref{eq:potential2s} also tells us that, in case of $r  \to \infty$, i.e., the two skyrmions are far away from each other, they become non-interacting objects. So that, product ansatz of the field configurations is a reasonable approach at long distance.

\subsection{The cubic crystal for nuclear matter}

\begin{figure}[htbp]\centering
\includegraphics[scale=0.5]{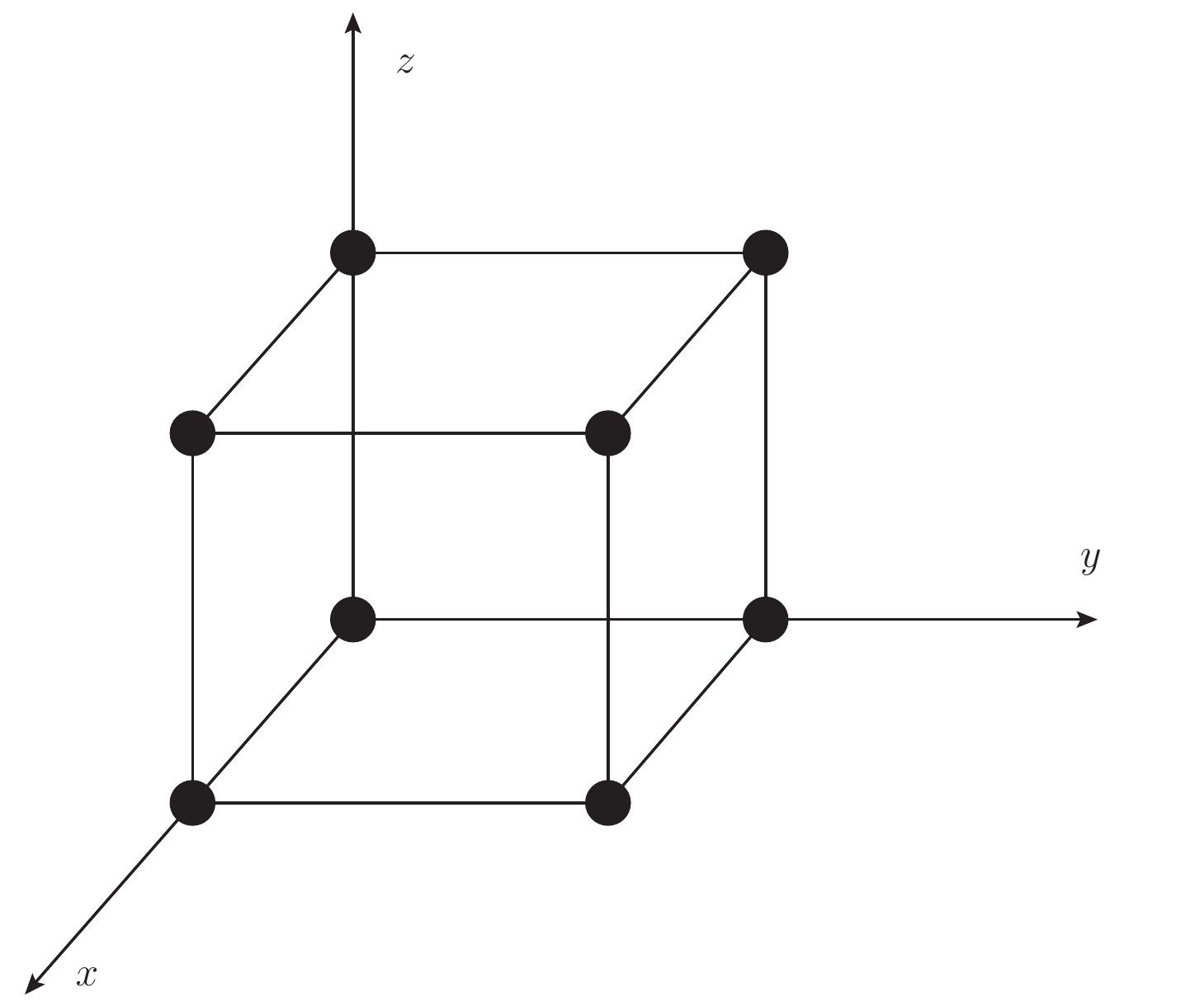}
\caption{Skyrmions on the cubic crystal lattice.}
\label{fig:skyrmioncc}
\end{figure}

Based on the lessons drowned from skyrmion-skyrmion interaction yielding the strongest attractive force, Klebanov
proposed that nuclear matter could be simulated by using the cubic crystal lattice~\cite{Klebanov:1985qi} with skyrmions sitting on the lattice vertices in a manner illustrated in Fig.~\ref{fig:skyrmioncc}.

Due to the crystal structure, for a specific skyrmion $U_0(\mathbf{ x})$, different sequences of two successive rotations $C_i(t)$ and $C_j(t)$ which preserve the lowest energy of the system should yield the same consequences, i.e.,
\begin{eqnarray}
C_i^\dag C_j^\dag U_0(\mathbf{ x}) C_j C_i & = & C_j^\dag C_i^\dag U_0(\mathbf{ x}) C_i C_j,
\end{eqnarray}
or, equivalently, in terms of collective coordinates
\begin{eqnarray}
\left(\textbf{a}_i\cdot{\bm \tau}\right)\left(\textbf{a}_j\cdot{\bm \tau}\right) U_0(\mathbf{x})\left(\textbf{a}_j\cdot{\bm \tau}\right)\left(\textbf{a}_i\cdot{\bm \tau}\right) & = & \left(\textbf{a}_j\cdot{\bm \tau}\right)\left(\textbf{a}_i\cdot{\bm \tau}\right) U_0(\mathbf{x})\left(\textbf{a}_i\cdot{\bm \tau}\right)\left(\textbf{a}_j\cdot{\bm \tau}\right),
\end{eqnarray}
where $i$ stands for the $i$-th rotation and the lessons from the two-skyrmion interaction on how to yield the lowest configuration energy have been used. This relation indicates that $\textbf{a}_i\cdot{\bm \tau}$ and $\textbf{a}_j\cdot{\bm \tau}$ should be either commute or anticommute. In order to have minimized crystal energy, we choose the latter. Typical choices are
\begin{eqnarray}
\textbf{a}_x & = & \hat{\textbf{y}}, \quad \textbf{a}_y = \hat{\textbf{z}}, \quad \textbf{a}_z = \hat{\textbf{x}}, \label{eq:choiceacc}
\end{eqnarray}
and their cyclic order.

\begin{figure}[htbp]\centering
\includegraphics[scale=0.5]{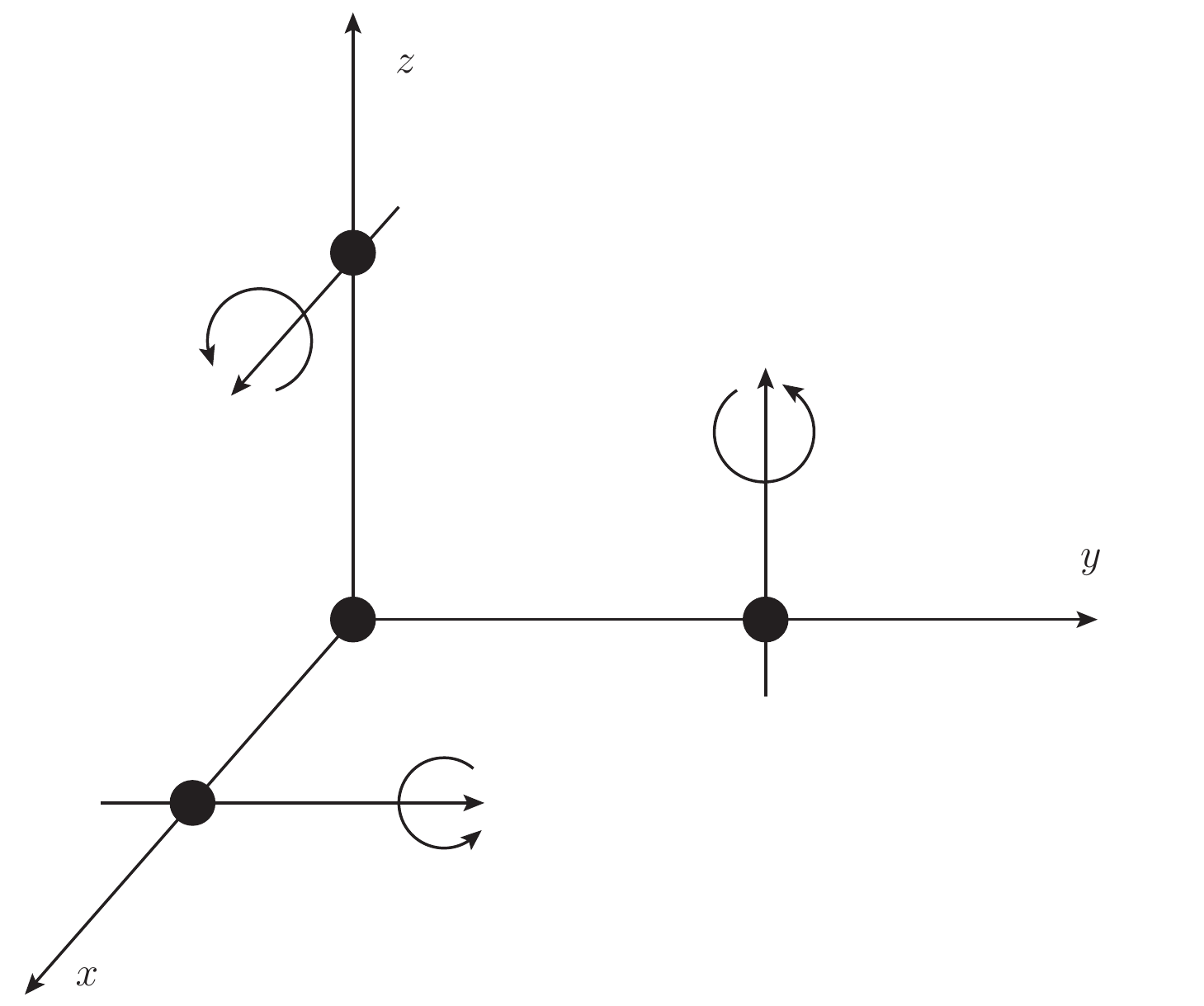}
\caption{Rotations of the nearest pair skyrmions in cubic crystal.}
\label{fig:ccBCs}
\end{figure}

Now let us consider the cubic lattice illustrated in Fig.~\ref{fig:skyrmioncc} with lattice size $2L$. With respect to the relative rotation of the nearest skyrmions inducing the lowest potential, we have the following boundary conditions for the nearest skyrmions:
\begin{eqnarray}
U_0(x,y,z) & = & \tau_y U_0(x+2L,y,z)\tau_y = \tau_z U_0(x,y+2L,z)\tau_z  = \tau_x U_0(x,y,z+2L)\tau_x . \label{eq:BCscc}
\end{eqnarray}
Supposing the $x = y = z = 0$, boundary conditions~\eqref{eq:BCscc} can be illustrated as Fig.~\ref{fig:ccBCs}.

Note that just boundary condition \eqref{eq:BCscc} is not enough for making numerical simulation since there is translation invariance in the configuration. So that, if one starts to make a numerical simulation by using the boundary condition \eqref{eq:BCscc} and constraints of the baryon number equal to one in a cubic, infinity degenerate solutions may be found. This infinity degeneracy could be avoided by imposing the coordinate reflection invariance accompanied by analogous reflections in the isospace on the configuration
\begin{eqnarray}
x & \to & {} - x, \tau_x \to {} - \tau_x;\quad y \to {} - y, \tau_y \to {} - \tau_y;\quad z \to {} - z, \tau_z \to {} - \tau_z . \label{eq:reflectioncc}
\end{eqnarray}
In addition to the indication that there should be a skyrmion sitting at the origin, this reflection invariance induces the following relation of the $U(x)$
\begin{eqnarray}
U_0(x,y,z) & = & \tau_x U_0^\dag(-x,y,z)\tau_x = \tau_y U_0^\dag(x,-y,z)\tau_y  = \tau_z U_0^\dag(x,y,- z)\tau_z , \label{eq:reflectionUcc}
\end{eqnarray}
where we have used for example
\begin{eqnarray}
\tau_xU_0(x,y,z)\tau_x & = & U_0^\dag(-x,y,z),
\end{eqnarray}
which can be easily understood by using the Hedgehog ansatz for $U(x)$.

\begin{figure}[htbp]\centering
\includegraphics[scale=0.5]{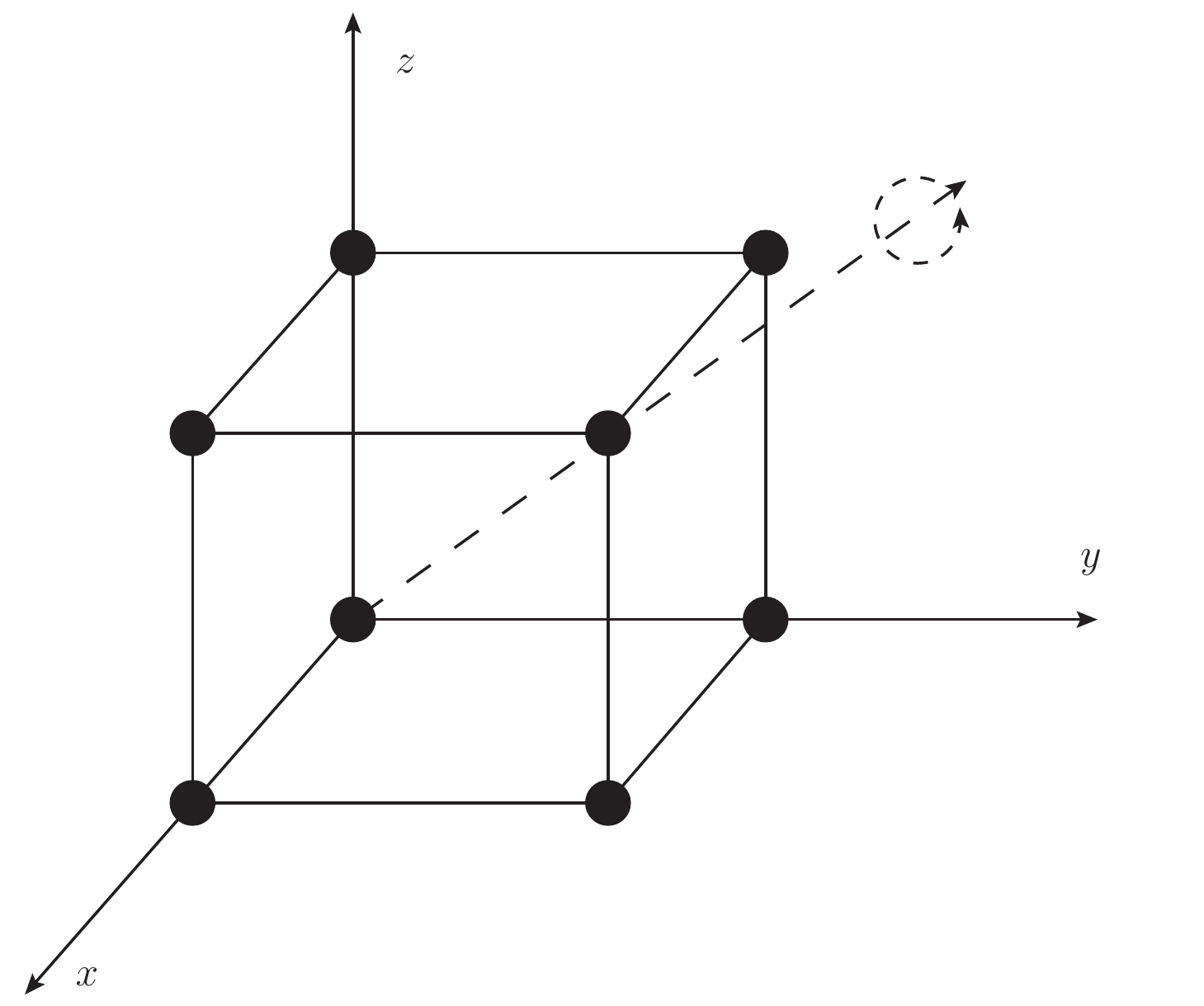}
\caption{Three-fold rotations about the main diagonals in the cubic crystal lattice.}
\label{fig:cubicmaindiag}
\end{figure}

In addition to the symmetry relations \eqref{eq:BCscc} and \eqref{eq:reflectionUcc} one can check that the system has a discrete symmetry of the three-fold rotations about the main diagonals
accompanied by the analogous rotations in isospace as illustrated in Fig.~\ref{fig:cubicmaindiag}.
Formally this could be realized through rotation
\begin{eqnarray}
U(x,y,z)  & = & A U_0(y,z,x) A^\dag; \quad A \tau_x A^\dag = \tau_y,\quad A \tau_y A^\dag = \tau_z,\quad A \tau_z A^\dag = \tau_x, \label{eq:3foldcc}
\end{eqnarray}
which could be derived by taking $A = \exp(-i\pi(\tau_x+\tau_y+\tau_z)/3\sqrt{3})$.

Now, let us consider a cell, $|x_i| \le L$. Eqs.~\eqref{eq:BCscc}, \eqref{eq:reflectionUcc} and \eqref{eq:3foldcc} together require $U_0 \in Z_2$ at the center since the sign of $U_0$ cannot be fixed. Moreover, each lattice cell contains one skyrmion. Since the starting configuration has a large enough size, we can choose the boundary condition to be $U_0= -1$ at the center and $U_0 = 1$ on the cell faces. Thus, the centers of skyrmions can be located at the points where $U_0 = -1$ and these centers form a cubic lattice. Note that this crystal lattice ansatz works in the low density limit in which $U_0(\mathbf{x})$ reduces to a product of spherically symmetric skyrmions centered
at the lattice sites satisfying \eqref{eq:BCscc}, \eqref{eq:reflectionUcc} and \eqref{eq:3foldcc}.

In summary, the cubic crystal lattice has the following symmetries (remember $U = \sigma + i\bm{\tau}\cdot {\pi}$)
\renewcommand{\theenumi}{\arabic{enumi}}
\renewcommand{\labelenumi}{(CC\theenumi)}
\begin{enumerate}
\item Translation: $(x,y,z) \to (x+2L, y, z)$, $(\sigma, \pi_1, \pi_2, \pi_3) \to (\sigma, -\pi_1, \pi_2, -\pi_3)$. This can be proved by using Eq.~\eqref{eq:BCscc} and the relations among Pauli matrices.

\item Reflection:  $(x,y,z) \to (-x, y, z)$, $(\sigma, \pi_1, \pi_2, \pi_3) \to (\sigma, -\pi_1, \pi_2, \pi_3)$ which can be obtained by using Eq.~\eqref{eq:reflectionUcc}.

\item Three-fold rotation: Rotation by $2\pi/3$ along the main diagonals:  $(x,y,z) \to (z, x, y)$, $(\sigma, \pi_1, \pi_2, \pi_3) \to (\sigma, \pi_3, \pi_1, \pi_2)$ which is due to the expression~\eqref{eq:3foldcc}.
\end{enumerate}
And $\sigma$ must be $\pm 1$ at the center and surface vertices and, to have a baryon number one configuration in each crystal size, $\sigma = -1$ at the origin and $\sigma = 1$ at the surface points.

We want to say that, in the crystal approach, since all the skyrmions are glued together, to quantize the system for taking into account the $\mathcal{O}(N_c^{-1})$ effect in the nuclear matter, the crystal should be rotated as a rigid object. We will not cover this aspect but refer to~\cite{Klebanov:1985qi}.

In the original paper~\cite{Klebanov:1985qi}, the relaxation method was used to explore the density, or equivalently the crystal size, dependence of the per-skyrmion energy in the crystal. In the following explicit calculation based on FCC crystal, we will apply another method, the Fourier serious expansion method, so that we will not discuss the relaxation method here.

After Klebanov proposed his pioneering idea that nuclear matter could be described by a cubic crystal of skyrmions, it was widely explored in the literature. In Ref.~\cite{Wuest:1987rc}, explicit simulation yielded that skyrmion was distorted at the minimum of the per-skyrmion energy and high density, and this distortion can be illustrated by the distribution of the baryon number density. The distribution of the baryon number density illustrated in Ref.~\cite{Wuest:1987rc} tells us the following facts: When two skyrmions are far away from each other the minimal field configuration has the symmetries of Klebanov's primitive cubic crystal of skyrmions. However when the two skyrmions become coincident and the minimal per-skyrmion energy is attained, skyrmion number density appears at the body center of the crystal cell. Therefore, the field configuration has an additional axial symmetry. In this case, skyrmions have lost their individual identities. Moreover, it was found that at very small crystal size $L$, a large baryon number density appears at the body center and there is a larger symmetry between the cubic crystal points and the points at the body centers.

\begin{figure}[htbp]\centering
\includegraphics[scale=0.5]{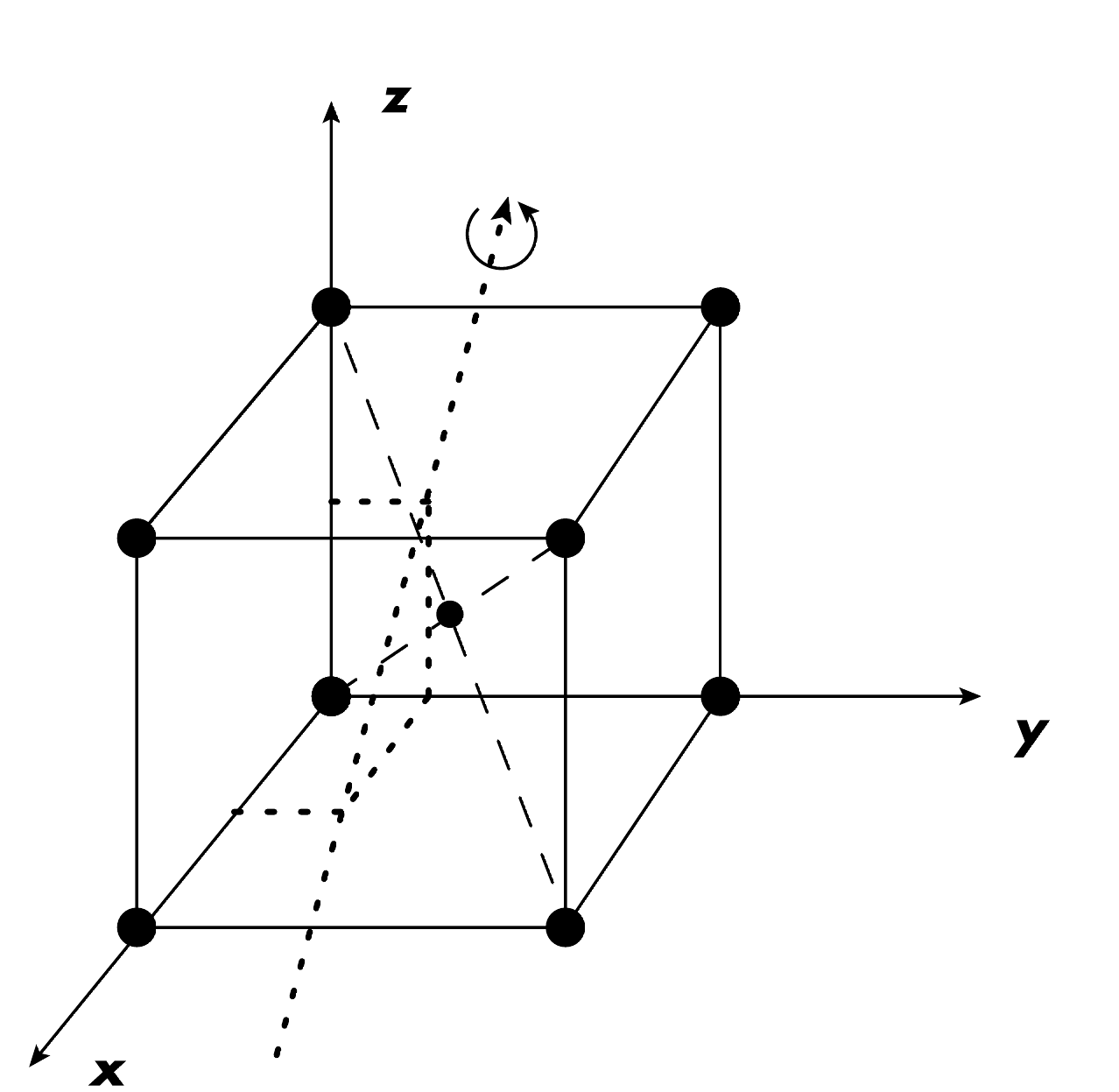}
\caption{Rotation of the additional symmetry in BCC.}
\label{fig:rotatebcc}
\end{figure}

The above observations tell us that, when two skyrmions become coincident and the large density is arrived, the skyrmion crystal has the following symmetries~\cite{Goldhaber:1987pb}
\renewcommand{\theenumi}{\arabic{enumi}}
\renewcommand{\labelenumi}{(BC\theenumi)}
\begin{enumerate}
\item \label{BCC1} Translation: $(x,y,z) \to (x+2L, y, z)$, $(\sigma, \pi_1, \pi_2, \pi_3) \to (\sigma, -\pi_1, \pi_2, -\pi_3)$; 

\item Reflection:  $(x,y,z) \to (-x, y, z)$, $(\sigma, \pi_1, \pi_2, \pi_3) \to (\sigma, -\pi_1, \pi_2, \pi_3)$;\label{BCC2}

\item Three-fold rotation: Rotation by $2\pi/3$ along the main diagonals:  $(x,y,z) \to (z, x, y)$, $(\sigma, \pi_1, \pi_2, \pi_3) \to (\sigma, \pi_3, \pi_1, \pi_2)$;\label{BCC3}

\item $(x,y,z)  \to (L-z, L-y, L-x); \quad (\sigma,\pi_1,\pi_2,\pi_3) \to (-\sigma, \pi_2, \pi_1, \pi_3)$.\label{BCC4}
\end{enumerate}
The first three are the same as Klebanov's primitive cubic crystal and the last one is the additional rotation symmetry observed for minimal energy and small crystal size. This additional symmetry is the rotation about the line passing through $(L,L/2,0)$ and $(0,L/2,L)$ with angle $\pi$ which, as illustrated in Fig.~\ref{fig:rotatebcc}, maps $(0,0,0)$ to $(L,L,L)$.

\begin{figure}[htbp]\centering
\includegraphics[scale=0.4]{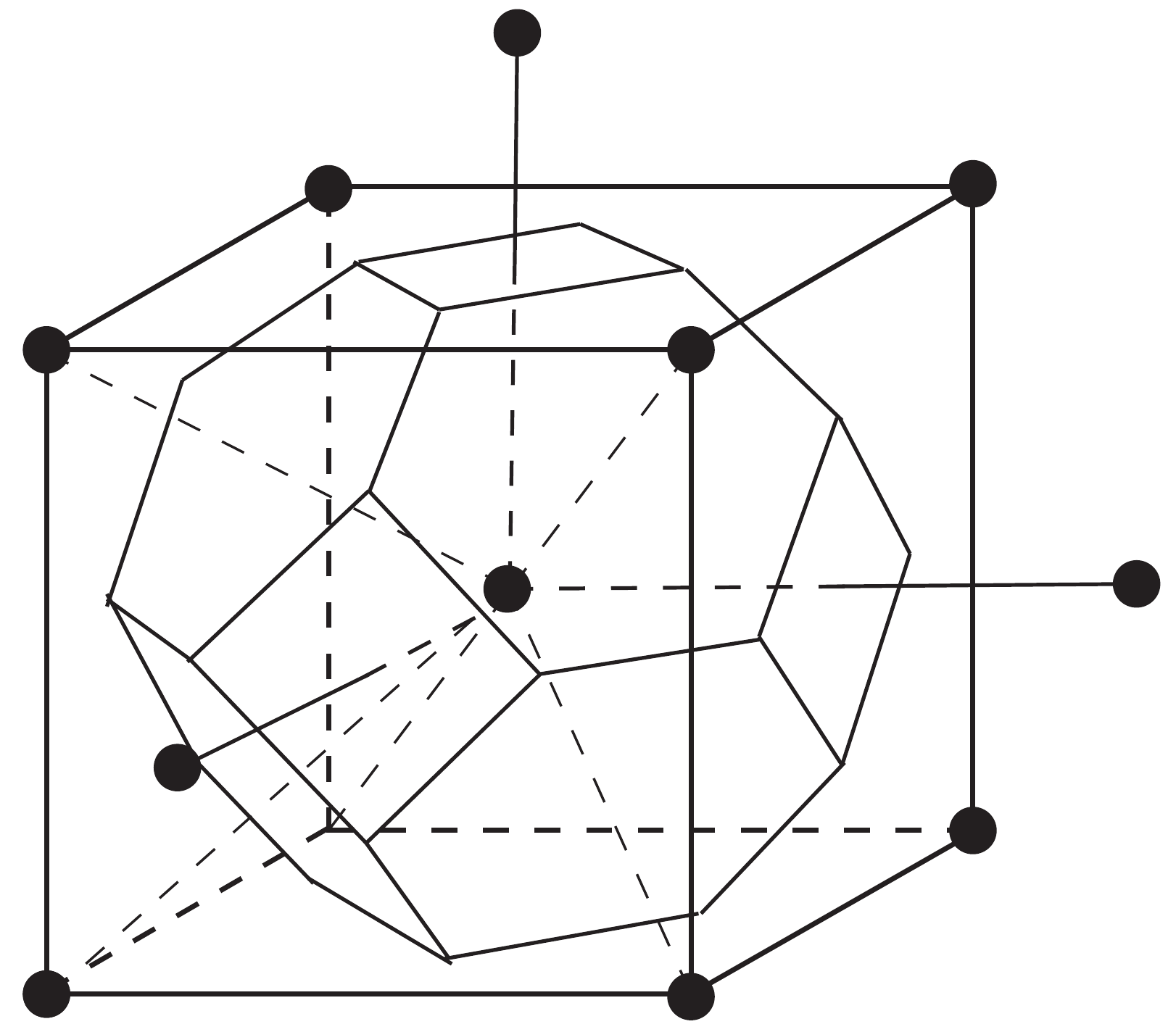}
\caption{Wigner-Seitz (W-S) cell.}
\label{fig:WScell}
\end{figure}

With respect to the new symmetries, it is convenient to divide the space into Wigner-Seitz (W-S) cells with volumes a half of the original cubic cells and, the W-S cells can be transformed to each other by transformations of the symmetry group. In this configuration, the center of each W-S cell is one point of the body-centered cubic crystal, and the region of a W-S cell is closer to that lattice point than any other. As shown in Fig.~\ref{fig:WScell} each W-S cell has $8$ hexagonal faces and $6$ square ones.

Concerning the extra symmetry (BC\ref{BCC4}), the skyrmion configurations are constrained and the field values at certain
points are determined as follows:
\renewcommand{\theenumi}{\arabic{enumi}}
\renewcommand{\labelenumi}{(BV\theenumi)}
\begin{enumerate}
\item \label{item:bccorigin } $\sigma = - 1$ at $(x,y,z)=(0,0,0)$.

This is determined by considering that before the skyrmions in the crystal become coincident, $(x,y,z)=(0,0,0)$ is an origin of a skyrmion with baryon number${}+ 1$.

\item $\sigma = + 1$ at $(x,y,z)=(L,L,L)$.

Because of the symmetry (BC\ref{BCC4}), the point $(x,y,z)=(L,L,L)$ can be obtained by rotating the point $(x,y,z)=(0,0,0)$ accompanied by $\sigma \to - \sigma$.

\item $\pi_1 =  1$ at $(x,y,z)=(L,0,\pm L/2)$.

Since this point is on the $x$-$z$ plane, one has $\pi_2=0$. Consider point $(x,y,z)=(L,0, L/2)$, after rotation (BC\ref{BCC4}), this point becomes $(L/2,L, 0)$ and correspondingly $(\pi_1,0,\pi_3) \to (0, \pi_1, \pi_3)$. Therefore $\pi_3 = 0$. So that we finally have $\pi_1 = 1$.

\item $\pi_1 = \pm \pi_2 = 1/\sqrt{2}$ at $(x,y,z)=(L,\pm L/2,0)$.

At point $(x,y,z)=(L,\pm L/2,0)$ we have $\pi_3 = 0$ because this point is in the $x$-$y$ plane. After rotation (BC\ref{BCC4}) this point becomes $(L,L/2,0)$ with $(\pi_2,\pi_1,\pi_3)$. Therefore we have $\pi_1 = \pi_2 = \sqrt{2}/2$.

\item $\pi_1 = \pi_2 = \pi_3 = 1/\sqrt{3}$ at $(x,y,z)=(L/2, L/2,L/2)$.

The point $(x,y,z)=(L/2, L/2,L/2)$ is in the middle of crystal vertex at the origin and the vertex at the body center therefore we should have $\sigma = 0$. Since the point $(x,y,z)=(L/2, L/2,L/2)$ has the same distances to $x, y$ and $z$ axes, $\pi_1 = \pi_2 = \pi_3 = 1/\sqrt{3}$ because of $\sigma^2 + \pi_1^2 + \pi_2^2 + \pi_3^2 = 3 \pi_1^2 = 3\pi_2^2 = 3\pi_3^2 = 1$.

\end{enumerate}

By imagining a body-centered cubic lattice of half-skyrmions - the skyrmion with a half-baryon number - one can understand the skyrmion crystal with extra symmetry (BC\ref{BCC4}) in the simpler way. In the hedgehog ansatz, the profile function $F(r)$ of a half-skyrmion decreases from $\pi$ at $r= 0$ to $\pi/2$ at certain radius $r_0$, or equivalently, $\sigma = -1$ at the center and $0$ at the boundary, and is undefined outside the radius $r_0$, so that the baryon number is $1/2$ within this vacuum. By using the transformation $(\sigma, {\bm \pi}) \to (-\sigma, -{\bm \pi})$, a half-skyrmion can be transformed into another type which also has skyrmion number $1/2$.

Note that, as further consequences of the symmetries, in the original crystal cell, the space average of $\sigma$ satisfies $\langle \sigma \rangle = 0 $ therefore $\langle \sigma \rangle$  can be regarded as the order parameter of the half-skyrmion phase.

From the field values discussed in (BV), one can determine the lattice size of half-skyrmions as $r_0 = \sqrt{3}L/4$ which is actually the distance between the body center and the hexagonal. The first type half-skyrmions are centered on the original lattice sites and have the same orientations as that of the original skyrmions. However, due to the overlaps of the original skyrmions, the centers of the second type half-skyrmions are the body centers. The orientations of the half-skyrmions can be chosen in such a way that the skyrmion fields are continuous at the points where the two types of half-skyrmions touch (e.g. at $(L/2, L/2, L/2)$) and the skyrmion field values at the touch points can be determined by the symmetry requirement. Note that to fill a W-S cell, each half-skyrmion should be distorted even though this distortion is fairly small as indeed numerically checked in Ref.~\cite{Goldhaber:1987pb}.

In summary, starting from the cubic crystal of skyrmions, the minimum energy crystal configuration is the body-centered cubic crystal of half-skyrmions.~\footnote{When we start from the face-centered cubic crystal of skyrmions at low density which will be considered in the next section, the half-skymrion phase at high density is the cubic crystal.}
It is difficult to say which pairs of half-skyrmions link up to form a full skyrmion when one release the space of body-centered cubic crystal lattice. Actually, the disappeared half-skyrmions divided equally their baryon numbers to their eight nearest neighbors.

\subsection{The face-centered cubic crystal for nuclear matter}

By using Klebanov's cubic crystal, it was found in the literature that, in the unit of baryon number $B$, the yielded minimal energy is $E = 1.08 B$ (in this unit the skyrmion energy in free space from the Skyrme model is $E = 1.23 B$). Later, Kugler and Shtrikman invented a different crystal structure at low density, face-centered cubic (FCC) crystal, which could yield a smaller minimal energy at high density which is just about 3.8\% above the lower bound and about 4\% below that obtained from Klebanov's cubic crystal~\cite{Kugler:1988mu,Kugler:1989uc}. Such a crystal configuration is known to give the lowest ground-state energy among the crystal symmetries studied so far.

\begin{figure}[htbp]\centering
\includegraphics[scale=0.4]{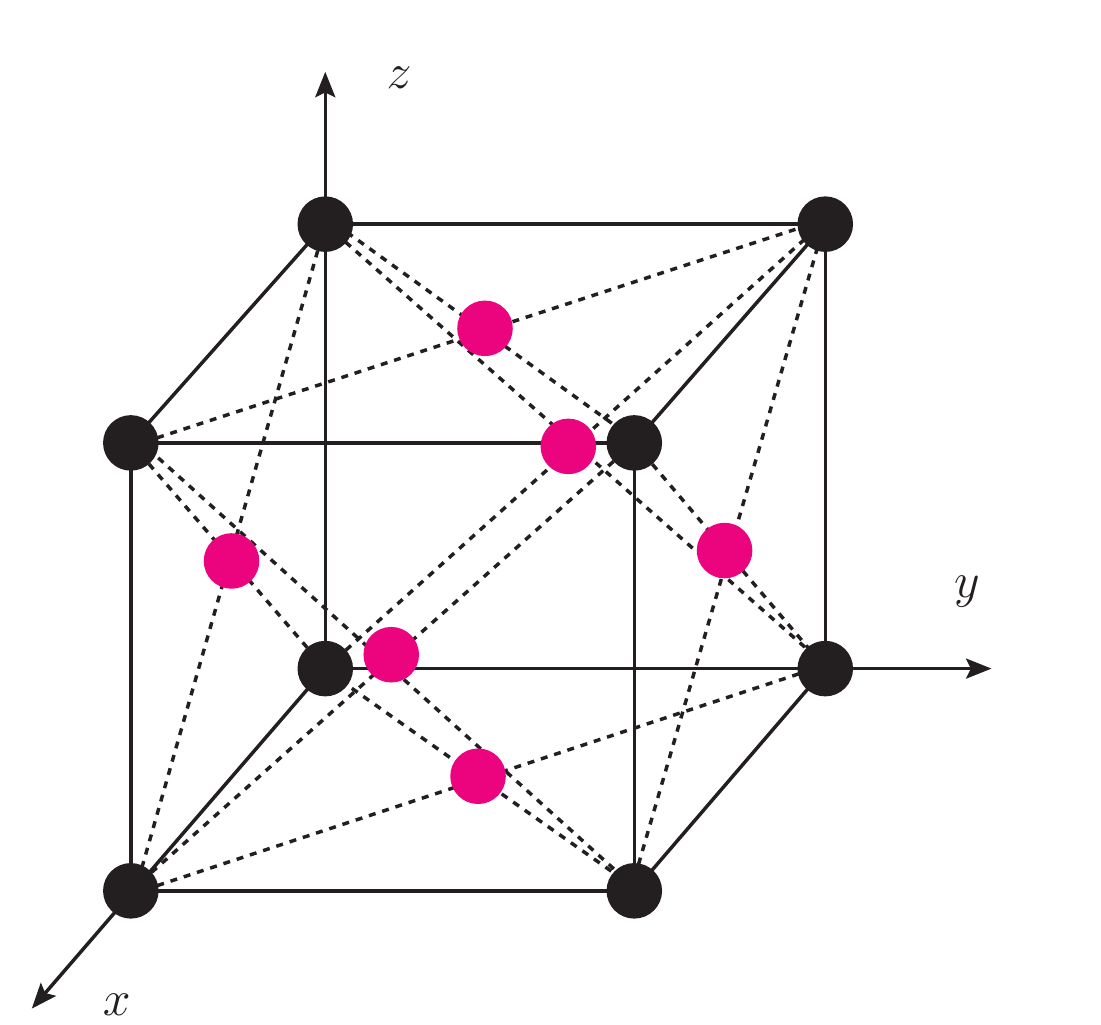}
\includegraphics[scale=0.2]{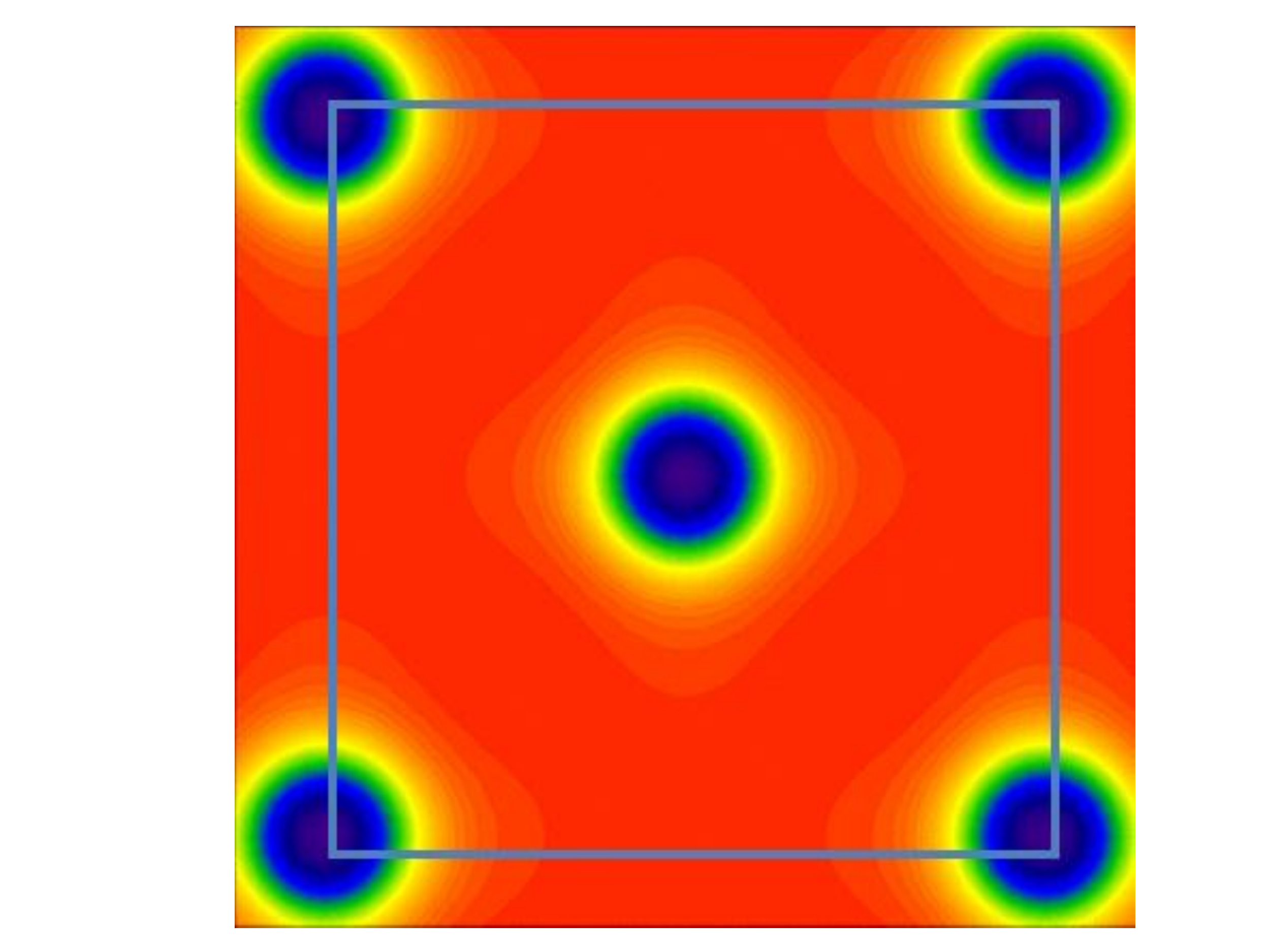}
\caption{ Arrangement of the skyrmions on the FCC crystal lattice and baryon density distribution.}
\label{fig:fccdis}
\end{figure}

The FCC crystal structure of skyrmions at low density together with the baryon density distribution are illustrated in Fig.~\ref{fig:fccdis} from which one can easily obtain the following features:  Compared with Klebanov's cubic crystal, FCC crystal has a combined translation invariance: $x_i=(x+L, y+L, z)$, and $n^\alpha = (\sigma, -\pi_1, -\pi_2, \pi_3)$ which can be obtained by a rotation angle $\pi$ around the $z$-axis, i.e., $U_0 \to \tau_z U_0 \tau_z$. One can easily count that in FCC configuration, each skyrmion is surrounded by twelve neighbors and, to have the maximum attractive force, the nearest two skyrmions should have a relative rotation in isospace.

In summary, considering a point in space $\vec{x}=(x,y,z)$ at which the fields are given by $(\sigma, \pi_1, \pi_2, \pi_3)$, the FCC configuration is defined by the following symmetries:
\renewcommand{\theenumi}{\arabic{enumi}}
\renewcommand{\labelenumi}{(F\theenumi)}
\begin{enumerate}
\item Under the reflection in space $\vec{x} \rightarrow (-x,y,z)$, the field is also reflected in isospin space according to $(\sigma, -\pi_1, \pi_2, \pi_3)$. \label{F1}

\item Under a rotation around the threefold axis in space
$\vec{x} \rightarrow (y,z,x)$, the field is simultaneously rotated in isospin space about the corresponding axis in isospin space according to $(\sigma, \pi_2, \pi_3, \pi_1)$. \label{F2}

\item Under a rotation around the fourfold axis in space $\vec{x} \rightarrow (x,z,-y)$, the field is rotated around the corresponding fourfold axis in isospin space according to $(\sigma, \pi_1, \pi_3, -\pi_2)$. \label{F3}

\item Under a translation from a corner of a cube to the center of a face $\vec{x} \rightarrow (x+L,y+L,z)$, the field becomes rotated by $\pi$ about the axis perpendicular to the face according to $(\sigma, -\pi_1, -\pi_2, \pi_3)$. \label{F4}
\end{enumerate}
Here, a single FCC unit cell has size $2L$ and contains 4 skyrmions and, thus, the baryon number density in the configuration is $\rho=1/2L^3$. If one regards the skyrmion matter as nuclear matter and takes the
normal nuclear matter density $\rho_0 =0.17/$fm$^3$ as input, the corresponding crystal size of the normal nuclear matter density is $L\sim 1.43$ fm.

After squeezing the crystal size, or equivalently increasing the density, the baryon number density distribution changes from FCC crystal of skyrmions to the cubic crystal of half-skyrmions.
Let us denote the skyrmion field at point $x_i=(x, y, z)$ as $n^\alpha = (\sigma, \pi_1, \pi_2, \pi_3)$. Then, one can define the symmetry of the crystal as a set of operations involving a change in $x_i$ and an appropriate change in $n^\alpha$.
It was found that the crytal structure of the half-skyrmion phase has the following properties : 
\renewcommand{\theenumi}{\arabic{enumi}}
\renewcommand{\labelenumi}{(FC\theenumi)}
\begin{enumerate}

\item Reflection: $x_i = ( -x, y, z)$, and $n^\alpha = (\sigma, - \pi_1, \pi_2, \pi_3)$.\label{FC2}

\item Three-fold rotation: $x_i = (y, z, x)$, and $n^\alpha = (\sigma, \pi_2, \pi_3, \pi_1 )$.\label{FC3}

\item Translation: $x_i = (x + L, y, z)$, and $n^\alpha= ( -\sigma, -\pi_1, \pi_2, \pi_3)$.\label{FC4}

\item Four-fold rotation: $x_i = (x, z, -y)$, and $n^\alpha = (\sigma, \pi_1, \pi_3, -\pi_2)$.\label{FC5}
\end{enumerate}

The physical meanings of these symmetries can be understood as follows (consider the origin $x_i = (0,0,0)$):
At the origin, $\pi_i = 0$ and we can take, without loss of generality, $\sigma = -1$. Because of the symmetries (FC\ref{FC2}) and (FC\ref{FC4}), $\sigma = 0$ on any surface specified by $x_i= \pm L/2$. A region surrounding the origin and bounded by surfaces $x_i= \pm L/2$ contains half a skyrmion with $\sigma < 0$. Because of symmetry (FC\ref{FC4}), $\sigma= 1$ at $(L, 0, 0)$. Near point $(L, 0, 0)$ one can put the second type of half-skyrmions with $\sigma > 0$. Thus, such a scenario of nuclear matter can be viewed as an ``antiferromagnetic'' arrangement of half-skyrmions and the half-skyrmions fill a cubic crystal with cell size $L$. The symmetry operations (FC) insure that the two types of half-skyrmions are appropriately rotated in isospin so that the fields are smoothly connected and the energy of the system keeps minimized.

The energy of the system can be evaluated by expanding the $\sigma$ and pions in terms of Fourier series due to the crystal structure~\cite{Kugler:1988mu,Kugler:1989uc}. By adjusting the Fourier coefficients the minimal energy of the system could be obtained. It was found that the minimum is $E = 1.038B$ which is just $3.8\%$ above the lower bound. By plotting the distribution of the baryon number density, in the half-skyrmion phase, it was found that, along the links which connect the centers of the half-skyrmions, the baryon number density is relatively large. However, the baryon number density vanishes along lines which are parallel to the links and going through the centers of the faces of this cube for symmetry reasons.

We next, following Ref.~\cite{Lee:2003aq}, make a concrete computation of the baryonic matter properties by putting skyrmions onto the FCC crystal and regarding the skyrmion matter as nuclear matter. Here we consider the Skyrme model in the chiral limit:
\begin{equation}
{\cal L}= -\frac{f_\pi^2}{4} \mbox{Tr} \left(U^\dagger \partial_\mu U
U^\dagger \partial^\mu U\right) + \frac{1}{32e^2} \mbox{Tr}
\left [U^\dagger \partial_\mu U, U^\dagger \partial_\nu U\right]^2.
\label{eq:Lsk}
\end{equation}
As pointed in Ref.~\cite{Lee:2003aq}, this Lagrangian could describe physical processes of free pions, free baryons, many baryon states, dense baryonic matter and moreover, the pions interacting with baryonic matter. With respect to this philosophy, the parameters of the Lagrangian can be fixed by one of these processes, for example by describing the free pion or nucleon systems. Based on this idea, we take the values of the parameters as~\footnote{The parameters determined here are different from that determined in last chapter where they are determined by fitting the nucleon mass and $\Delta$-$N$ mass splitting and also that used in Ref.~\cite{Lee:2003aq}.}
\begin{eqnarray}
f_\pi & = & 92.4~{\rm MeV},\quad e = 5.93,
\end{eqnarray}
where the parameter $e$ is estimated from the hidden local symmetry approach in the next chapter.

The nuclear matter properties from the FCC crystal can be simulated by making a Fourier mode expansion of the relevent fields. For this purpose,  we introduce ``{\em unnormalized}" fields
$(\bar{\phi}_0, \bar{\phi}_1, \bar{\phi}_2, \bar{\phi}_3)$ which have the Fourier series expansions as~\footnote{ Instead of expanding the introduced unnormalized modes, one can make a Fourier expansion of the normalized modes $\phi_\alpha (\alpha = 0,1,2,3)$ defined by Eq.~\eqref{eq:definenorphi}. In such a case, the structures of the series and the relations among the expansion coefficients are the same as those of the unnormalized ones since they are obtained from the crystal structure~\cite{Suenaga:2014sga}.
}
\begin{equation}
\bar{\phi}_0 = \sum_{a,b,c} \bar{\beta}_{abc} \cos(a\pi x/L)
\cos(b\pi y/L) \cos(c\pi z /L),
\label{sigma}
\end{equation}
and
\begin{eqnarray}
\bar{\phi}_1 &=& \sum_{h,k,l} \bar{\alpha}_{hkl} \sin(h\pi x/L)
\cos(k\pi y/L) \cos(l\pi z/L),
\label{pi1} \\
\bar{\phi}_2 &=& \sum_{h,k,l} \bar{\alpha}_{hkl} \cos(l\pi x/L)
\sin(h\pi y/L) \cos(k\pi z/L),
\label{pi2} \\
\bar{\phi}_3 &=& \sum_{h,k,l} \bar{\alpha}_{hkl} \cos(k\pi x/L)
\cos(l\pi y/L) \sin(h\pi z/L),
\label{pi3}
\end{eqnarray}
where the expansion $\bar{\phi}_2$ and $\bar{\phi}_3$ are obtained from $\bar{\phi}_1$ by using the symmetry relation (FC\ref{FC3}). By using these quantities we can define the following normalized fields
\begin{eqnarray}
\phi_\alpha & = & \frac{\bar{\phi}_\alpha}{\sqrt{\sum_{\beta=0}^3 \left(\bar{\phi}_\beta\right)^2}}, \quad (\alpha,\beta = 0,1,2,3).\label{eq:norm}
\end{eqnarray}

In terms of $\phi_\alpha$, we can define the field $U_0(\mathbf{x})$ as
 \begin{eqnarray}
 U_0 & = & \phi_0+i \vec{\tau}\cdot\vec{\phi}. \label{eq:definenorphi}
\end{eqnarray}
By taking the Fourier coefficients $\bar{\beta}$ and $\bar{\alpha}$ as free parameters and varying them the minimal value of per-energy at a specific crystal size,  the density dependence of the per-skyrmion energy can be obtained. Although this straightforward procedure works, it takes a long computing time. To save the computing time, we can resort to some relations among the Fourier coefficients due to the symmetries of crystal structure. From FCC symmetry relations (F\ref{F1})-(F\ref{F4}) the modes appearing in Eqs.~(\ref{sigma}-\ref{pi3}) are restricted as follows:
\renewcommand{\theenumi}{\arabic{enumi}}
\renewcommand{\labelenumi}{(M\theenumi)}
\begin{enumerate}
\item From (F\ref{F2}) one has $\bar{\beta}_{abc}=\bar{\beta}_{bca}=\bar{\beta}_{cab} =\bar{\beta}_{acb}=\bar{\beta}_{cba}=\bar{\beta}_{bac}$. \label{M1}

\item From  (F\ref{F3}) one obtains $\bar{\alpha}_{hkl}=\bar{\alpha}_{hlk}$. \label{M2}

\item From  (F\ref{F4}) one can prove the following conclusion:
\begin{itemize}
  \item $a,{}b,{}c$ are all even numbers or odd numbers.
  \item When $h$ is even, $k,{}l$ are both odd.
  \item When $h$ is odd, $k,{}l$ are both even.
\end{itemize}
 \label{M3}
\end{enumerate}
Note that the normalization process (\ref{eq:norm}) does not spoil any symmetries that the unnormalized fields have, while the expansion coefficients $\alpha_{hkl}$ and $\beta_{abc}$ lose their meaning as Fourier coefficients in the normalized fields.

Without loss of generality, we locate the skyrmions to the FCC crystal as illustrated in Fig.~\ref{fig:fccdis} by letting ${\phi}_0=-1$ and ${\phi}_i(i=1,2,3)=0$ at the face centers and vertices of the lattice. To have a baryon number one configuration at per site, the skyrmion field components should satisfy the constraints ${\phi}_0=+1$ and ${\phi}_i(i=1,2,3)=0$ at some points such as those in the middle of the lines connecting two skyrmions. Such a configuration indicates that $\bar{\beta}_{abc}$ satisfies \begin{equation}
\sum_{a,b,c=\mbox{\scriptsize even}} \bar{\beta}_{abc}=0,
\end{equation}
which can be easily read off by considering ${\phi}_0=-1$ at $(L,0,0)$.

Note that if the mode appearing in the expansion are only that $h$ is odd and $a,{}b,{}c$ are all odd, the configuration is invariant under the translation $\vec{x}\rightarrow (x+L,y,z)$ accompanied by the field rotating under $O(4)$ by $\pi$ in the $\sigma, \pi_1$ plane. This configuration is nothing but the CC crystal of half-skyrmions. This additional symmetry indicates that, around the points where $\sigma$ takes values $\pm 1$, some physical quantities, such as the  baryon number density and the energy density are identical. Thus, the distribution of the baryon number is splitted and one-half of it is concentrated at the centers of skyrmions at the original FCC while the other half is concentrated in the middle of the links connecting FCC vertices.
In the middle of the links, $\sigma$ takes the value $+1$ and the baryon number density is high in the FCC configuration. As a consequence of this new distribution of baryon number, the space average value $\langle \sigma \rangle$ vanishes and we will see that this phenomenon signals that, in the dense medium, chiral symmetry is globally restored. It is important to stress that it is the precise structure of the ground state which is responsible for the restoration.  In the calculation, all the modes satisfying (M\ref{M1}-M\ref{M3}) are included and the half-skyrmion crystal arises as the stable ground state at high density where the expansion coefficients associated with other modes are suppressed.

In the Fourier series expansion method, although there are infinite numbers of modes, only a few of them are necessary since the expansions converge rapidly. The expansion coefficients are used as variational parameters and determined by minimizing the energy of the configuration. The coefficients depend on the box size $L$. In Table~\ref{Tab:mode}, we list a few modes below $E=16(\pi/L)^2$ and we use only the modes with $E \le 10(\pi/L)^2$ in our calculation.

The energy and the degeneracy of the modes listed in Table~\ref{Tab:mode} can be understood as follows: The energy of the modes of $\phi_0$ and $\phi_i$ can be calculated as $E_{\phi_0} = (a^2 + b^2 + c^2)(\pi/L)^2$ and $E_{\phi_i} = (h^2 + k^2 + l^2)(\pi/L)^2$, respectively. Since $a, b, c$ should be all even numbers or odd numbers, the lowest energy mode must be given by $(a, b, c) = (0,0,0)$ and the degeneracy is $1$ and energy of the mode is $0$. The next energy is given by the mode with $(a, b, c) = (1,1,1)$ with $E_{\phi_0} = (1^2 +1^2 + 1^2)(\pi/L)^2 = 3 (\pi/L)^2$ with degeneracy $d =1$. For the mode $(a, b, c) = (2,0,0)$, the energy is $E_{\phi_0} = (2^2 +0^2 + 0^2)(\pi/L)^2 = 5 (\pi/L)^2$ and, because of the property~(M\ref{M1}), $d = 3$. Similar argument can be applied to the modes of $\phi_i$.

\begin{table}
\caption{\label{Tab:mode}A few modes used in the Fourier series expansion
coefficients $\alpha_{hkl}$ and $\beta_{abc}$ in
eqs.~(\ref{sigma}-\ref{pi3}). $E$ is the energy of the modes in
units $(\pi/L)^2$  and $d$ is the degeneracy of the mode.}
\begin{center}
\begin{tabular}{ccccc|ccccc}
\hline
\; $h$\; &\; $k$ \;&\; $l$ \;&\; $E$ \;&\; $d$ \;&\; $a$ \;&\; $b$ \;&\; $c$ \;&\; $E$ \;&\; $d$\; \\
\hline
 1 & 0 & 0 & 1 & 1 & 0 & 0 & 0 & 0 & 1 \\
 1 & 2 & 0 & 5 & 2 & 1 & 1 & 1 & 3 & 1 \\
 2 & 1 & 1 & 6 & 1 & 2 & 0 & 0 & 4 & 3 \\
 1 & 2 & 2 & 9 & 1 & 2 & 2 & 0 & 8 & 3 \\
 3 & 0 & 0 & 9 & 1 &   &   &   &    &   \\
\hline
 3 & 2 & 0 & 13 & 2 & 3 & 1 & 1 & 11 & 3 \\
 2 & 3 & 1 & 14 & 2 & 2 & 2 & 2 & 12 & 1 \\
\hline
\end{tabular}
\end{center}
\end{table}

\begin{figure}[htbp]\centering
\subfigure[]
{
\includegraphics[scale=0.27]{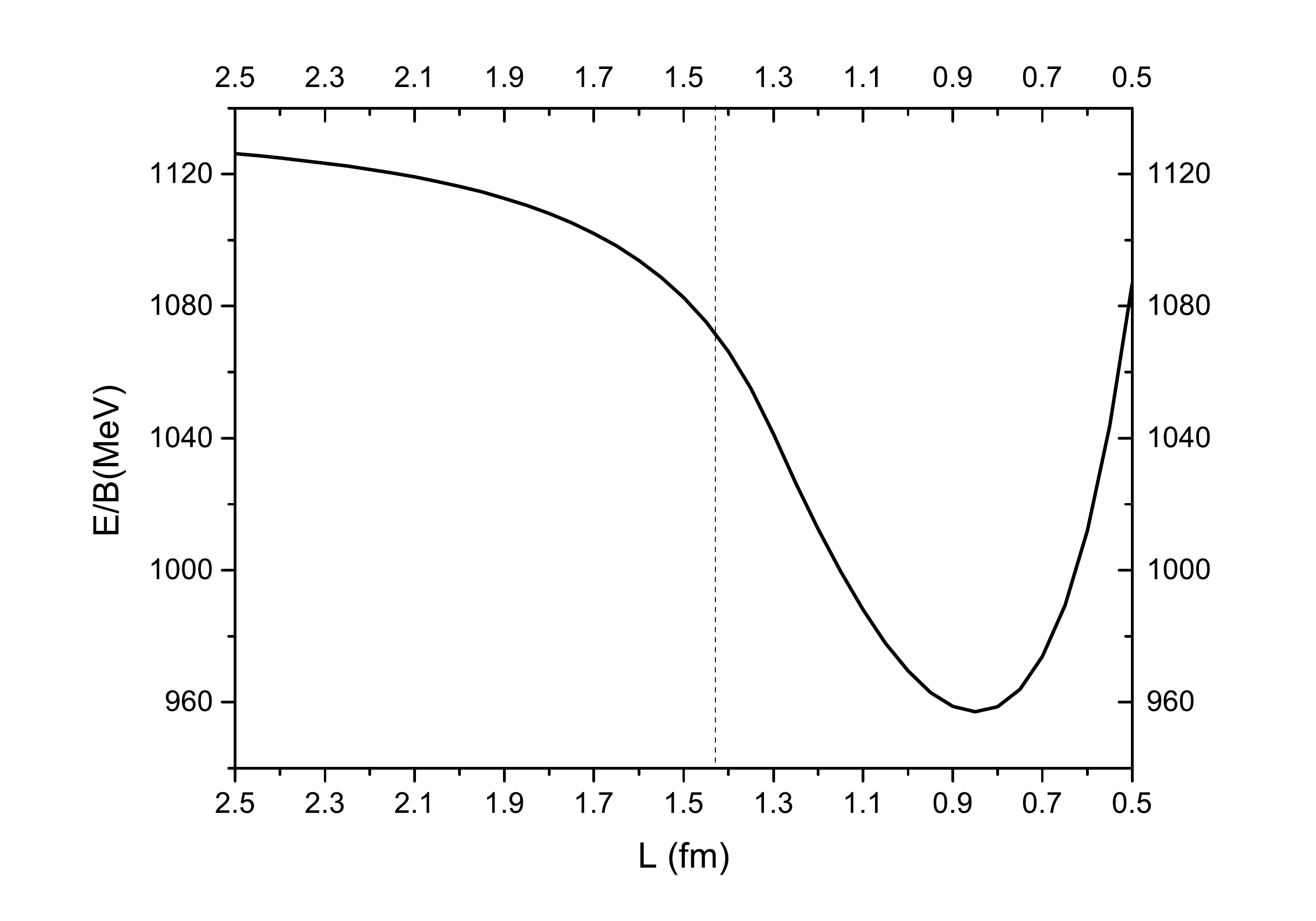}
}
\subfigure[]
{
\includegraphics[scale=0.27]{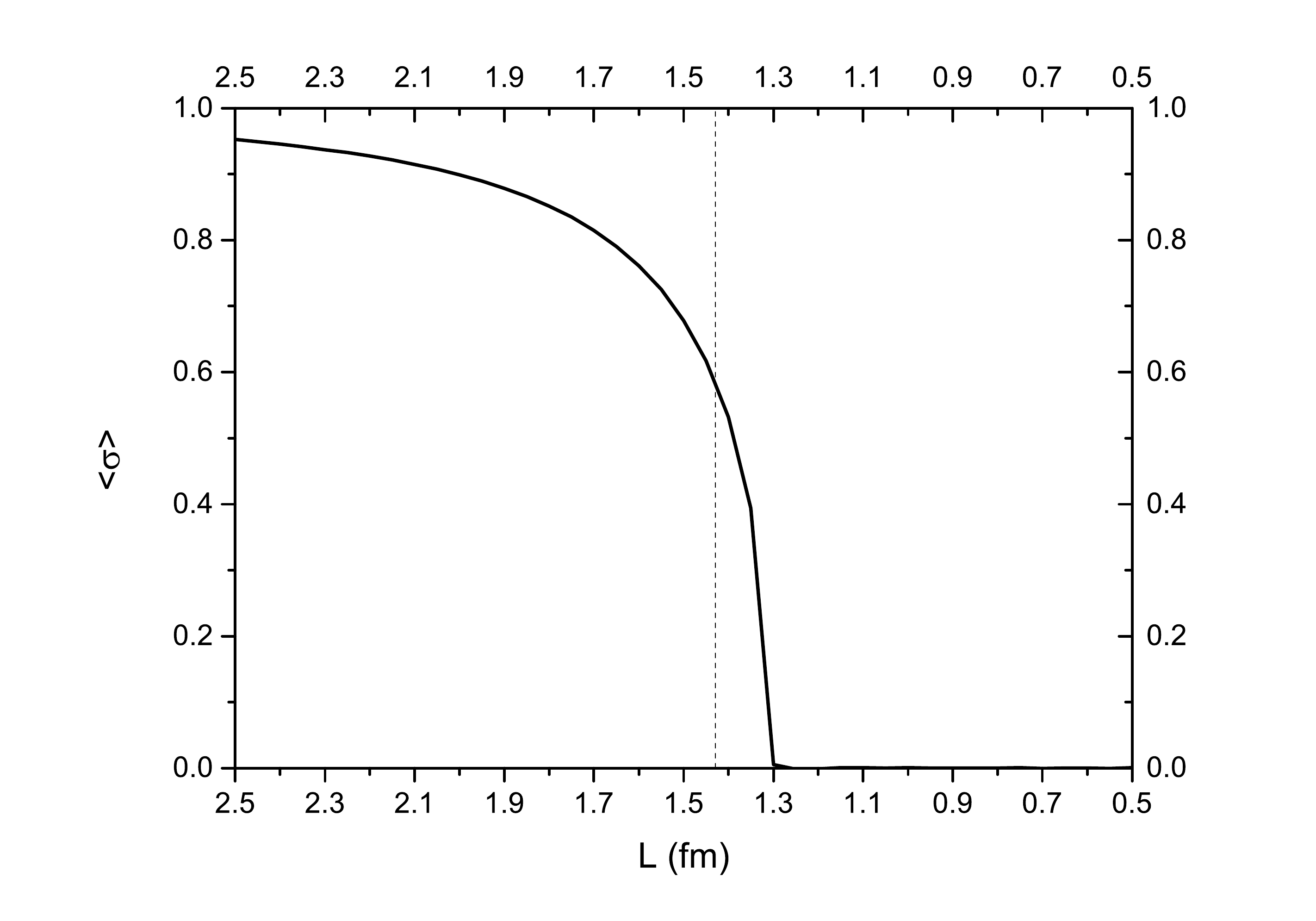}
}
\caption{Per-skyrmion energy $E/B$ and $\langle\sigma\rangle$ as a function of crystal size $L$.}
\label{fig:fccfuncL}
\end{figure}

In Fig.~\ref{fig:fccfuncL} we show the per-skyrmion energy $E/B$ and $\langle\sigma\rangle$ as a function of crystal size $L$. The vertical line indicates the normal nuclear density. The figure shows that as we squeeze the system from $L=2.5$\,fm to around $L=1.3$\,fm, the skyrmion system undergoes a phase transition from the FCC skyrmion configuration to the CC half-skyrmion configuration.~\footnote{Exactly speaking, this is not a phase transition but a topology transition~\cite{Harada:2016tkf}.}
 The system has a minimum energy at $L\sim0.85$\,fm
with the energy per baryon $E/B\simeq 957$\,MeV. From $\langle \sigma \rangle$ we see that the average value of $\sigma$ over space rapidly drops as the system shrinks. It reaches  zero at $L\sim 1.3~$fm where the system goes to the half-skyrmion phase. This phase transition can be interpreted, once the pion fluctuations are incorporated, as a signal for global chiral symmetry restoration although locally the system is still in the chiral symmetry breaking phase.~\footnote{The value at which the skyrmion matter transfer to the half-skyrmion phase obtained here is different from that obtained in Ref.~\cite{Lee:2003aq} beacuse of the different Skyrme parameters applied. If we take the empirical value $e = 4.75$, the skyrmion to half-skyrmion phase transition appears above the normal nuclear density.}

\subsection{Pion properties in the skyrmion matter}

As we emphasized before, one of the advantages of the Skyrme model is that, by using this model, the medium modified meson properties can be explored~\cite{Lee:2003aq,Lee:2003eg}. Here we give a concrete calculation of the medium modified pion decay constant $f_\pi^{\ast}$.

Taking the skyrmion crystal solution as background classical fields, we can
interpret the fluctuating fields on top of it as the corresponding mesons in
dense baryonic matter. Following the procedure proposed in Ref.~\cite{Ma:2013ela}, we write the minimum energy solution as $U_{(0)}$ and introduce the fluctuating fields as
\begin{eqnarray}
U^{} &=&u_{(0)}\check{U}u_{(0)}, \label{eq:fieldfluct}
\end{eqnarray}
where $\check{U} = \exp (2i \tau_a \check{\pi}_a/f_\pi)$ stands for the corresponding fluctuating field and $u_{(0)}^2 = U_{(0)}$. It is worth to note that the decomposition given in Eq.~\eqref{eq:fieldfluct} guarantees that to each order of the fluctuation the chiral invariance of the model is preserved. By substituting the fields in Eq.~\eqref{eq:fieldfluct} into the Skyrme model Lagrangian one can obtain the Lagrangian for pion in medium.

To define the pion decay constant in the skyrmion matter, we consider the
axial-vector current correlator
\begin{eqnarray}
i G_{\mu\nu}^{ab}(p) & = & i \int d^4 x \, e^{ip\cdot x}
\left\langle 0 \mid TJ_{5\mu}^a(x)J_{5\nu}^b(0) \mid 0\right\rangle .
\label{eq:defaacorr}
\end{eqnarray}
This correlator can be evaluated from the medium modified Lagrangian by introducing
the corresponding external source by gauging the chiral symmetry, i.e., substituting
the derivative defined in Eq.~\eqref{eq:Lsk} with
\begin{eqnarray}
D_\mu U & = & \partial_\mu U - i \mathcal{L}_\mu U + i U \mathcal{R}_\mu ,
\end{eqnarray}
where ${\cal L}_\mu$ and ${\cal R}_\mu$ are introduced as the gauge fields of the
chiral symmetry. The external source for the axial-vector current $J_{\mu 5}$ is a combination
$(\mathcal{R}_\mu - \mathcal{L}_\mu)/2$.

\begin{figure*}[htbp]\centering
\includegraphics[scale=0.6]{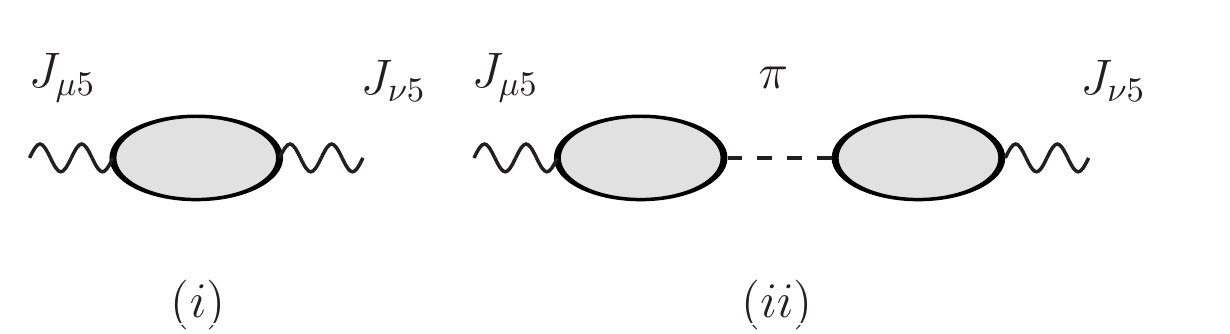}
\caption{Two types of contributions to the correlator of
Eq.~\eqref{eq:defaacorr}: (i) the contact diagram and (ii) the pion exchange diagram. Shaded blobs stand
for interaction vertices in the skyrmion matter.  }
\label{fig:Correlator}
\end{figure*}

In the present calculation, we do not consider the contributions from the loop diagrams of the fluctuation fields to the correlator~\eqref{eq:defaacorr}. Therefore, as illustrated in Fig.~\ref{fig:Correlator}, there are two types of contributions: (i) the contact diagram, and (ii) the pion exchange diagram. In the present evaluation of the correlator, to show our idea, we only consider the matter effect from $u_{(0)L,R}$. In such an approximation two types of contributions are expressed as
\begin{eqnarray}
&\mbox{(i)} : & i f_\pi^2 g_{\mu\nu}^{} \delta^{ab}
\left(1-\frac{2}{3}\left\langle\bm{\phi}_\pi^2\right\rangle \right) ,
\nonumber\\
&\mbox{(ii)} : &
- i f_\pi^2 \frac{p_\mu^{} p_\nu^{}}{p^2} \delta^{ab}\left(1-\frac{2}{3}\left\langle\bm{\phi}_\pi^2\right\rangle \right) .
\end{eqnarray}
Summing over the above two types of contributions, one concludes that the
axial-vector current correlator \eqref{eq:defaacorr} is gauge invariant and
therefore can be decomposed into the longitudinal and transverse parts as
\begin{eqnarray}
G_{\mu\nu}^{ab}(p) &=& \delta^{ab} \left[ P_{T\mu\nu} G_{T}(p)
+ P_{L\mu\nu} G_{L}(p) \right],
\end{eqnarray}
where the polarization tensors $P_{L,T}$ are defined as
\begin{eqnarray}
P_{T\mu\nu}^{} &=&
g_{\mu i}^{} \left( \delta_{ij} - \frac{p_i^{} p_j^{}}{|\bm{p}|^2} \right) g_{j\nu} ,
\nonumber\\
P_{L\mu\nu}^{} &=& - \left( g_{\mu\nu} - \frac{p_\mu^{} p_\nu^{}}{p^2}
\right) - P_{T\mu\nu}^{}.
\end{eqnarray}
We next define the medium modified pion decay constant through the longitudinal
component in the low energy limit
\begin{eqnarray}
f_\pi^{\ast 2} & \equiv & {} - \lim_{p_0 \to 0} G_{L}(p_0,\bm{p}=0) = f_\pi^2 \left [1 - \frac{2}{3}
\left( 1-   \left\langle \sigma^2_{(0)}\right\rangle \right)\right] ,
\label{eq:mediumfpi}
\end{eqnarray}
where the intrinsic density dependence is brought in by the minimal energy solution $\sigma_{(0)}^2$, and the relation
$\sigma_{(0)}^2 + {\bm \phi}_\pi^2 = 1$ has been used. Equation \eqref{eq:mediumfpi} shows the direct relation between the medium modified pion decay constant $f_\pi^{\ast}$ and the parameter $\langle\sigma\rangle$ which signalled the phase transition in the case of skyrmion matter.

\begin{figure*}[htbp]\centering
\includegraphics[scale=0.27]{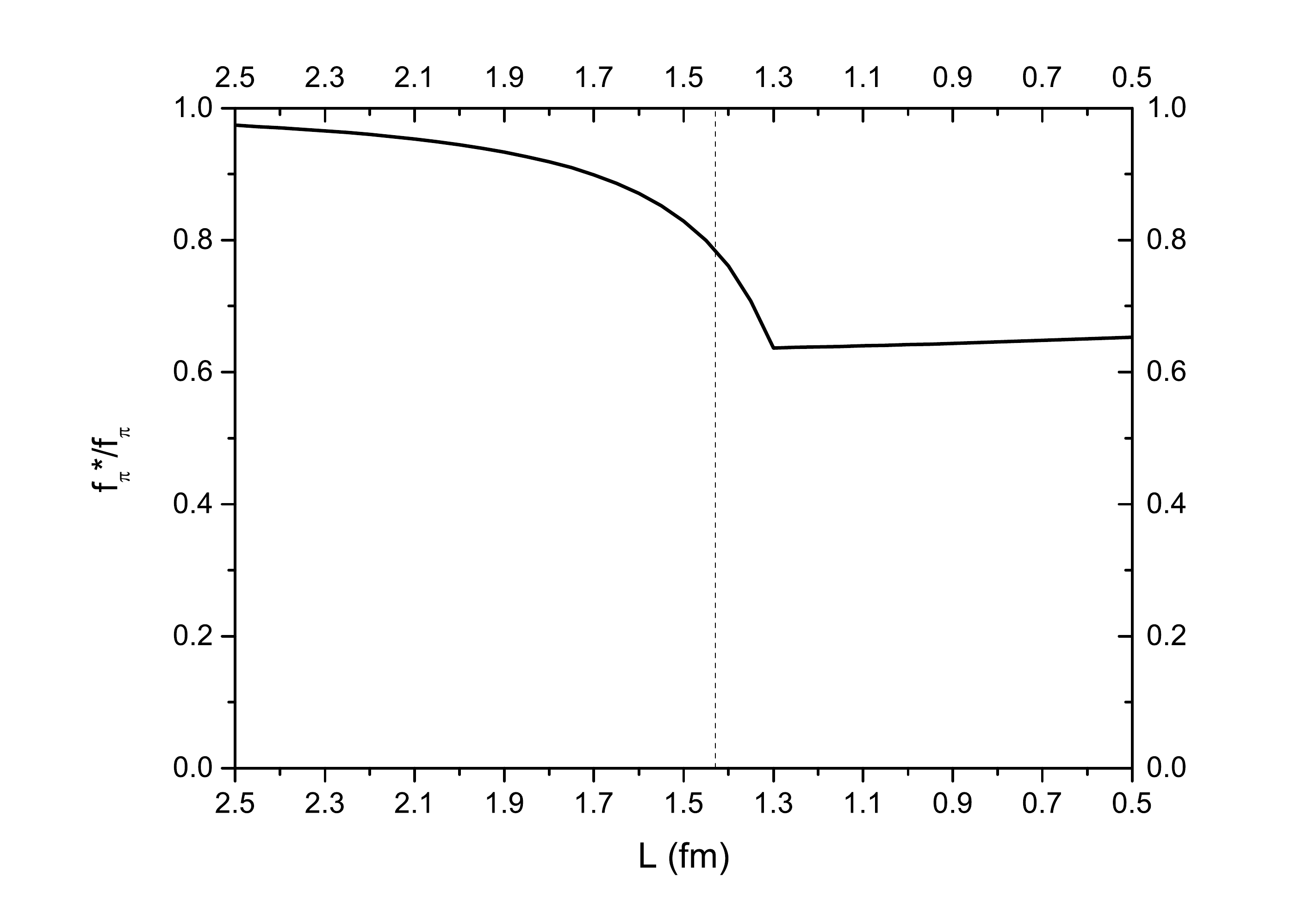}
\caption{$f^*_\pi/f_\pi$ as a
function of the crystal size $L$ .}
\label{fig:inmediumpion}
\end{figure*}

In Fig.~\ref{fig:inmediumpion} we plot the crystal size dependence of $f^*_\pi/f_\pi$. From this plot we see that, as the density
increases, $f^*_\pi$ decreases to $\sim 0.65 f_\pi$ at the density where half-skyrmion phase appear and in the half-skyrmion phase it stays as a nearly constant.

Finally, we want to point out that, although the chiral symmetry is globally restored, it is still locally violated since pion are still there and the pion decay constant is nonzero. Actually, the simulation of the position dependence of the quark-antiquark condensate explicitly shows the local chiral symmetry breaking and also the magnitude of the breaking~\cite{Harada:2015lma}.

\newpage

\section{Baryons as solitons from the Hidden Local Symmetry}

\label{sec:skyrhls}

Although the chiral symmetry breaking scale is estimated to be $\Lambda_\chi \simeq 1.1~$GeV~\cite{Manohar:1983md}, chiral perturbation theory cannot describe the low energy processes of pions quite well in the whole energy region below $\Lambda_\chi$, even the loop corrections are included. For example, in the $P$-wave
$\pi$-$\pi$ scattering, the data shows a sharp peak at about $770$~MeV which indicates the existence of $\rho$ meson.
However, it is difficult to reproduce such a peak at one-loop level in the chiral perturbation theory.
This observation shows that to describe the $P$-wave
$\pi$-$\pi$ scattering data below $\Lambda_\chi$, not only the pion but
also other hadrons such as the rho meson should be included.

Moreover, lessons from nuclear physics tell us that the vector mesons, such as the rho and omega mesons, play essential roles in the nuclear force, for example the tensor force and repulsive force between nucleons (for a recent review, see, e.g., Ref.~\cite{Rho:2014dea}). So that, if one wants to explore the nuclear properties by using the Skyrme model, the vector meson effects should be included. This also motivates us to include vector mesons in the pion models.

In the literature, there are several theoretical frameworks for exploring the hadron dynamics including vector mesons~\cite{Harada:2003jx}. Here, we shall concern one of them, the hidden
local symmetry (HLS) approach~\cite{Bando:1984ej}. We first discuss the main idea of the HLS which was reviewed in Refs.~\cite{Bando:1987br,Harada:2003jx}. Then we study the skyrmion properties by using the leading order Lagrangian of the HLS.

\subsection{Basics of the hidden local symmetry}

The idea of the HLS is the following: A nonlinear sigma model based on the manifold $G/H$ is gauge equivalent to a model having the symmetry $G_{\rm global}\times H_{\rm local}$. After dynamical breaking of the local symmetry $H_{\rm local}$, the gauge boson of the local symmetry $H_{\rm local}$ is identified as the massive vector mesons. Here we consider the case in which $G_{\rm global} = [{\rm SU}(2)_L \times {\rm SU}(2)_R]_{\rm global}$ and $H_{\rm local} = [{\rm SU}(2)_V]_{\rm local}$. Integrating out the vector mesons from this model and keeping only the minimal number derivative terms one can get the two-flavor nonlinear sigma model.

In the HLS, we decompose the field $U(x)$ as
\begin{equation}
U(x)=\xi_L^\dag(x)\xi_R(x).
\end{equation}
Due to this decomposition, one can insert a gauge symmetry $H_{\rm local}$ by requiring that $\xi_{L,R}$ have the transformation
\begin{equation}
\xi_{L,R}(x)\mapsto\xi_{L,R}^\prime(x) =
h(x)\xi_{L,R}(x)g_{L,R}^\dag,
\label{hidden-trans}
\end{equation}
where $h(x) \in H_{\rm local}$. The variables $\xi_{L,R}$ can be parameterized as
\begin{equation}
\xi_{L,R}(x)=e^{i\sigma(x)/(2f_\sigma)}e^{\pm
i\pi(x)/(2f_\pi)},~~~~\pi(x)=\pi^aX^a,~~~~\sigma(x)=\sigma^\alpha
S^\alpha , \label{parameHidden}
\end{equation}
where $S^a$ is the generator of the unbroken subgroup $H$ of $G_{\rm
global}$, here ${\rm SU}(2)$, while $X^a$ is the broken generator.
$\pi$ and $\sigma$ denote the Nambu-Goldstone bosons associated with the spontaneous breaking of $G_{\rm global}$ and $H_{\rm local}$, respectively, and $f_\pi$ and $f_\sigma$ are the corresponding decay constants.

Corresponding to the HLS gauge symmetry ${\rm SU}(2)$ one can introduce the gauge
field $V_\mu(x)$ which transforms as
\begin{eqnarray}
V_\mu(x) \rightarrow V_\mu(x)^\prime = h(x)V_\mu(x) h^\dag(x) - i
\partial_\mu h(x) \cdot h^\dag(x),~~~~h(x)\in {\rm SU}(2)_V.
\end{eqnarray}
With quantities $\xi_{L,R}$ and the introduced gauge bosons $V_\mu$
one can define the following two $1$-forms:
\begin{eqnarray}
\hat{\alpha}_{\parallel\mu} & = & \frac{1}{2i}(D_\mu \xi_R \cdot
\xi_R^\dag +
D_\mu \xi_L \cdot \xi_L^\dag), \nonumber\\
\hat{\alpha}_{\perp\mu} & = & \frac{1}{2i}(D_\mu \xi_R \cdot
\xi_R^\dag - D_\mu \xi_L \cdot \xi_L^\dag)
\ , \label{eq:1form}
\end{eqnarray}
where the covariant derivative is defined as $D_\mu \xi_{R,L} = (\partial_\mu - i V_\mu)\xi_{R,L}$, and both of these quantities transform as $\hat{a}_{\parallel,\perp}^{\mu} \rightarrow h(x) \hat{a}_{\parallel,\perp}^{\mu} h(x)^\dag$.
For the gauge field $V_\mu$ we have the field strength tensor
\begin{eqnarray}
V_{\mu\nu}(x) = \partial_\mu V_\nu(x)-\partial_\nu
V_\mu(x)-i[V_\mu(x),V_\nu(x)],\label{eq:hiddentensor}
\end{eqnarray}
with the transformation property $V_{\mu\nu}(x) \rightarrow h(x) V_{\mu\nu}(x) h(x)^\dag$.

With the two 1-forms defined by Eq.~(\ref{eq:1form}) and field strength tensor (\ref{eq:hiddentensor}), one can construct a Lorentz invariant Lagrangian with the minimal number of derivatives as
\begin{eqnarray}
{\cal L}_{\rm HLS} & = & f_\pi^2 {\rm
Tr}[\hat{a}_{\perp\mu}\hat{a}_{\perp}^{\mu}] + f_\sigma^2{\rm
Tr}[\hat{a}_{\parallel\mu}\hat{a}_{\parallel}^{\mu}] -
\frac{1}{2g^2}{\rm Tr}[V_{\mu\nu}V^{\mu\nu}]. \label{eq:lagrhls}
\end{eqnarray}
In this Lagrangian, vector bosons are massless therefore cannot be identified as the massive $\rho$ mesons in nature. To generate the masses of the gauge bosons, we use the Higgs mechanism and take the the unitary gauge
\begin{eqnarray}
\xi_L^\dag & = & \xi_R \equiv \xi = e^{i\pi/(2f_\pi)}, \;\;\;\; U(x)
= \xi^2(x). \label{eq:hiddenbreaking}
\end{eqnarray}
Then, we finally have the unitary gauged Lagrangian as
\begin{eqnarray}
{\cal L}_{\rm HLS} & = & \frac{f_\pi^2}{4} {\rm
Tr}\left[\partial_\mu U\partial U^\dag \right] + \frac{f_\sigma^2}{4}{\rm Tr}\left[\left(\partial_\mu \xi \cdot \xi^\dag + \partial_\mu
\xi^\dag \cdot \xi -2iV_\mu \right)\left(\partial_\mu \xi \cdot \xi^\dag +
\partial_\mu \xi^\dag \cdot \xi -2iV_\mu \right)\right] \nonumber\\
& & - \frac{1}{2g^2}{\rm
Tr}\left[V_{\mu\nu}V^{\mu\nu}\right]\ .\label{eq:hlsfix}
\end{eqnarray}
This explicitly shows that the gauge bosons $V_\mu$ acquires mass
\begin{eqnarray}
m_V^2 = g^2 f_\sigma^2,
\label{eq:massvhls}
\end{eqnarray}
which has the standard form of the gauge boson mass from the Higgs mechanism.

To explicitly see that, in the HLS, the gauge boson mass arises from the Higgs
mechanism and the Nambu-Goldstone boson eaten by the gauge boson is
$\sigma(x)$ but not $\pi(x)$, we expand $\xi_L$ and $\xi_R$ as
\begin{eqnarray}
\xi_L & = & 1+i\sigma(x)/(2f_\sigma) + i\pi(x)/(2f_\pi) + \cdots,\\
\xi_R & = & 1+i\sigma(x)/(2f_\sigma)  -i\pi(x)/(2f_\pi)+\cdots,
\end{eqnarray}
which leads to
\begin{eqnarray}
f_\sigma^2{\rm Tr}[\hat{a}_{\parallel\mu}\hat{a}_{\parallel}^{\mu}] & = &{} - f_\sigma^2{\rm Tr}\left[V_\mu+{1\over
f_\sigma}\partial_\mu\sigma\right]^2+\cdots.
\end{eqnarray}
From this derivation one can see that there is a $V_\mu$ and $\sigma$ mixing term $V_\mu\partial_\mu\sigma$ in the Lagrangian. This mixing term can be removed from the Lagrangian by a gauge fixing such as the unitary gauge used above. This is the standard Higgs mechanism.

For constructing an effective theory, a consistent power counting mechanism is essential. Here we discuss the power counting mechanism of the HLS following Ref.~\cite{Harada:2003jx}.

Since the masses of $\rho$ mesons are smaller than the chiral symmetry breaking scale $\Lambda_\chi$, the effective theory consisting of vector mesons can be expanded with respect to the ratio $m_\rho/\Lambda_\chi$ and therefore to study the physics at the scale slightly above the vector mesons. It was pointed by H.~Georgi~\cite{Georgi:1989gp,Georgi:1989xy} that, due to the gauge invariance, the systematic expansion including vector meson loops can be made to the HLS, especially when the vector mesons are light. And the practical calculation shows that, although the expansion parameter in the real-life QCD
\begin{equation}
\frac{m_\rho^2}{\Lambda_\chi^2} \sim 0.5 \ \label{expar}
\end{equation}
is not so small, it is valid in reality.

Keeping this discussion in mind, we can summarize the power counting mechanism of the HLS as follows: Similarly to the chiral perturbation theory, the derivative operator is ${\cal O}(p)$, i.e.,
\begin{eqnarray}
\partial_\mu \sim {\cal O}(p).
\end{eqnarray}
To make the power counting of the covariant derivative consistent, the vector field $V_\mu \equiv g \rho_\mu$ should be ${\cal O}(p)$
\begin{equation}
V_\mu = g \rho_\mu \sim {\cal O}(p) \ . \label{V:order}
\end{equation}
Considering the expansion parameter (\ref{expar}) and in the HLS
\begin{equation}
m_\rho^2 = g^2 f_\sigma^2 \sim {\cal O}(p^2) \ ,
\end{equation}
one can regard the gauge coupling constant $g$ as ${\cal O}(p)$, i.e.,
\begin{equation}
g \sim {\cal O}(p) \ , \label{g:order}
\end{equation}
and the power of the vector meson field $\rho_\mu$ as ${\cal O}(1)$.

From the power counting mechanism of the HLS, one can conclude that the Lagrangian we constructed in Eq.~(\ref{eq:lagrhls}) is the leading order Lagrangian. And, one can construct the Lagrangian to the higher orders in case of necessary.

Finally, we study the relation between the HLS and the chiral perturbation theory. To this purpose, one should integrate out the vector mesons from the HLS using the their equations of motion derived from Eq.~(\ref{eq:hlsfix}),
\begin{eqnarray}
2V^b_\mu {\rm Tr}(T^a T^b)=-i{\rm
Tr}\{T^a(\partial_\mu\xi\cdot\xi^\dag+\partial_\mu\xi^\dag\cdot\xi)\}
+ \mathcal{O}(p^3),
\end{eqnarray}
where the $\mathcal{O}(p^3)$ term is from the vector meson kinetic term.~\footnote{When we integrate out the vector meson fields from the model, it intrinsically means that the vector mesons are heavy objects so that $M_V^2$ can not be regarded as an $\mathcal{O}(p^2)$ quantity but a large constant without chiral order. So that, due to Eq.~\eqref{eq:massvhls}, the gauge coupling $g$ does not carry chiral order and therefore the last term in Lagrangian \eqref{eq:lagrhls} is $\mathcal{O}(p^4)$ instead of the original $\mathcal{O}(p^2)$.} Without including the $\mathcal{O}(p^3)$ term one has
\begin{eqnarray}
2V_\mu & = & -
i(\partial_\mu\xi\cdot\xi^\dag+\partial_\mu\xi^\dag\cdot\xi) \label{eq:EOMvector}.
\end{eqnarray}
Substituting Eq.~(\ref{eq:EOMvector}) into the Lagrangian (\ref{eq:hlsfix}) with neglecting the kinetic term of the vector meson, we obtain
\begin{eqnarray}
{\cal L} & = & {f_\pi^2\over4}{\rm Tr}\left\{\partial_\mu
U\partial^\mu U^\dag\right\},
\end{eqnarray}
which is the leading order of the chiral perturbation theory. In this sense, the linear model based on the manifold $G_{\rm global}\times H_{\rm local}$ is gauge equivalent to the nonlinear model based on the manifold $G/H$.

\subsection{Baryons in the Hidden Local Symmetry}

Now let us study baryon physics from the Skyrme model with HLS at the leading order of the chiral counting written by \eqref{eq:lagrhls}~\footnote{From now on, we will rewrite $f_\sigma^2$ as $af_\pi^2$.}
\begin{eqnarray}
{\cal L}_{\rm HLS} & = & f_\pi^2 {\rm
Tr}\left[\hat{\alpha}_{\perp\mu}\hat{\alpha}_{\perp}^{\mu}\right] +
a f_\pi^2{\rm
Tr}\left[\hat{\alpha}_{\parallel\mu}\hat{\alpha}_{\parallel}^{\mu}\right]
- \frac{1}{2g^2}{\rm Tr}\left[V_{\mu\nu}V^{\mu\nu}\right].
\end{eqnarray}
In this subsection, we will take the unitary gauge (\ref{eq:hiddenbreaking}).

First, let us establish the relation between the leading order HLS Lagrangian (\ref{eq:lagrhls}) and the Skyrme model Lagrangian (\ref{eq:lagrskyrme}). Since in the Skyrme model Lagrangian (\ref{eq:lagrskyrme}), there is no vector meson field, one should integrate out the vector meson fields from the HLS Lagrangian (\ref{eq:lagrhls}) by using their equations of motion (\ref{eq:EOMvector}). The calculation in the previous subsection shows that the first two terms of Lagrangian gives the first term of the Skyrme model Lagrangian (\ref{eq:lagrskyrme}). To obtain the Skyrme term, we should substitute the EOM (\ref{eq:EOMvector}) into the third term of the HLS Lagrangian (\ref{eq:lagrhls}). Explicit calculation yields the following result
\begin{eqnarray}
- \frac{1}{2g^2}{\rm Tr}\left[V_{\mu\nu}V^{\mu\nu}\right] & = &
\frac{1}{32 g^2}{\rm Tr}\left[L_\mu, L_\nu\right]^2,
\end{eqnarray}
which has the same structure as the Skyrme term and the relation between the HLS gauge coupling constant $g$ and the Skyrme term parameter $e$ is $e=g$.

After the standard derivation, the Hamiltonian of the HLS can be obtained as
\begin{eqnarray}
\mathcal{H}_{\rm HLS} & = & \frac{\partial {\cal L}_{\rm
HLS}}{\partial \partial_0\xi}\partial_0 \xi + \frac{\partial {\cal
L}_{\rm HLS}}{\partial \partial_0 V_\mu}\partial_0 V_\mu - {\cal
L}_{\rm HLS}.
\end{eqnarray}
Then the energy of the system is decomposed as
\begin{eqnarray}
E^{\rm HLS} & = & \int d^3 x \mathcal{H}_{\rm HLS} \equiv E^{\rm
HLS}_{\rm rotation} + E^{\rm HLS}_{\rm static},
\end{eqnarray}
with
\begin{eqnarray}
E^{\rm
HLS}_{\rm static} & = & \int d^3x \left\{ f_\pi^2 {\rm Tr}\left[ \alpha_{\perp i} \alpha_{\perp i}\right] + a f_\pi^2 {\rm Tr}\left[ \alpha_{\parallel i} \alpha_{\parallel i}\right] + \frac{1}{2g^2}{\rm Tr}\left[ F_{ij}F_{ij}\right]\right\},\nonumber\\
E^{\rm HLS}_{\rm rotation} & = & \int d^3x \left\{ f_\pi^2 {\rm
Tr}\left[ \alpha_{\perp 0} \alpha_{\perp 0}\right] + a f_\pi^2 {\rm
Tr}\left[ \alpha_{\parallel 0} \alpha_{\parallel 0}\right] +
\frac{1}{g^2}{\rm Tr}\left[
\partial_0 V_i \partial_0 V_i
\right]\right\}.\label{eq:energyhls}
\end{eqnarray}

Similarly to the Skyrme model, to study the soliton properties in the HLS, we take the Hedgehog ansatz (\ref{eq:hedgehog}) for the pseudoscalar field, i.e.,
\begin{eqnarray}
\xi_c(\mathbf{x}) & = &
\exp\left[i\bm{\tau}\cdot\hat{\mathbf{x}}\frac{F(r)}{2}\right],
\;\;\;\;\; U_c(\mathbf{x})=\xi_c^2(\mathbf{x}), \ \label{eq:ansatzxi}
\end{eqnarray}
with $\tau_i$ being the Pauli matrices and the subscript $c$ standing for the classical solution.
For the vector mesons, their profile functions can be parameterized as
\begin{eqnarray}
\rho_{i,c}^a = \frac{1}{gr}\epsilon_{ija}\hat{\mathbf{x}}_j G(r)
\label{eq:ansatzv},
\end{eqnarray}
with the boundary conditions
\begin{eqnarray}
G(0) & = & 2, \;\;\;\;\;\; G(\infty) = F(\infty) = 0.
\label{eq:boundaryG}
\end{eqnarray}
The parametrization of \eqref{eq:ansatzv} and boundary conditions \eqref{eq:boundaryG} can be understood by substituting ansatz \eqref{eq:ansatzxi} into Eq.~\eqref{eq:EOMvector}. From the hedgehog ansatz (\ref{eq:hedgehog}) and profile functions (\ref{eq:ansatzv}) we express the quantities $\hat{\alpha}_\perp^{\mu}$ and $\hat{\alpha}_\parallel^{\mu}$ as
\begin{eqnarray}
\hat{\alpha}_{\perp}^\mu & = & \left(0,\bm{a}_\perp \right) =
\left(0, \frac{1}{2}\left[a_1(r)\bm{\tau} +
a_2(r)\left(\bm{\tau}\cdot \hat{\mathbf{x}}\right)
\hat{\mathbf{x}}\right]\right), \nonumber\\
\hat{\alpha}_{\parallel}^\mu & = & \left( 0,\bm{a}_\parallel\right)
= \left(0, \varphi(r)\hat{\mathbf{x}} \times \bm{\tau} \right),
\label{eq:hedg1forms}
\end{eqnarray}
with
\begin{eqnarray}
a_1(r) & = & \frac{\sin F(r)}{r}, \;\;\;\;\; a_2(r) = F^\prime(r) -
\frac{\sin F(r)}{r}, \nonumber\\
\varphi(r) & = & \frac{1}{r}\sin^2\frac{F}{2} - \frac{1}{2r}G(r).
\label{eq:coeffaparaperpen}
\end{eqnarray}

\begin{figure}[htbp]\centering
\subfigure[]
{
\includegraphics[scale=0.27]{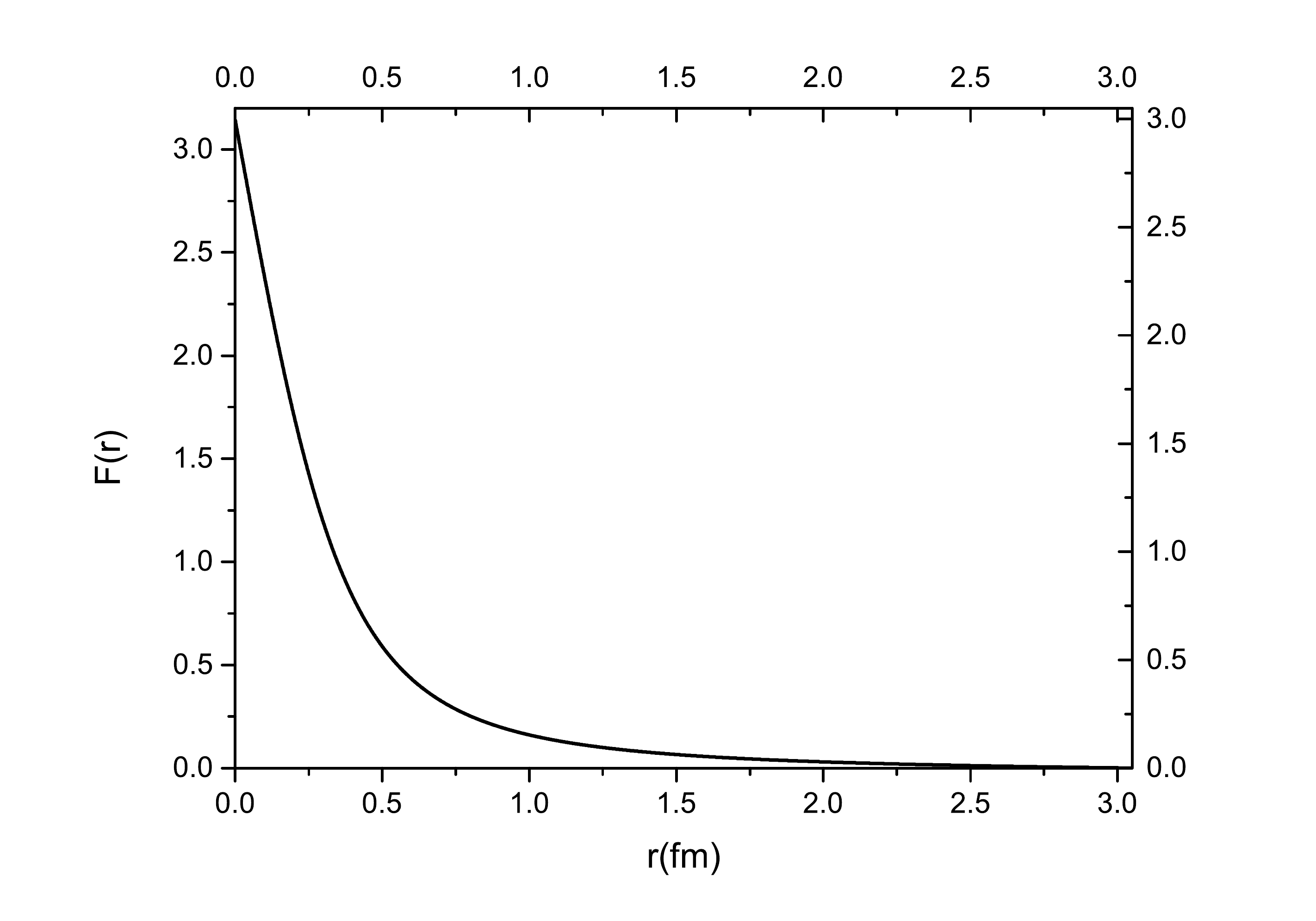}
}
\subfigure[]
{
\includegraphics[scale=0.27]{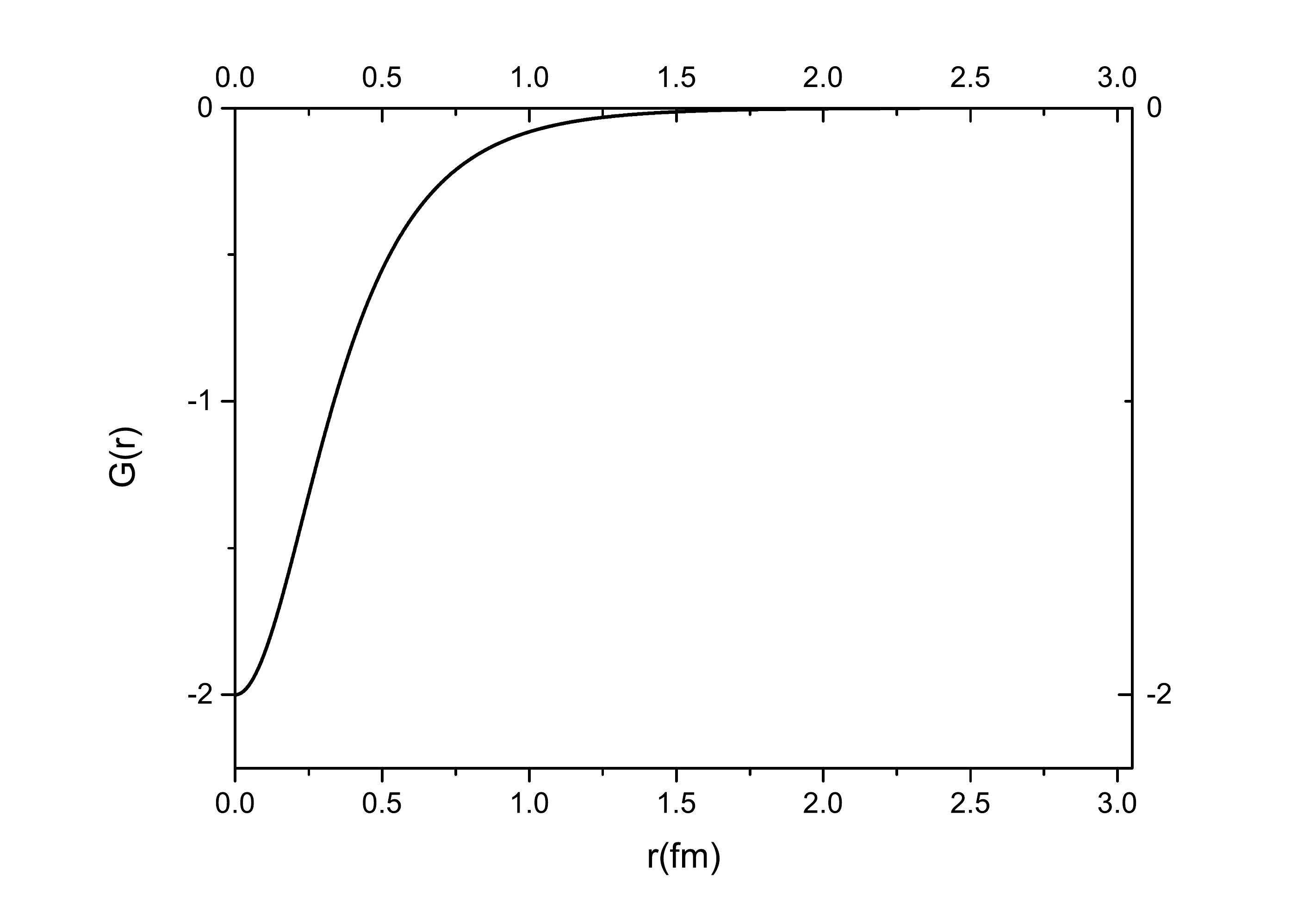}
}
\caption{
Profile functions $F(r)$ and $G(r)$ obtained from $O(p^2)$ HLS.}
\label{fig:SolFG}
\end{figure}

Using the ansatz (\ref{eq:hedgehog}, \ref{eq:ansatzv}) the static energy of HLS which was given in Eq.~(\ref{eq:energyhls}) can be expressed in terms of $F$ and $G$. Explicit derivation yields
\begin{eqnarray}
M_{\rm static}^{\rm HLS} & = & 4\pi\int_0^\infty dr
\Big\{\frac{F_\pi^2}{2}(r^2F^{\prime \, 2} + 2\sin^2F) + aF_\pi^2(G
- 1 + \cos F)^2 \nonumber\\
& & \;\;\;\;\;\;\;\;\;\;\;\;\;\;\;\;\;\; +
\frac{1}{2g^2}\Big[2G^{\prime \, 2} + \frac{1}{r^2}G^2(G -
2)^2\Big]\Big\},\label{eq:statichidden}
\end{eqnarray}
which is the soliton mass including vector meson contribution. By minimizing the skyrmion mass \eqref{eq:statichidden} we obtain the EoMs of the profile functions as
\begin{eqnarray}
r^2 F^{\prime\prime} + 2r F^\prime + (a-1)\sin2F + 2a(G-1)\sin F & =
& 0, \nonumber\\
r^2 G^{\prime\prime} - (m_\rho^2 r^2 + 2)G + 3G^2 - G^3 +
m_\rho^2r^2(1-\cos F) & = & 0, \label{eq:eompv}
\end{eqnarray}
where the vector meson mass relation (\ref{eq:massvhls}) has been used. Note that these two equations are coupled equations so that they should be solved numerically. We plot the numerical solutions in Fig.~\ref{fig:SolFG}.

\begin{figure}[htbp]\centering
\subfigure[]
{
\includegraphics[scale=0.27]{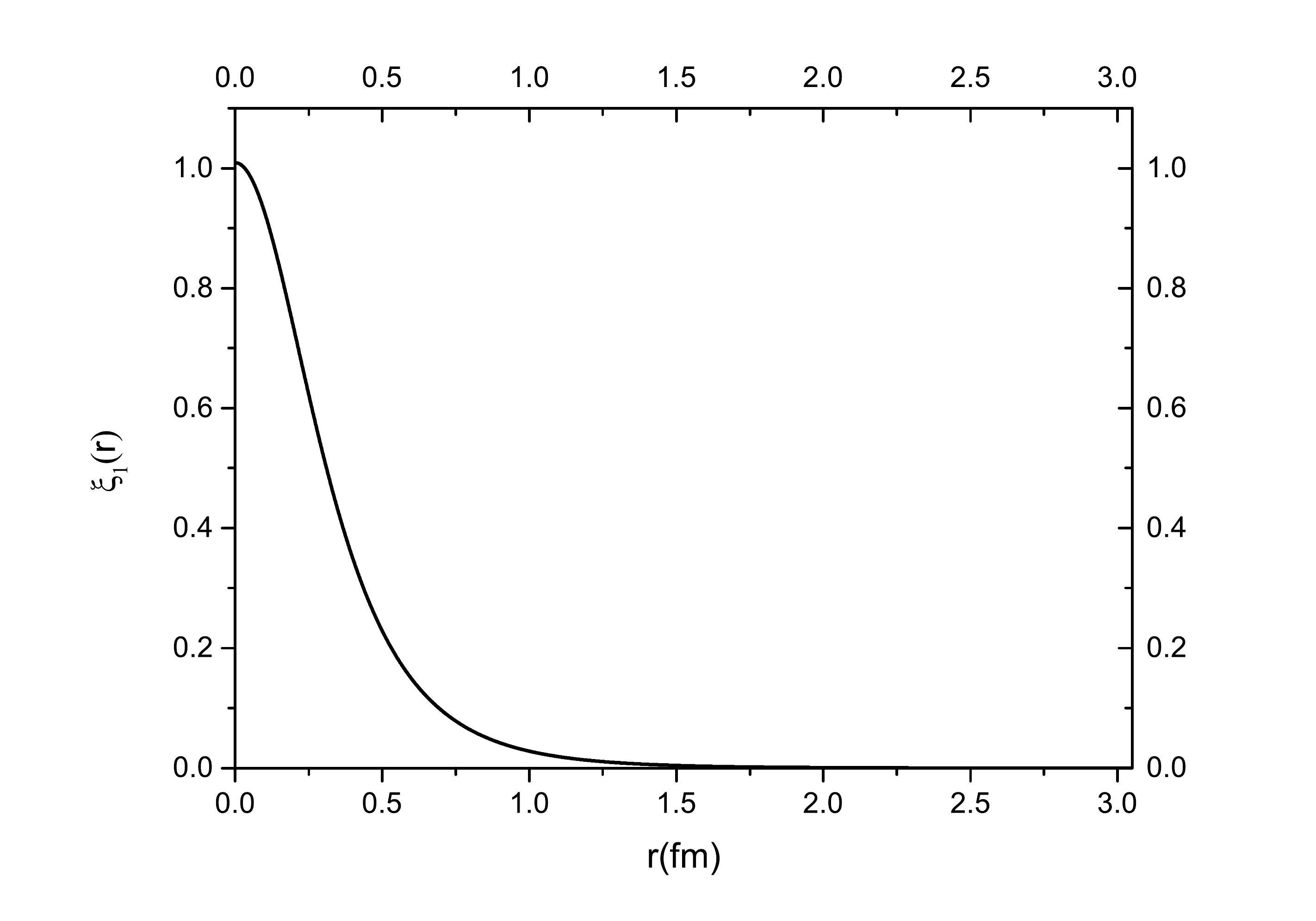}
}
\subfigure[]
{
\includegraphics[scale=0.27]{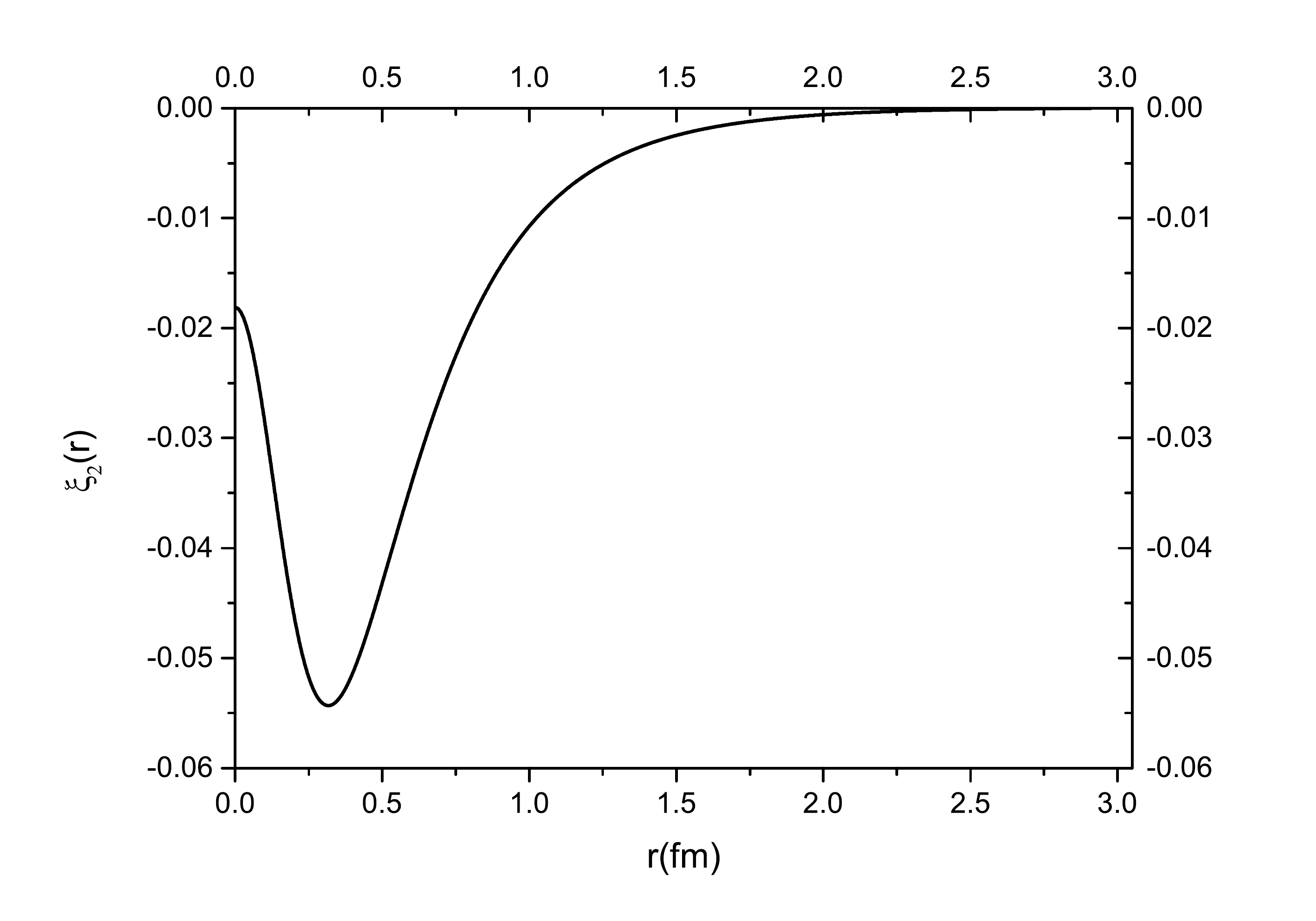}
}
\caption{
Wave functions $\xi_1(r)$ and $\xi_2(r)$ obtained from $O(p^2)$ HLS.}
\label{fig:Solxi}
\end{figure}

To endow the solitons with definite quantum numbers, in addition to the pseudoscalar fields, the vector meson fields should also be collectively rotated by substituting
\begin{eqnarray}
V_{\mu,c}(\mathbf{x}) \rightarrow V_{\mu}(\mathbf{x},t) & = & C(t)
V_{\mu,c} C^\dag(t).\label{eq:collectiverotv}
\end{eqnarray}
where $C(t)$ is a time-dependent SU(2) matrix. We define the angular velocity $\bm{\Omega}$ of the collective coordinate rotation as
\begin{eqnarray}
i \bm{\tau} \cdot \bm{\Omega} & \equiv & C^\dagger(t) \partial_0 C(t).
\label{eq:angularvelocity2}
\end{eqnarray}
Under the rotation (\ref{eq:collectiverotv}), the time component of the $\rho$ meson field gets excited.
The most general form for the vector-meson excitations are written as
\begin{eqnarray}
\rho^0 (\bm{r},t) &=& A(t) \frac{2}{g}\left[ \bm{\tau} \cdot \bm{\Omega} \,
\xi_1^{}(r)
+ \hat{\bm{\tau}} \cdot \hat{\bm{r}} \, \bm{\Omega} \cdot \hat{\bm{r}} \,
\xi_2^{}(r)
\right] A^\dagger(t) .
\label{eq:VM_excited}
\end{eqnarray}
With these discussions we can write the rotation induced energy formally as
\begin{eqnarray}
E_{\rm rotation}^{\rm HLS} & = & \frac{1}{2} \mathcal{I}_{\rm HLS}
\Omega^2,\label{eq:energhlsrot}
\end{eqnarray}
where the angular velocity $\Omega_i$ is defined in Eq.~(\ref{eq:angularvelocity2}) and $\mathcal{I}_{\rm HLS}$ is the moment of inertia of the soliton configuration in HLS which is given by
\begin{eqnarray}
\mathcal{I}_{\rm HLS} & = & 4\pi \int_0^\infty dr
\left\{\frac23 f_\pi^2  r^2\sin^2 F
+ \frac13 a f_\pi^2 r^2 \left[ \left( \xi_1^{} + \xi_2^{} \right)^2
+ 2 \left( \xi_1^{} - 2 \sin^2 \frac{F}{2} \right)^2 \right] \right. \nonumber \\
&& \left. \qquad\qquad\quad \mbox{}
- \frac16 a g^2 f_\pi^2 \varphi^2
- \frac16 \left( \varphi'^2 + \frac{2\varphi^2}{r^2} \right)
+ \frac{r^2}{3g^2} \left( 3 \xi_1'^2 + 2 \xi_1' \xi_2' + \xi_2'^2 \right) \right. \nonumber \\
&& \left. \qquad\qquad\quad \mbox{}
+ \frac{4}{3g^2} G^2 \left( \xi_1^{} - 1 \right)
\left( \xi_1^{} + \xi_2^{} - 1 \right)
+ \frac{2}{3g^2} \left( G^2 + 2 G+2 \right) \xi_2^2\right\}.\label{eq:inemohls}
\end{eqnarray}
From this moment of inertia, we have the following EoMs of the excited fields $\xi_1(r)$ and $\xi_2(r)$ as
\begin{eqnarray}
\xi_1'' &=& - \frac{2}{r} \xi_1' + a g^2 f_\pi^2 \left( \xi_1^{}
- 2 \sin^2 \frac{F}{2} \right)
+ \frac{G^2}{r^2} \left( \xi_1^{} - 1 \right) - \frac{2}{r^2} (G+1) \xi_2,
\\
\xi_2'' & = & - \frac{2}{r} \xi_2' + a g^2 f_\pi^2
\left( \xi_2^{} + 2 \sin^2 \frac{F}{2} \right)
+ \frac{G^2}{r^2} \left( \xi_1^{} + 2 \xi_2^{} - 1 \right)
+ \frac{6}{r^2} (G+1) \xi_2 .
\end{eqnarray}
The boundary conditions imposed on the excited fields are
\begin{eqnarray}
\xi_1'(0) & = & \xi_1^{} (\infty) = 0 , \nonumber\\
\xi_2'(0) & = & \xi_2^{} (\infty) = 0 , \label{eq:BCsexcitation}
\end{eqnarray}
and $\xi_1^{}(r)$ and $\xi_2^{}(r)$ at $r=0$ satisfy the constraint,
\begin{eqnarray}
2\xi_1^{} (0) + \xi_2^{} (0) & = & 2 .
\end{eqnarray}
With the boundary conditions~\eqref{eq:BCsexcitation}, the field excitations can be solved numerically and plotted as Fig.~\ref{fig:Solxi}. By using the numerical solutions of the excited fields, we can obtain the soliton mass and moment of inertia from the $O(p^2)$ HLS as
\begin{eqnarray}
M_{\rm static}^{\rm HLS} & = & 1206.8~{\rm MeV} , \nonumber\\
\frac{1}{\mathcal{I}_{\rm HLS}} & = & 812.5~{\rm MeV}.
\end{eqnarray}

After the standard quantization procedure as was done in the
Skyrme model, the baryon masses can be expressed as
\begin{eqnarray}
m_B & = & M_{\rm static}^{\rm HLS} + \frac{j(j+1)}{2
\mathcal{I}_{\rm HLS}}.
\end{eqnarray}
After numerical calculations of $ M_{\rm static}^{\rm HLS}$ and
$\mathcal{I}_{\rm HLS}$, we can get the numerical values of the
baryon masses.

For the physical applications of the Skyrme model with vector meson
effects, one can follow the method applied in the pure Skyrme model provided
in the previous sections. We will not discuss the details here.

\newpage

\section{Recent Development and Remarks}

\label{sec:remark}

In this lecture notes, we are trying to explain the basic idea of the Skyrme model and some of its physical applications. Our discussion only concerns the two light flavors, the up and down quarks. The topology in this case is clear. However, there are many other topics of the Skyrme model that are not covered in this note due to the purpose of the lecture.

In the literature, the two-flavor Skyrme model is extended to include the third flavor, i.e., the strange quark. Unfortunately, in case that the two-flavor model was naively extended to three-flavor one including the finite quark mass corrections, the Skyrme model runs into difficulties. One finds that the mass relations among the lightest hyperons are wrong. This is maybe because the strange quark is much heavier than the up and down quarks. With respect to the difficulties and concerning the badly broken ${\rm SU}(3)$ symmetry, Callan and Klebanov proposed a bound state approach to the strangeness in the Skyrme model~\cite{Callan:1985hy}. In this approach, the Kaon is bound to the Skyrme soliton and the bound state is collectively quantized as a whole.

Based on the Callan-Klebanov's bound state approach, the properties of the heavy baryons including a heavy quark, such as $\Lambda_c$, can also be studied. In this approach, one regards the heavy-light mesons (mesons include one charm quark or bottom quark) and soliton as the constituents of heavy baryons and can treat the soliton as an infinitely heavy background in the sense of large $N_c$ limit (see, e.g., Ref.~\cite{Harada:2012dm} and references therein). In the study of the heavy baryons with the bound state approach, both the heavy quark symmetry and the chiral symmetry could be imposed in the model construction.

In the original Skyrme model, there are only pions. To take into account the higher resonance contributions, the model should be extended. Moreover, from the effective theory point of view, the Skyrme model consists of the leading order term of ChPT and one of the next to leading order terms of ChPT, so that to make a more precise prediction the Skyrme model should be extended to include the higher order terms of the effective theory.  All these extensions will inevitably introduce more low energy constants to the model which could not be fixed phenomenologically at this moment and therefore the predictions from these extended models are highly parameter dependent. The recently developed AdS/CFT correspondence provides a possible way to fix these many model parameters with a few holographic QCD model parameters which can be fixed by empirical physical values and thus the effects of the higher resonances and higher chiral order terms can be self-consistently analysed~\cite{Ma:2012kb,Ma:2012zm,Ma:2013zpa,Ma:2013ooa}. So far, the systematical exploration of the higher resonance and higer chiral order terms perfomed in Refs.~\cite{Ma:2012kb,Ma:2012zm,Ma:2013zpa,Ma:2013ooa,Ma:2013ela} are in the chiral limit. The expression of the low energy constants in chiral effective theories including explicit chiral symmetry breaking in terms of the holographic QCD model was given in Ref.~\cite{Harada:2014lza} and the exploration of the skyrmion and skyrmion matter properties are in progress.

As mentioned before, the advantage of the Skyrme model is that, by putting skyrmions on to the crystal lattice, both hadron properties in free space and in medium can be investigated simultaneously~\cite{Ma:2013zpa,Ma:2013ooa,Ma:2013ela,Suenaga:2014dia,Suenaga:2014sga}. Therefore one can use this model to study the chiral symmetry restoration~\cite{Ma:2013ooa,Ma:2013ela} and the equation of state (EoS) of baryonic matter and also the neutron star properties~\cite{Lee:2013nca}.

Lessons from nuclear physics tell us that the scalar meson which is an iso-singlet is essential for providing the binding force between nucleons. However, how to include this scalar is highly nontrivial~\cite{Ma:2013ela}. In addition, in the minimal Skyrme model, it was found that Casimir force induced by the loop correction in $S$-channel $\pi$-$\pi$ scattering could contribute about $- 500$\,MeV to the skyrmion mass therefore the calculated nucleon mass from the minimal Skyrme model agrees with its physical value~\cite{Meier:1996ng}. So that it deserved to estimate the Casimir energy in models including higher resonances.

We end up this lecture by saying that although the Shyrme model is proposed long time ago, it is still a powerful tool nowadays in the study of
physics of baryons and baryonic matter.

\newpage
\acknowledgments

\label{ACK}

We would like to thank Hyun-Kyu Lee, Yongseok Oh, Byung-Yoon Park and Mannque Rho for their valuable discussions.
The work of Y.-L.~M. was supported in part by National Science Foundation of China (NSFC) under
Grant No.~11475071, 11547308 and the Seeds Funding of Jilin University. M.~H. was supported in part by the JSPS Grant-in-Aid for Scientific Research (C) No. 24540266 and No. 16K05345.


\appendix




\end{document}